\documentclass[useAMS,usenatbib,usegraphicx]{mn2e}
\usepackage{times}

\def\aap{A\&A} 
\def\aaps{A\&AS}
\def\aj{AJ}
\def\apj{ApJ}
\def\apjs{ApJS}
\def\apjl{ApJL}
\def\araa{ARA\&A}
\def\baas{BAAS}
\def\mnras{MNRAS}
\def\nat{Nature}
\def\pasp{PASP}  

\def\kms{\,km\,s$^{-1}$}

\def\hi{{\sc H\,i}}
\def\hii{{\sc H\,ii}}
\def\oiii{{\sc [O\,iii]}}
\def\halpha{H\,$\alpha$}
\def\nii{{\sc [N\,ii]}}

\title[A kinematic survey of PNe in M31]{A deep kinematic survey of planetary nebulae in the Andromeda Galaxy using the Planetary Nebula Spectrograph}
\author[Merrett et al.]
{H. R. Merrett$^{1}$, 
M. R. Merrifield $^{1}$,
N. G. Douglas$^{2}$,
K. Kuijken$^{3,2}$,
A. J. Romanowsky$^{4,1}$,
\newauthor
N. R. Napolitano$^{5,2}$,
M. Arnaboldi$^{6,7}$,
M. Capaccioli$^{8}$,
K. C. Freeman$^{9}$,
O. Gerhard$^{10}$,
\newauthor
L. Coccato$^2$,
D. Carter$^{11}$,
N. W. Evans$^{12}$, 
M. I. Wilkinson$^{12}$, 
C. Halliday$^{13}$, 
T. J. Bridges$^{14}$
\\
$^{1}$School of Physics and Astronomy, The University of Nottingham, NG7 2RD, UK\\
$^{2}$Kapteyn Astronomical Institute, PO Box 800, NL-9700 AV Groningen, The Netherlands\\
$^{3}$Leiden Observatory, PO Box 9513, NL-2300 RA Leiden, The Netherlands\\
$^{4}$Departamento de F\'\i sica, Universidad de Concepci\'on, Casilla
      160-C, Concepci\'on, Chile \\
$^{5}$INAF, Osservatorio di Capodimonte, Via Moiariello 16, Naples
      80131, Italy\\
$^{6}$ESO, Karl-Schwarzschild Str. 2, 85748 Garching bei Munchen, Germany\\
$^{7}$INAF, Osservatorio Astronomico di Torino, Strada Osservatorio
      20, I-10025 Pino Torinese, Italy \\
$^{8}$ Department of Physical Sciences, University of Naples Federico
      II, via Cinthia, 80126 Napoli, Italy\\ 
$^{9}$Research School of Astronomy and Astrophysics, Australian National 
    University, Canberra ACT 2601, Australia\\
$^{10}$Astronomisches Institut, Universit\"at Basel, Venusstrasse 7, 
    CH 4102 Binningen, Switzerland\\
$^{11}$Astrophysics Research Institute, Liverpool John Moores University, 
       Twelve Quays House, Egerton Wharf, Birkenhead CH41 1LD, UK \\
$^{12}$Institute of Astronomy, Madingley Road, Cambridge CB3 OHA, UK\\
$^{13}$INAF, Osservatorio Astrofisico di Arcetri, Largo E. Fermi 5,
       I-50125, Firenze, Italy\\
$^{14}$Department of Physics, Queen's University, Kingston, Ontario,
       K7L 3N6, Canada
}

\begin{document}

\date{Accepted 2006 ????? ??. Received 2006 ?????? ??; in original form 2006 January ??}

\pagerange{\pageref{firstpage}--\pageref{lastpage}} \pubyear{2006}

\maketitle

\label{firstpage}


\begin{abstract}
We present a catalogue of positions, magnitudes and velocities for
3300 emission-line objects found by the Planetary Nebula Spectrograph
in a survey of the Andromeda Galaxy, M31. Of these objects, 2615 are
found likely to be planetary nebulae (PNe) associated with M31. The
survey area covers the whole of M31's disk out to a radius of
1.5\degr. Beyond this radius, observations have been made along the
major and minor axes, and the Northern Spur and Southern Stream
regions.  The calibrated data have been checked for internal
consistency and compared with other catalogues.  With the exception of
the very central, high surface brightness region of M31, this survey
is complete to a magnitude limit of $m_{5007}\sim23.75$, 3.5
magnitudes into the planetary nebula luminosity function.

We have identified emission-line objects associated with M31's
satellites and other background galaxies.  We have examined the data
from the region tentatively identified as a new satellite galaxy,
Andromeda~VIII, comparing it to data in the other quadrants of the
galaxy. We find that the PNe in this region have velocities that
appear to be consistent with membership of M31 itself. 

The luminosity function of the surveyed PNe is well matched to the
usual smooth monotonic function.  The only significant spatial
variation in the luminosity function occurs in the vicinity of M31's
molecular ring, where the luminosities of PNe on the near side of the
galaxy are systematically $\sim 0.2$ magnitudes fainter than those on
the far side.  This difference can be explained naturally by a modest
amount of obscuration by the ring.  The absence of any difference
in luminosity function between bulge and disk suggests that the sample
of PNe is not strongly populated by objects whose progenitors are more
massive stars.  This conclusion is reinforced by the excellent
agreement between the number counts of PNe and the R-band light.  

The number counts of kinematically-selected PNe also allow us to probe
the stellar distribution in M31 down to very faint limits.  There is
no indication of a cut-off in M31's disk out to beyond four
scale-lengths, and no signs of a spheroidal halo population in excess
of the bulge out to 10 effective bulge radii.  

We have also carried out a preliminary analysis of the kinematics of
the surveyed PNe.  The mean streaming velocity of the M31 disk PNe is
found to show a significant asymmetric drift out to large radii.
Their velocity dispersion, although initially declining with radius,
flattens out to a constant value in the outer parts of the galaxy.
There are no indications that the disk velocity dispersion varies with
PN luminosity, once again implying that the progenitors of PNe of all
magnitudes form a relatively homogeneous old population.  The
dispersion profile and asymmetric drift results are shown to be
mutually consistent, but require that the disk flares with radius if
the shape of its velocity ellipsoid remains invariant.
\end{abstract}



\section{Introduction}

Observations of the distribution of stellar velocities in the solar
neighbourhood of the Milky Way reveal a wealth of complex structure
\citep{dehnen98}.  For a dynamical entity such as the Galaxy, this
structure is as fundamental as the spatial arrangement of its stars,
and should contain a mass of information as to the system's current
state, as well as archaeological clues as to how it formed.
Unfortunately, such data from a single locality are difficult to
interpret in terms of the global structure of the galaxy.  Although
kinematic observations of more distant stars in the Milky Way have
been made \citep[see, for example,][]{sevenster97}, the complexity of
disentangling the geometry of our own galaxy make such data difficult
to interpret.  We also clearly need more than one example before any
general conclusions about the dynamics of disk galaxies can be drawn.
An obvious next target is therefore the Milky Way's nearby sister
system, the Andromeda Galaxy, M31.

However, studying the detailed kinematics of stars in even a nearby
external galaxy is quite challenging.  In fact, the first problem
arises from its very nearness: M31 subtends such a large angle on the
sky that many spectroscopic instruments are unable to survey the whole
system within a realistic time.  Further, even at what by
extragalactic standards is the very modest distance of $785\pm25\,{\rm
kpc}$ \citep{mcconnachie05}, obtaining high-quality spectra of
individual stars is challenging; it is only relatively recently that
the requisite observations of even bright red giant branch (RGB) stars
have been made \citep{reitzel02, ibata04, kalirai06}.

Planetary nebulae (PNe) have long been recognised as a potentially
simpler dynamical tracer of the stellar population.  Stars evolve very
rapidly but quite gently from the RGB phase to become PNe, so the
kinematics of these two populations should be essentially identical.
Further, the presence of strong emission lines in PNe make them quite
easy to identify, and render the measurement of their Doppler shifts
fairly trivial.  More than twenty years ago, \citet{nolthenius84}
reported on kinematic measurements of 34 PNe in M31, but it was not
until recently with the development of wide-field multi-object
spectrographs that it has become possible to start to obtain the large
sample necessary for dynamical studies \citep{hurleyk04, halliday06}.
Even with such instruments, these studies involve both narrow-band
imaging to identify candidate PNe and subsequent spectral follow-up,
and this complex two-stage process has limited astronomers' enthusiasm
for such projects.  However, the recently-commissioned Planetary
Nebula Spectrograph has reduced this process to a single stage,
simultaneously identifying PNe and measuring their velocities.  We
have therefore used this instrument to carry out a deep kinematic
survey of PNe over $\sim6$s square degrees, covering the bulk of M31's
visible area.

This paper describes the survey, tabulates the final data set of 3300
detected emission-line objects (of which 2730 are probably PNe, 2615
in M31), and presents some initial analysis of these data.  In
Section~\ref{Sobs}, we briefly describe the instrument and the
observations made. A detailed description of the data analysis is
presented in Section~\ref{Sdata}. Section~\ref{Scomp} compares the
photometric and kinematic results with those from previous smaller
surveys, while Section~\ref{Sconcom} explores the contamination of the
data by objects other than PNe and the sample's completeness.  The
catalogue resulting from all this analysis is presented in
Section~\ref{Scat}.  In Section~\ref{Ssats}, we identify PNe
associated with M31's satellites and other galaxies in the field, both
to study these systems and to create a ``clean'' M31 sample.  We end
the paper with a preliminary analysis of the resulting sample of M31 PNe:
Section~\ref{Spnlf} investigates their luminosity function;
Section~\ref{Sscale} compares their spatial distribution with that of
the starlight in the galaxy; Section~\ref{Salpha} calculates the
luminosity specific number density of PNe; and Section~\ref{Skin}
examines the basic 
kinematic properties of M31's disk out to large radii that can be
gleaned from these data.  We draw conclusions and look to the future
in Section~\ref{Sconc}.


\section{Instrumentation and observations}\label{Sobs}

\subsection{The Planetary Nebula Spectrograph}

\begin{figure*}
  \includegraphics[width=\textwidth]{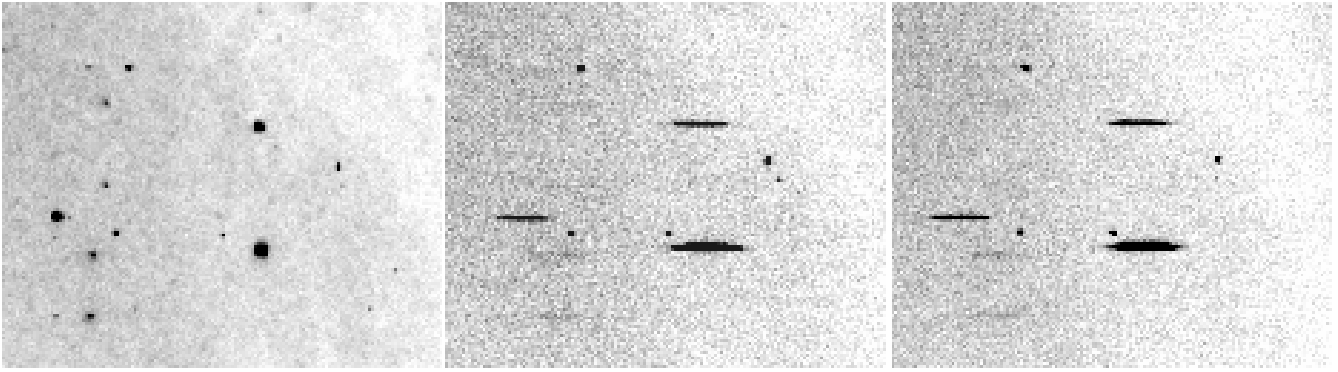}
  \caption{A sample of PN.S data.  The left hand panel shows a
    conventional \oiii\ 
    image of a small section of M31 \citep{massey02}, while the two 
    right-hand panels show the same field as seen through the two arms
    of PN.S.  Note how continuum objects are dispersed into 
    ``star trails'' while \oiii\ emission-line objects remain as
    shifted point sources. 
  }\label{pns_images}
\end{figure*}

The Planetary Nebula Spectrograph (PN.S), described in detail in
\citet{douglas02}, was designed specifically to measure the shifted
wavelength (and hence velocity) of the 5007~\AA\ \oiii\ emission line
that dominates the spectra of PNe. It makes this measurement using a
technique called ``counter-dispersed imaging.''  Light from the galaxy
initially passes through a narrow-band filter ($\sim 35$~\AA\ wide)
centred on the emission line.  The light is then divided into two
beams, with each beam dispersed by a grating before being imaged onto
separate CCDs, termed the left- and right-arm images.  As illustrated
in Figure~\ref{pns_images}, we thus obtain a pair of slitless
spectrographic images in which stellar continua appear as elongated
streaks, or ``star trails'', and PNe are visible as dots from their
\oiii\ emission lines. However, the two gratings are arranged to 
disperse the light in opposite directions, so the dots are offset from
the true location of the PN in opposite directions as well.  Thus, in
essence, the mean location of the dots in the two images gives the
position of a PN, while their separation yields its velocity.

The instrument's ability to obtain all the necessary information from a
single observation, combined with its relatively large field of
view,\footnote{Mounted on the William Herschel Telescope using the
Isaac Newton Group's EEV CCDs, the PN.S has a field of 10\farcm39 by
11\farcm35, with pixel scales of 0\farcs301 per pixel in the dispersed
direction and 0\farcs272 per pixel in the non-dispersed direction, and
a spectral dispersion of 0.775~\AA\ per 13.5~\micron\ pixel.}
make it a very efficient tool for studying PN kinematics in external
galaxies.  Its main drawback relative to the conventional approach of
optical identification and follow-up spectroscopy is that the
attainable velocity accuracy is somewhat lower at $\sim$15--20~\kms,
compared with the $\sim$2--10~\kms\ obtained from traditional spectroscopy
\citep{halliday06, hurleyk04, ciardullo04}.  However, this level of
accuracy is good enough to study even quite subtle kinematic
properties of external galaxies.

\subsection{Observations}

\begin{figure*}
  \includegraphics[width=0.8\textwidth]{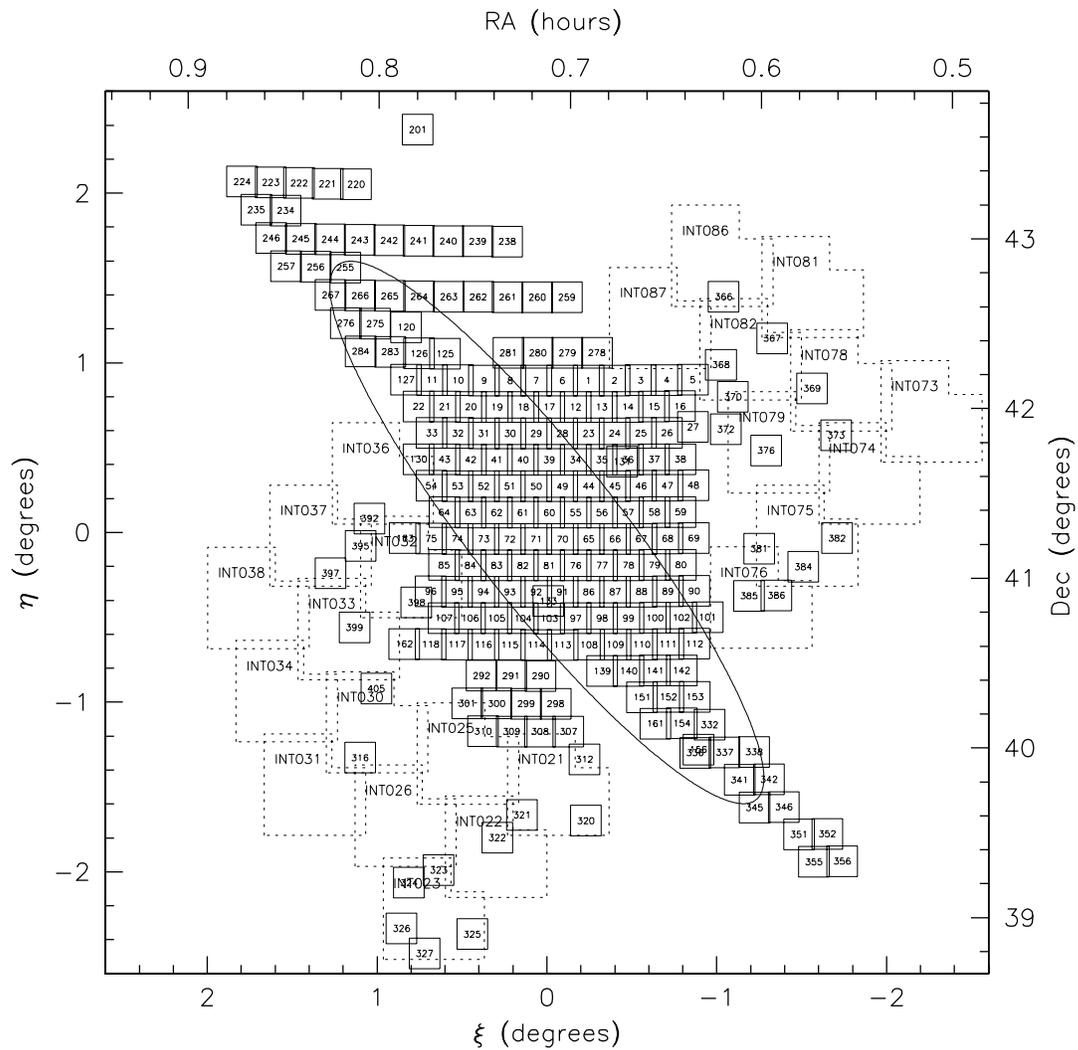}
  \caption{Fields observed in M31. The small square fields are PN.S
  field locations, with numbers 1-163 observed on the first PN.S run,
  the rest on the second; the larger fields with dotted outlines are
  the INT WFC fields. The ellipse marks a 2\degr\ (27.4~kpc) disk
  radius.}
  \label{pns_survey}
\end{figure*}

The PN.S observations for this survey were carried out on the William
Herschel Telescope (WHT) in La Palma on two runs: 2002 October 8 -- 13
and 2003 September 29 -- October 5.  In M31's halo, where PNe are
scarce, supplementary imaging was used to optimise the choice of
fields to be observed with PN.S. This imaging was obtained using the
Wide Field Camera on the Isaac Newton Telescope (INT), also in La
Palma, on 2003 August 16 -- 21. Figure~\ref{pns_survey} shows the
locations of the fields observed with both instruments.

Fifteen minute exposures were used for the majority of PN.S fields,
although multiple observations were made for a few fields. INT WFC
exposures were twenty minutes long through the narrow-band \oiii\
filter and two minutes with the off-band g$'$ filter.  For the PN.S
observations, we used the instrument's `A' filter at a 0\degr\ tilt
angle \citep{douglas02}. This configuration provides a central
wavelength of 5002.2~\AA\ and a 36.5~\AA\ bandwidth, allowing PNe to
be detected with velocities between $-1369$~\kms and $817$~\kms.  This
range is well matched to the velocity range of components of M31,
which lie between $\sim-600$~\kms and 0~\kms, but also allows a
significant margin for the detection of any unexpected high-velocity
features.

The adopted survey strategy was to tile the entire area of M31's disk
out to a 2 degree radius with overlapped PN.S pointings.  We also
fully mapped regions of particular interest at larger radii, such as
the Northern Spur and Southern Stream \citep{ibata01}.  However, at
larger radii PNe become sufficiently scarce that most fields would be
completely empty, so we used the INT narrow-band imaging to identify
candidate PNe, and only made PN.S observations of fields where at
least one object had been detected. In total, 226 PN.S fields were
observed.

\begin{figure}
  \includegraphics[width=0.475\textwidth]{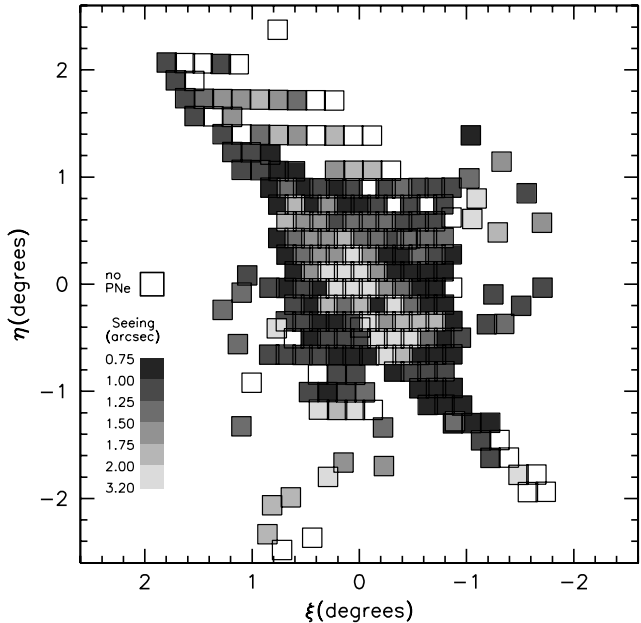}
  \caption{Seeing as measured for different survey fields. Darker
  squares represent better seeing, with values in arcseconds as given
  on the accompanying scale.  Empty squares contained no detected PNe.
  }\label{seeing} 
\end{figure}

Weather conditions varied considerably during the PN.S runs with two
half nights lost to cloud cover during the first run, and four nights
lost in the second. The seeing also changed significantly over the
runs, with values between 0\farcs8 and 3\farcs1 recorded.  To make
optimum use of these varying conditions, we observed the central
fields in the worst seeing conditions (see Figure~\ref{seeing}): PNe
are sufficiently plentiful in these fields that we did not need to get
so far down the luminosity function to obtain a useful dynamical
sample, and the high continuum surface-brightness in this region meant
that we would in any case not be able to achieve completeness at
fainter magnitudes (see Section~\ref{Sconcom}). Confusion with
discrete objects is not a significant problem in surveys of this type
as stars must be fairly bright to leave a visible trail. Even the very
brightest stars at the distance of M31 only leave faint stellar trails
and in the bright central regions they are not visible above the
background light.

\section{Data analysis}\label{Sdata}

\subsection{Wide Field Camera data reduction}

The INT Wide Field Camera data frames were debiased and flat-fielded
using standard data reduction packages in {\sl IRAF}. The g$'$-band
images were remapped to match the coordinates of the \oiii\ images,
and {\sl SExtractor} \citep{bertin96} was used in double-image mode to
find sources in the \oiii\ image and simultaneously measure the flux
of counterparts in the g$'$ band. Objects with more than five times as
many \oiii\ counts as g$'$ counts were selected as potential PN
candidates for PN.S follow-up.  These candidates were then visually
confirmed individually, and their approximate positions were
calculated; more accurate astrometry was not required as the PN.S has
a large field of view.

\subsection{Planetary Nebula Spectrograph data reduction}
Since the PN.S is such an unconventional instrument, we have had to
produce a customised data reduction package to convert the raw data to
a scientifically-usable form.  This package has been constructed
within {\sl IRAF}, supplementing the standard tasks with {\sl Fortran}
and {\sl PGPLOT} extensions.  Given the non-standard nature of this
procedure, we now describe the reduction package and its application
to the M31 observations in some detail.

\subsubsection{Initial pipeline}\label{Spipeline}

Images were first debiased by subtracting a surface function fit to
the pre- and over-scan regions using the {\sl IRAF} task {\sl
imsurfit}. There was very little structure in the bias of the EEV CCDs,
so the function subtracted was close to flat.

To map out and remove bad regions on the CCDs, zero-exposure bias
frames were used to create a map of pixels where charge transfer
problems have occurred.  Skyflats were employed to locate other
low-sensitivity pixels (by comparing to a median smoothed version;
pixels with greater than 9$\sigma$ difference between the two were
taken to be bad).  These maps were then combined with a list of known
bad pixel regions to make a mask which is fed to the standard {\sl
IRAF} routine {\sl fixpix} to correct all the flats and science
images.

Normalised pixel response maps were made by dividing flats by a
median-filtered image (to remove spatial structures) then combining
the flats using weights (to keep the noise Poissonian) and rejection
(to eliminate cosmic ray events). The science and calibration images were
flatfielded by dividing through by this response map.

The next step is to remove any residual cosmic rays: if they were left
in through the next stages that remap pixels, then they would become
smeared out, and could well mimic the point-spread function of a PN,
so it is important to eliminate them at this stage.  The cosmic rays
were identified and removed using the Laplacian cosmic ray
identification routine, {\sl lacos\_im} \citep{vandokkum01}, with five
iterations and a contrast limit between the cosmic rays and the
underlying object of 1.2. This produces fairly clean images and any
residual cosmic ray events are unlikely to be a problem as there would
have to be a pair of residuals in the two images in locations such that
they could be a PN.

\begin{figure}
  \includegraphics[width=0.475\textwidth]{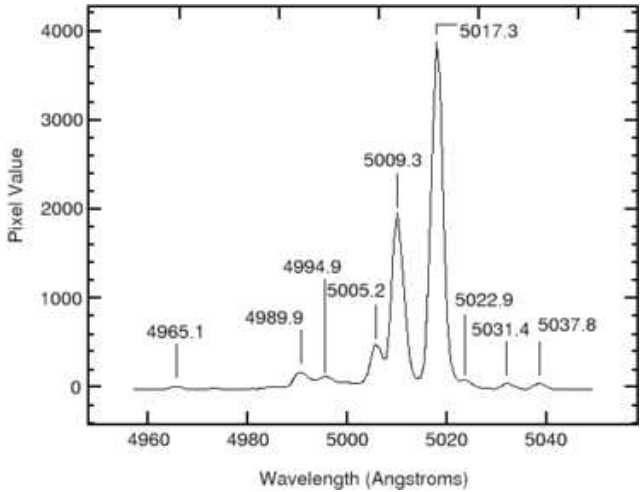}
  \caption{Detailed arc spectrum seen through filter-A set at 0\degr.
  The wavelengths of all the detected lines within the
  bandpass are annotated.}\label{arc_spec}
\end{figure}

The resulting cleaned science images still contain curvatures and
rotations which must be eliminated in order to match up the left-right
image pairs.  The distortions involved are somewhat complicated for a
slitless spectrograph like the PN.S, since they involve both imaging
distortions and the wavelength calibration of the dispersed light.  To
determine the requisite mapping, we obtained frequent calibration
images in which a mask containing 178 regularly-spaced holes is
introduced into the beam \citep[as described in detail in][]
{douglas02}.  By illuminating this mask with a
copper-neon-argon arc lamp, we obtained an array of short
filter-limited spectra on the CCD, allowing both the spatial mapping
and the wavelength calibration of the instrument to be determined.  As
Figure~\ref{arc_spec} illustrates, even the small wavelength range
admitted by the narrow-band filter contains a number of lines.  We used
the brightest of these lines (at 5017~\AA) to measure any departures in
the locations of 
these dots from the regular array of holes in the mask, thus
determining the spatial mapping.  These distortions were then removed
from the accompanying science images using {\sl IRAF's} standard
geometrical mapping routines, {\sl geomap} and {\sl geotran}.  Two
other lines at 4990~\AA\ and 5009~\AA\ were then used in conjunction
with the 5017~\AA\ line to characterise the wavelength calibration
across the image by interpolating a quadratic solution between these
points.  A database of these solutions for the array of points on each
CCD was produced; by interpolating between adjacent solutions, we can
map directly from the locations of an object detected in the two arms,
$\{x_L, y_L, x_R, y_R\}$ to a true position on the sky and an emission
wavelength, $\{x_0,y_0,\lambda\}$.

Following the spatial correction step, it is also possible to co-add
images in those cases where multiple observations of the same field
have been taken. Any offsets between images were determined by measuring
the positions of dispersed brighter stars in the field \citep[selected from
the USNO-B catalogue;][]{monet03}, and the necessary shifts were
made.  As a slight refinement, we actually performed both the image
undistortion and this shift in a single transformation of the original
image, so that there is only one interpolation of the data.  We then
combined the matched image frames using a weight determined by the
quality of each image.  These weights were calculated by using the {\sl
IRAF} routine {\sl fitprofs} to fit a one-dimensional Gaussian across
each star trail used in determining the shifts; the parameters of 
these fits were used to quantify the transparency, seeing and background
level of each exposure.  The appropriate weighting for each exposure
was then given by
\begin{equation}
  W = \frac{({\rm exposure\ time})\times({\rm transparency})}
            {({\rm seeing})^{2} \times ({\rm background})}.
\end{equation}

\subsubsection{Identification of emission-line objects}

Emission-line objects in the pipeline-processed images were identified
by a semi-automated routine. The undistorted images from both left and
right arm were median-subtracted to remove unresolved continuum from
M31, and {\sl SExtractor} was run on the image to detect point sources.
In order to eliminate the incorrect identification of features in star
trails as point sources, {\sl SExtractor} was rerun on the same image
after it had been convolved with an elliptical Gaussian function
elongated in the dispersion direction; any sources originating from
the stellar spectra were still found as this blurring has
essentially no effect on star trails, so these false detections can be
eliminated. 

Sources independently identified from the two arms of the spectrograph
were initially automatically matched in left--right pairs by selecting
detections that have the same coordinate in the undispersed $y$
directions to within two pixels, and whose coordinates in the
dispersed $x$ direction differ by an amount consistent with the
bandpass of the filter and the wavelength calibration determined
above.  These candidate pairs were passed on to a custom-written {\sl
Fortran} routine which allows the two images to be examined
side-by-side.  Any false detections, such as confused points within
extended structures, were eliminated at this point. Further manual
inspection then added back in any detections that have been missed by
the automated routine, usually due to proximity to the field edge or a
star. The edited list of emission-line pairs was then passed to the
{\sl IRAF} {\sl phot} routine to improve the position determination
and to measure the brightness of the source [using an aperture radius
equal to the mean full width at half maximum (FWHM) seeing for that
field to minimise sky contamination].  The FWHM of each source was
determined using the Moffat parameter returned by the {\sl IRAF} {\sl
imexam} task, as this turned out to be the simplest robust metric.

\begin{figure}
  \includegraphics[width=0.475\textwidth]{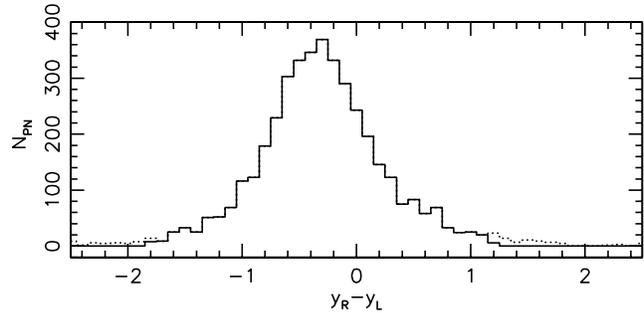}
  \caption{Difference in undispersed $y$ coordinates of matched pairs
  of M31 point sources detected in the two arms of the PN.S.  The
  dotted line shows those few pairs deemed spurious by the application
  of further cuts at $Y_R-Y_L=-1.80$ and 1.17.
  }  \label{dy_cuts}
\end{figure}

Once the source positions on both arms have been refined in this way,
we can apply a slightly more stringent cut in matching up the spatial
coordinates of sources to eliminate chance alignments: as
Figure~\ref{dy_cuts} shows, the distribution of differences in the $y$
coordinate between the two arms is significantly tighter than the
initial cut of $\pm 2$ pixels, so we deemed the objects in tails of the
distribution to be chance alignments, and eliminated them from the
source list.

\begin{figure}
  \includegraphics[width=0.475\textwidth]{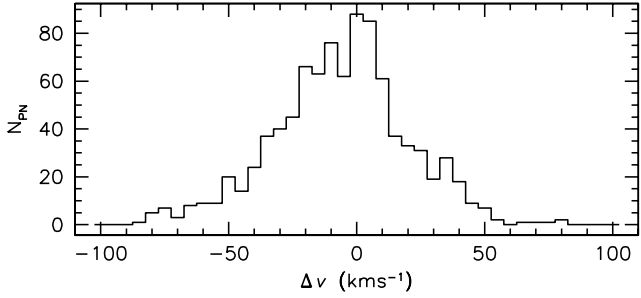}
  \caption{Differences between the velocities obtained for repeated
  observations of PNe in the overlaps between fields. A Gaussian fit
  gives a dispersion from PN.S of $17$~\kms.
  }\label{self_vel} 
\end{figure}

At this point, we have the definitive list of pairs of objects, and by
applying the calibration determined in Section~\ref{Spipeline} we can
transform these coordinates into a spatial location and a wavelength.
Identifying the emission with the 5007~\AA\ \oiii\ line, we can
translate the observed wavelength into a velocity, which we correct to
a heliocentric value on a field-by-field basis using {\sl IRAF}'s {\sl
rvcorrect} routine.  A measure of the internal consistency of these
velocities can be obtained using the PNe that lie in the overlaps
between fields for which we have multiple velocity measurements.
There are 732 objects detected in more than one field (643 in two
fields, 88 in three fields, 1 in four fields), so we have quite a
large sample to play with.  Figure~\ref{self_vel} shows the
distribution of velocity differences between pairs of measurements. A
Gaussian fit to these data gives a dispersion of 24~\kms, implying an
error on each individual measurement of $17$~\kms. This value is
clearly an over-simplification since the distribution appears somewhat
non-Gaussian and possibly even multi-modal.  This complex shape can be
attributed to the complexity involved in the wavelength calibration of
slitless spectroscopy.  However, the repeat observations used in
Fig.~\ref{self_vel} all lie close to the edges of the instrument's
field of view, where distortions in the wavelength solution are
greatest and the amount of calibration data is smallest.  The quoted
error should therefore be viewed as conservative, but with some
systematic residual errors on the scale of $\sim 5-10$~\kms\ not ruled
out. Nonetheless, even these pessimistic errors are entirely adequate
in a study of the large-scale kinematics of a massive galaxy like M31.

\subsubsection{Astrometry}

The measured coordinates must next be transformed into a standard
reference frame.  This process was carried out using the same basic
routines used for image stacking (see Section~\ref{Spipeline}).  The
approximate location of each field was determined from its nominal
pointing, and star positions in the vicinity were read in from the
USNO-B catalogue.  The corresponding stars trails in the PN.S images
were found and locations calculated using {\sl IRAF}'s {\sl xregister}
task; the resulting coordinates were fed to the {\sl ccmap} task to
calculate the astrometric solution for the field.  In the very central
fields, the bright background means there are very few stars
available. In these fields, PN positions that have already been
calibrated in overlapping neighbouring fields have been added to the
list of coordinates, in order to bootstrap the astrometric solution
across the central region.

\begin{figure}
  \includegraphics[width=0.475\textwidth]{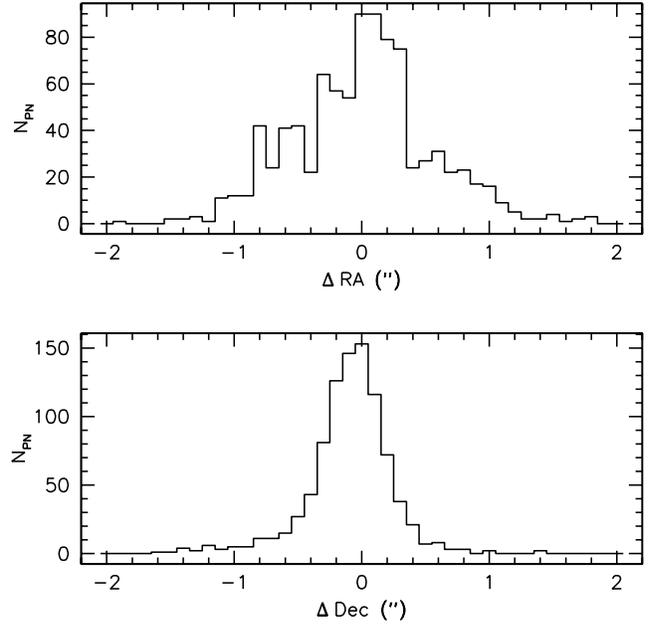}
  \caption{Differences between the positions obtained for repeated
  observations of PNe in the overlaps between fields. Gaussian fits to
  these give measurement uncertainties of $\sigma_{\rm RA}=0\farcs34$
  and $\sigma_{\rm Dec}=0\farcs16$.
  }\label{self_pos} 
\end{figure}

The PNe in the overlap regions between fields also provide an internal
check on the accuracy of the astrometry, by measuring offsets between
the coordinates for the same source in different fields.  As with the
velocity consistency check described above, this will tend to be a
conservative measure of accuracy, since spatial distortions are
greatest near the edges of fields where overlaps exist, so one might
expect the astrometric solution to be poorest in these regions.  As
Figure~\ref{self_pos} shows, the astrometric accuracy is somewhat better
in declination than right ascension.  This difference
arises because right ascension 
corresponds to the direction in which the calibrating stars are
dispersed, so centroiding their emission in this direction is
fundamentally less accurate.  However, the absolute positional
accuracy is more than adequate for this large-scale kinematic survey
of M31.

The astrometric matching of fields also allows us to make a final pass
through the list of detected sources to eliminate the duplicate
detections from field overlaps.  Where the duplicate observations were
obtained under similar seeing conditions, the positions, fluxes and
velocities of the two detections were simply averaged; where the seeing
conditions vary by more than 0\farcs5, just the better-quality data
were used.

For this survey, it is also useful to obtain the coordinates in an
M31-based reference frame.  We have therefore followed the geometric
transformations of \citet{huchra91}, which define right ascension and
declination offsets relative to the centre of M31, $\{\xi,\eta\}$, and
coordinates aligned with the system's major and minor axes, $\{X,Y\}$,
via the relations

\begin{equation} \label{eq_geomxi}
  \xi = \sin ({\rm RA} - {\rm RA}_0) \cos ({\rm Dec}),
\end{equation}
\begin{eqnarray} \label{eq_geometa}
  \eta &=& \sin ({\rm Dec}) \cos ({\rm Dec}_0) \nonumber \\
   &&-\cos({\rm RA} - {\rm RA}_0)\cos({\rm Dec})\sin({\rm Dec}_0),
\end{eqnarray}
\begin{equation}
  X = -\xi \sin ({\rm PA}) - \eta \cos ({\rm PA}),
\end{equation}
\begin{equation}
  Y =  -\xi \cos ({\rm PA}) + \eta \sin ({\rm PA}).
\end{equation}
here, we have taken the centre of M31 to lie at ${\rm RA}_0 = 00^{\rm
h}42^{\rm m}44\fs3$, ${\rm Dec}_0 = 41\degr 16\arcmin 09\arcsec$
(J2000.0), and have adopted a position angle for the major axis of
$\rm PA=37\fdg7$ \citep{devauc58}.  With this choice of coordinates,
positive $X$ is located southwest of the centre of M31 and positive
$Y$ lies to the northwest.

\subsubsection{Flux calibration}

\begin{table}
\centering
\begin{minipage}{0.475\textwidth}
\caption{Instrumental efficiencies calculated from spectrophotometric
  standard stars.} \label{stdstars}
\begin{tabular}{@{}l r c c c c@{}}
  \hline
  Star        &  F$_{5007}$ & Date        & \multicolumn{3}{c}{Efficiency} \\
  \multicolumn{3}{c}{(erg s$^{-1}$\AA$^{-1}$)} &  Left   &  Right  &  Total \\
  \hline                                  
  LDS 749B$^a$  &    154.759 & 2002-10-10 & 0.1424 & 0.1467 & 0.2891  \\
  BD+33 2642$^b$&   6367.221 & 2002-10-10 & 0.1305 & 0.1322 & 0.2628  \\
  G193-74$^c$   &     63.438 & 2002-10-13 & 0.1533 & 0.1393 & 0.2926  \\
  BD+17 4708$^c$&  18303.018 & 2003-09-29 & 0.1454 & 0.1435 & 0.2890  \\
  BD+28 4211$^c$&   8708.624 & 2002-10-08 & 0.1365 & 0.1325 & 0.2691  \\
  BD+28 4211$^c$&   8708.624 & 2002-10-10 & 0.1331 & 0.1290 & 0.2621  \\
  BD+28 4211$^c$&   8708.624 & 2002-10-11 & 0.1328 & 0.1299 & 0.2626  \\
  BD+28 4211$^c$&   8708.624 & 2002-10-11 & 0.1322 & 0.1312 & 0.2634  \\
  BD+28 4211$^c$&   8708.624 & 2002-10-12 & 0.1331 & 0.1298 & 0.2629  \\
  BD+28 4211$^c$&   8708.624 & 2003-09-29 & 0.1438 & 0.1414 & 0.2852  \\
  BD+28 4211$^c$&   8708.624 & 2003-09-30 & 0.1371 & 0.1292 & 0.2662  \\
  \hline
\end{tabular}
\flushleft
References:  $^a$ \citet{oke74}, $^b$ \citet{oke83}, $^c$ \citet{oke90}
\end{minipage}
\end{table}

\begin{table}
\centering
\begin{minipage}{0.475\textwidth}
\caption{Extinction per unit airmass in $r'$ from the Carlsberg
  Meridian Telescope, and weather notes for the nights observed.} \label{extinction} \centering
\begin{tabular}{@{}c r@{}l c r@{}l c l@{}}
  \hline
  Night & \multicolumn{2}{c}{Ext$_{r'}$} & Error & \multicolumn{2}{c}{Ext$_{5000}$$^a$} & Error & Comments \\
  \hline                                  
  2002-10-08   & 0.101&   & 0.004   & 0.172&   & 0.007  & Rain at end  \\
  2002-10-09   & 0.129&:  & 0.029   & 0.219&:  & 0.049  & Cloud first half \\
  2002-10-10   & 0.106&   & 0.006   & 0.180&   & 0.010  &   \\
  2002-10-11   & 0.107&   & 0.005   & 0.182&   & 0.009  &   \\
  2002-10-12   & 0.106&   & 0.005   & 0.180&   & 0.009  &   \\
  2002-10-13   & 0.112&   & 0.009   & 0.190&   & 0.015  &   \\
  \hline                                  
  2003-09-29   & 0.234&:  & 0.024   & 0.398&:  & 0.041  &   \\
  2003-09-30   & 0.213&:  & 0.012   & 0.362&:  & 0.020  & Cloudy at start  \\
  2003-10-01   & 0.317&:  & 0.068   & 0.539&:  & 0.116  & Mostly cloudy \\
  2003-10-02   & - &  & - & - &  & - &  Mostly rain \\
  2003-10-03   & - &  & - & - &  & - &  Rain and cloud\\
  2003-10-04   & - &  & - & - &  & - &  Rain\\
  2003-10-05   & - &  & - & - &  & - &  Rain\\
 \hline
\end{tabular}
\flushleft
$^a$ The conversion to Ext$_{5000}$ is calculated from Table~\ref{stdstars}
in RGO/La Palma Technical Note 31 to be a factor 1.70. 5000~\AA\ is the
closest listed value to the filter central wavelength.
\end{minipage}
\end{table}

Although not primarily designed as a photometric instrument, the PN.S
does provide information on the magnitudes of detected PNe through the
brightness of the spots in the two arms.  However, the conversion of
count rates into fluxes is not entirely straightforward.  The PN.S has
a rather slow shutter that takes some 15 seconds to fully open or
close, so the effective vignetting in short exposures is significantly
different from that in long exposures.  This limitation means that we
could not use sky flats to determine the overall vignetting, so instead
used long exposure dome flats obtained during the day.  To obtain an
optimal vignetting function, dome flats obtained at many position
angles were averaged to render the illumination of the field as uniform
as possible.  Since our science exposures are 15 minutes long, we have
not applied any correction for the small variation in effective
exposure time due to the slow shutter.

We measured the total system efficiency for PN.S in the observational
setup described using observations of spectrophotometric standard
stars. The flux per unit wavelength was measured from a narrow column
through the brightest region of the stellar trail and compared with
the published value. As Table~\ref{stdstars} shows, the total system
efficiency estimates are found to be surprisingly consistent, with a
mean value of ${\rm Eff_{total}} = 27.3 \pm 0.4 \%$, with much of the
scatter attributable to the non-photometric conditions.  As only a few
standard stars observations were made during the run this level of
consistency may be artificially low.  However, since the science goals
of this kinematic survey do not require high precision photometry,
this uncertainty is of little significance.

Fluxes of the detected sources were then calibrated using the formula 
\begin{equation}
F_{5007} = \frac{(C_L.g_L + C_R.g_R) E_{5007}}{\rm Eff_{total}} A
  T_{\rm exp} 10^{\left(\frac{a_{\rm sky} a_{\rm Gal}} {2.5}
  \right)},
\end{equation}
where $C_L$ and $C_R$ are the total counts from the left and right
arms; $g_L$ and $g_R$ are the gains of the CCDs on each arm;
$E_{5007}$ is the energy of a photon at 5007~\AA; $A$ is the effective
geometric collecting area of the telescope and is equal to 13.85
m$^2$; $T_{\rm{exp}}$ is the exposure time; ${a_{\rm sky}}$ is the
nightly sky extinction as recorded by the Carlsberg Meridian Telescope
(see Table~\ref{extinction}) multiplied by the airmass at which the
field was observed; and ${a_{\rm Gal}}$ the Galactic extinction taken
from \citet{schlegel98} and corrected to our filter central wavelength
using the relation from \citet{cardelli89} giving a value of
${a_{\rm Gal}} = 0.226$.  The conversion to PN-specific magnitudes
follows the \citet{jacoby89} relationship,
\begin{equation}
m_{5007} = -2.5 \log (F_{5007}) - 13.74,
\end{equation}
where the zero-point is chosen such that an emission line object would
have the same apparent magnitude as would be recorded if observed
through a V-band filter. To give a sense of this zero-point for
objects in our catalogue, the resulting magnitudes range from a fairly
bright \hii\ region at 17.8 mag to the faintest PNe at $\sim26$ mag.

\begin{figure}
  \includegraphics[width=0.475\textwidth]{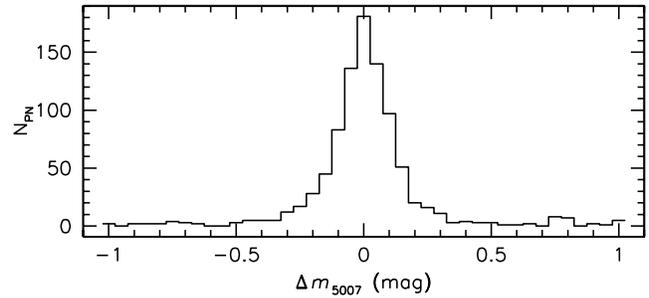}
  \caption{Magnitude variations within the PN.S data set from PNe in
  field overlaps. A Gaussian fit to this gives a measurement
  uncertainty of $\sigma_{m_{5007}}=0.07$\,mag.
  } \label{self_mag} 
\end{figure}

Once again, we can obtain an internal estimate of the uncertainty in
magnitudes by comparing the values derived for duplicate observations
in the overlaps between fields.  Figure~\ref{self_mag} shows the
differences between magnitudes in these repeated observations; a
Gaussian fit to this distribution gives a rather small measurement
uncertainty of 
$\sigma_{m_{5007}}=0.07$\,mag.  Magnitudes for extended objects will be
systematically underestimated by this technique as no allowance has
been made in the choice of aperture size for objects that are larger
than the field point-spread function.  However, since the main focus of
this survey is to determine the properties of unresolved PNe, this
issue is also not a major concern.


\section{Comparison with other data sets}\label{Scomp}

One concern with any data set from a novel instrument like the PN.S is
that it may contain unpredicted systematic errors.  Fortunately, a
number of smaller data sets obtained using more conventional
instrumentation already exist, so we can compare the new data with
these subsets to check for such systematic effects.

\subsection{Velocity comparison}\label{Svcomp}
The largest existing kinematic survey of PNe in M31 is that by
\citet[hereafter H06]{halliday06}, which determined velocities for 723
PNe using the conventional approach of narrow-band imaging and
fibre-fed spectroscopy.  Of these PNe, we find 715 objects within
4\arcsec of a PN.S object; the remaining 8 are likely obscured behind
stellar trails in the PN.S images, but a 99\% recovery rate is very
respectable, and already gives a measure of the completeness for these
brighter PNe.

\begin{figure}
  \includegraphics[width=0.475\textwidth]{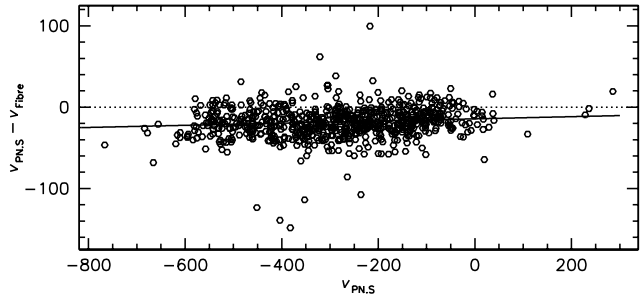}
  \caption{Comparison of initial PN.S velocities to \citet{halliday06} 
  fibre-spectrograph velocities. The solid line shows a fit to the
  data, while the dotted line is at 0~\kms.}
\label{comp_halliday} 
\end{figure}

\begin{figure}
  \includegraphics[width=0.475\textwidth]{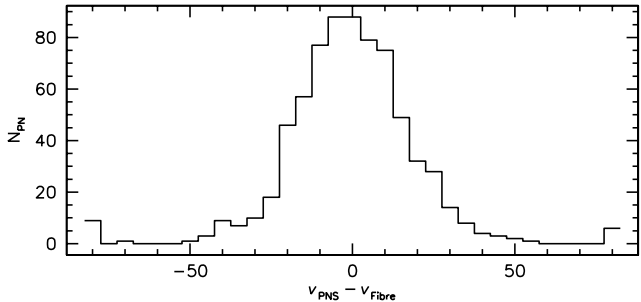}
  \caption{Plot of the difference between final PN.S velocities and
  \citet{halliday06} fibre-spectrograph velocities. A gaussian fit gives
  $\overline{\Delta v}=-0.7\pm0.4$~\kms\ and $\sigma=15.4\pm0.4$~\kms.
  }  \label{comp_hal_hist}
\end{figure}

If we plot the difference between H06 and PN.S velocities against the PN.S
velocity, as in Figure~\ref{comp_halliday}, we find a small zero-point
offset and a very slight linear dependence on velocity, fitted by
the relation 
\begin{equation}\label{vcorr}
v_{\rm PN.S}-v_{\rm H06} = 0.0151 v_{\rm PN.S} - 17.0.
\end{equation}

This small systematic effect presumably arises from the limitations of
the restricted range of arc lines available for PN.S wavelength
calibration (see Figure~\ref{arc_spec}). There is no large scale
spatial dependence in this velocity difference, though there appears
to be some small variation within the PN.S field. It makes very little
difference to any of the scientific results, but for consistency we
have applied the correction implicit in equation~(\ref{vcorr}) to our
complete data set and for all subsequent analysis. The residual
differences in velocity between PN.S and H06 data are shown in
Figure~\ref{comp_hal_hist}.  Fitting a Gaussian to this distribution
gives a dispersion of 15.4~\kms; H06 quote an error of $6$~\kms\ for
their fibre spectroscopy, implying that the PN.S data have an
uncertainty of $14$~\kms. This value is somewhat smaller than the
error deduced from self-calibration for the reasons discussed above,
and is probably a more accurate measure of the true uncertainty in the
PN.S velocities. 

An estimate of M31's system velocity for the PN.S data was made by
averaging the mean velocities from radial bins on either side of the
major axis for data within 1\degr\ ($\sim2.3$ scale lengths) of the
centre so as to exclude warps and asymmetries that have been observed
at large radial distances, leading to a value of $-309$~\kms. This
measurement is in reasonable agreement with the standard value of
$-300\pm4$~\kms\ \citep{devauc91} and in good agreement with H06's
value of $-309$~\kms.

\begin{figure*}
  \includegraphics[width=0.8\textwidth]{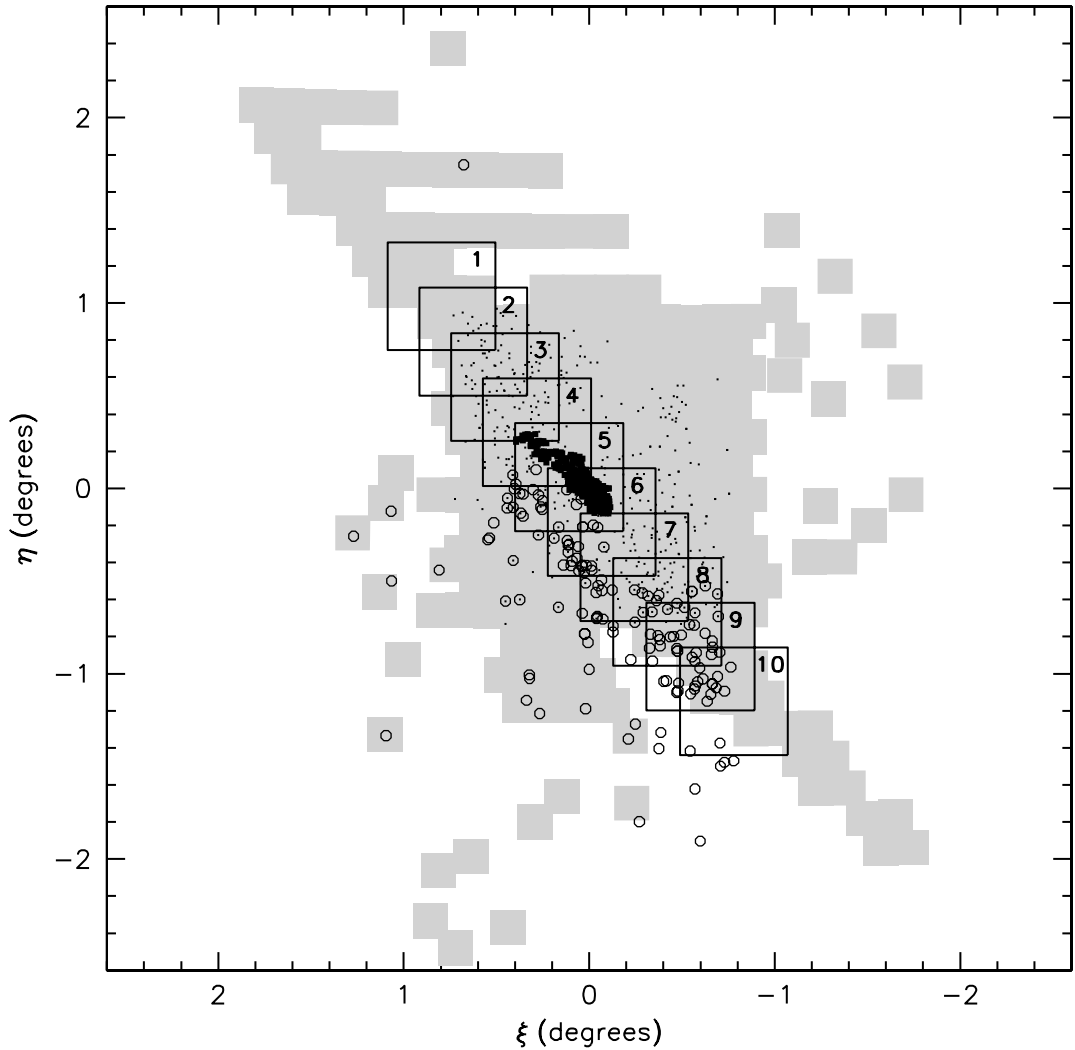}
  \caption{Positions of other surveys with respect to the PN.S survey.
    The PN.S survey area is shown in grey;
    Local Group Survey fields \citep{massey02} are the large fields
    with solid outlines;
    the \citet{ciardullo89} data are the solid black points near the
    centre;
    the open circles are the \citet{hurleyk04} data;
    and the small points are H06's fibre spectroscopy data.}\label{surveys} 
\end{figure*}

\begin{figure}
  \includegraphics[width=0.475\textwidth]{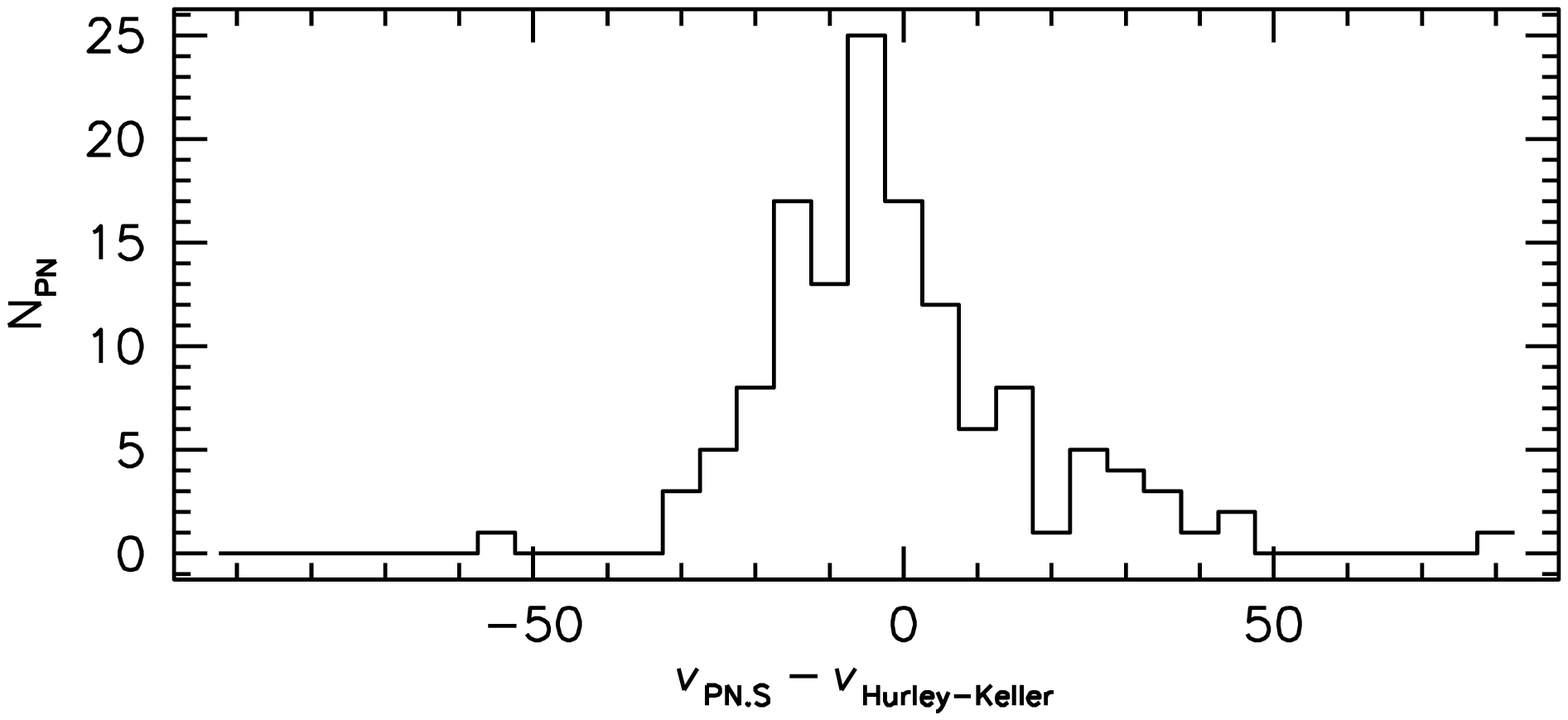}
  \caption{Plot of the difference between the final PN.S and the
  \citet{hurleyk04} velocities. A gaussian fit gives
  $\overline{\Delta v}=-5\pm1$~\kms\ and $\sigma=12\pm1$~\kms.} 
  \label{comp_hk}
\end{figure}

As a test of the robustness of this velocity cross-calibration, we can
also compare our results with a smaller but differently located survey
carried out by \citet{hurleyk04}.  They surveyed one quadrant of M31's
halo, finding a total of 146 emission-line objects, 135 PNe associated
with M31, eight in M32 and three \hii\ regions in Andromeda~IV (see
Figure~\ref{surveys}). We have covered the majority of their survey
area, finding 126 of their emission-line objects. The distribution of
differences in velocities is shown in Figure~\ref{comp_hk}.  A
Gaussian fit to this distribution gives a mean velocity difference of
$-5\pm1$~\kms\ and a combined dispersion of 12~\kms, lower than the
claimed PN.S velocity error. However, the distribution of errors here
is clearly non-Gaussian, so not too much weight should be given to the
exact value.  Nonetheless, the errors are clearly small, well below a
level that would compromise this kinematic study of M31.

\subsection{Astrometric comparison}

\begin{figure}
  \includegraphics[width=0.475\textwidth]{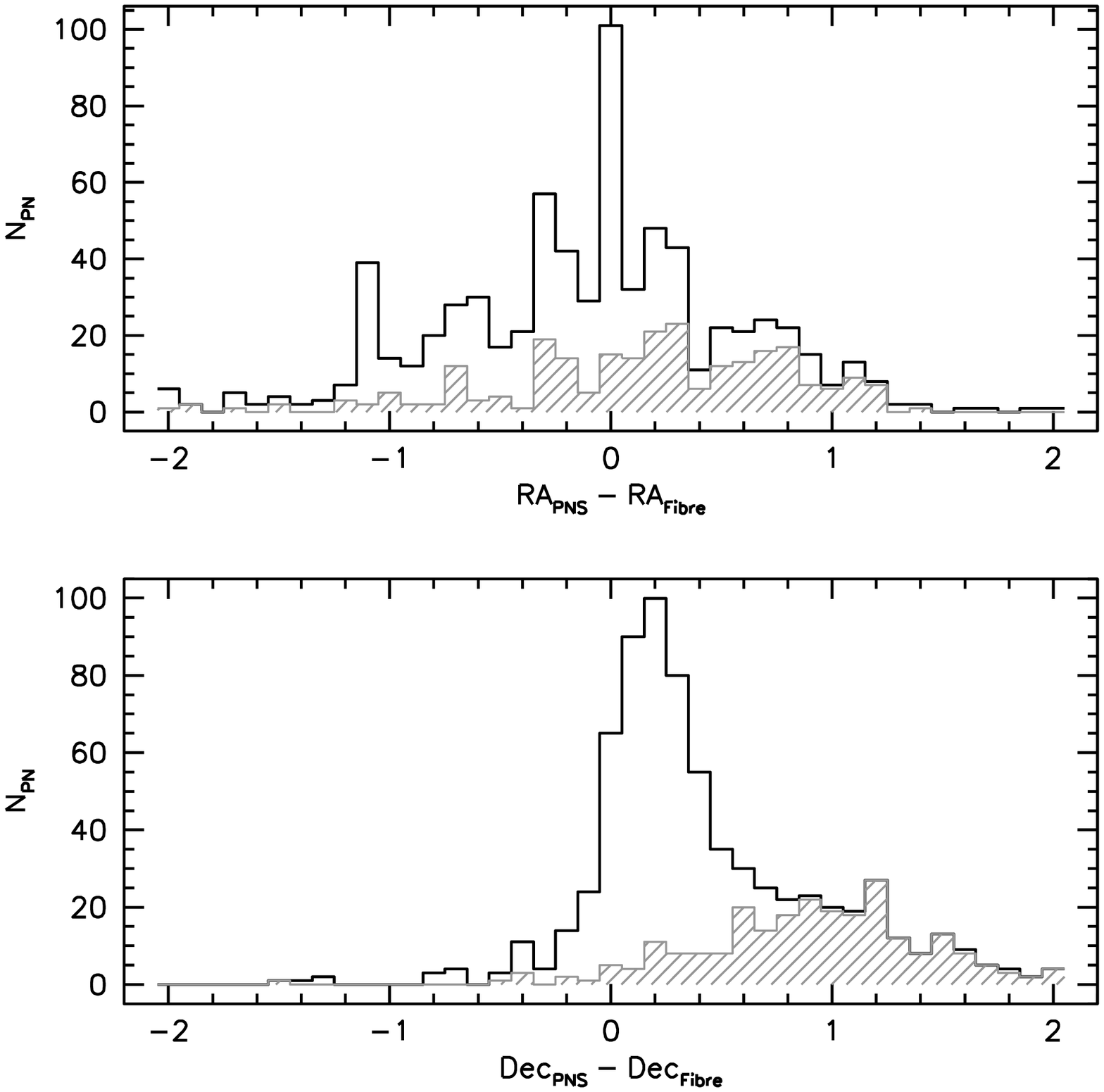}
  \caption{Comparison between PN.S and H06 astrometry, with differences 
  measured in arcseconds.  The black line
  represents all the PNe found in both catalogues; the grey shaded
  area shows those PNe in the central PN.S fields (60, 61, 70, 71)}
  \label{halliday_pos}
\end{figure}

The PN.S astrometry can also be compared with the H06 data set.  The
distribution of differences in coordinates are shown in
Figure~\ref{halliday_pos}.  The right ascension distribution is
similar to that found internally within the PN.S data, with a combined
dispersion of $\sim 0\farcs6$.  In declination there is a small
systematic offset of +0\farcs18, with a long tail to positive values.
As Figure~\ref{halliday_pos} shows, the tails of the distribution are 
almost entirely populated by PNe from the four central PN.S fields.
As discussed above, the lack of measurable stars in these fields
compromises the PN.S astrometric solution somewhat, so such larger
errors are not surprising. Excluding these fields, the combined
dispersion in declination is $\sim 0\farcs2$. Given the enormous size
of M31, such uncertainties are completely negligible for kinematic
purposes.

\subsection{Photometric comparison}

Although not principally a photometric survey, we have been able to
obtain fairly consistent magnitudes from the PN.S data, so once again
it would be useful to compare against other data to obtain an external
measure of the data quality.  Fortunately, M31 has been extensively
imaged as part of the Local Group Survey \citep[hereafter
M02]{massey02}, which includes narrow band studies of both the \oiii\
line and \halpha\,+\,\nii\ lines.  As Figure~\ref{surveys} shows, M02
imaged 10 fields over the majority of the disk of M31 and hence a
large portion of the PN.S survey region (covering 2745 of the objects
in the PN.S survey).

\begin{figure}
  \includegraphics[width=0.475\textwidth]{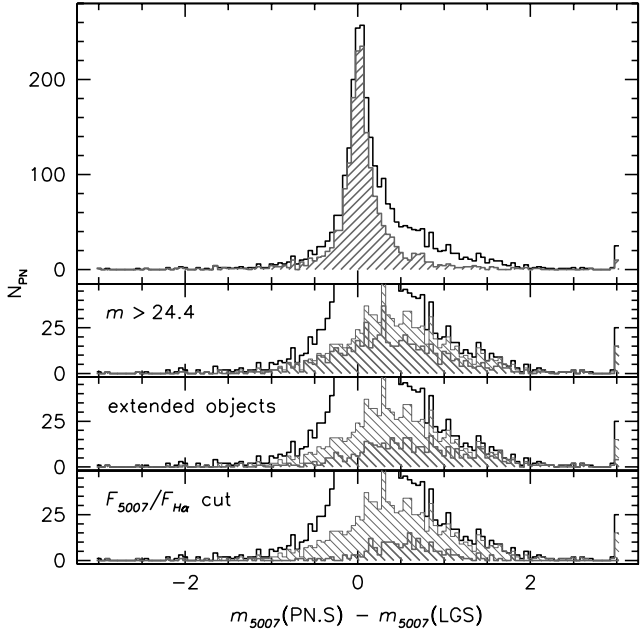}
  \caption{Photometric comparison between the magnitudes derived from
    the PN.S data and those from the \citet{massey02} imaging data.
    The black outline represents all the objects that are within both
    surveys. In the upper panel the shaded area represents PN
    remaining after a number of cuts have been applied to the data.
    The lower panels show the distribution of objects for the cuts
    indicated.  The larger, light grey area is the total cut made
    while the smaller dark grey histograms show the particular objects
    being excluded by each criterion.
  }\label{pns_lgs_mag}
\end{figure}

We therefore analysed the M02 survey images by obtaining fluxes at the
positions of PN.S sources using the {\sl IRAF} {\sl phot} task with a
fixed aperture of 2\arcsec -- a value somewhat larger than the typical
survey seeing ($\sim0\farcs8 - 1\farcs5$) to allow for positional
uncertainties, but not so large as to introduce significant source
confusion.  The photon counts were converted to fluxes using factors
of $3.92\times 10^{-16}$ and $1.79\times 10^{-16}$ for the \oiii\ and
\halpha\,+\,\nii\ data respectively (Massey, private communication).
Galactic extinction corrections were made using data from
\citet{schlegel98} and \citet{cardelli89}, as for the PN.S data,
yielding extinctions of 0.225~mag and 0.164~mag for \oiii\ and
\halpha\,+\,\nii\ respectively.

Figure~\ref{pns_lgs_mag} shows the comparison between the imaging
magnitudes derived in this way and the PN.S data.  If we remove the
faintest PNe for which photometry is rather uncertain, and probable
\hii\ regions for which this aperture photometry is inappropriate,
there is a good overall agreement between the data sets.  Fitting a
Gaussian to the remaining 1769~PNe gives negligible zero-point offset
of 0.02~mag, and a combined dispersion of 0.16~mag.  There is no
evidence for varying offsets in different fields or for data taken on
different nights.

\begin{figure}
  \includegraphics[width=0.475\textwidth]{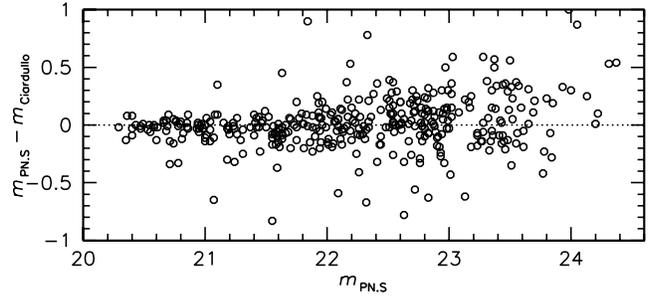}
  \caption{Comparison between PN.S magnitudes and those published in
  \citet{ciardullo89}. Examining this plot by eye there appears
  to be a slight trend, with PN.S magnitudes being fainter than the
  corresponding \citet{ciardullo89} magnitudes for faint PNe. However,
  gaussian fits in half magnitude bins show this to be small
  ($<0.1$~mag) and of low significance.
}  \label{comp_ciard}
\end{figure}

As a further cross-check, Figure~\ref{comp_ciard} shows the comparison
between the PN.S magnitudes and the smaller data set published by
\citet{ciardullo89}.  As expected, the scatter increases somewhat for
the fainter PNe, but the overall agreement is very good.  A Gaussian
fit to the magnitude differences yields a mean of $0.00$~mag with a
combined dispersion of 0.13~mag, and even at the faintest magnitude
the combined dispersion is less than 0.3~mag.  Once again, there is no
evidence of varying offsets for different fields or data taken on
different nights.  Since we are only going to use the photometric data
to make quite crude cuts in the luminosity function of PNe, which we
have measured over a range of more than three magnitudes, photometry
with an error of less than 0.3~magnitudes is all that is required.

\section{Contamination and completeness}\label{Sconcom}

The next issue to be addressed is to try to identify those emission
line sources that are not PNe, and to estimate what fraction of PNe we
may have missed in the survey.  

\begin{figure}
  \includegraphics[width=0.475\textwidth]{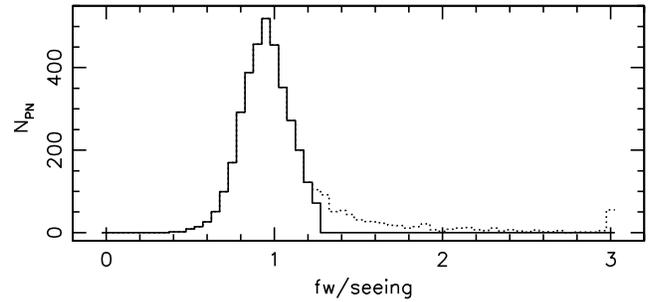}
  \caption{A plot of the FWHM of sources detected by the PN.S survey,
  normalised by the seeing of each observed field.  Sources in the
  dotted long tail toward large sizes have been flagged as contaminating
  \hii\ regions. The cut between extended and point sources is made at
  fw/seeing~=~1.25.
}   \label{fw_cuts} 
\end{figure}

With only a single emission line detected by PN.S, any object that
emits \oiii\ may form part of the data set (or, indeed, any object
along the line of sight with a redshift that places an unrelated
emission line at this wavelength).  The dominant contaminating objects
are likely to be \hii\ regions, which emit strongly in \oiii, and
exist in large numbers in disk galaxies like M31.  Most of these
objects will already have been excluded, since their extended nature
means that they will have been cut from the data set at the
source-identification stage (even the largest PNe, with diameters of
up to 3~pc, will be essentially unresolved in these data).  However,
\hii\ regions have a broad size distribution, and the more compact
ones will be barely resolved so will still be in the sample.
Figure~\ref{fw_cuts} shows the distribution of source sizes,
normalised to the seeing, of the objects remaining in the survey;
there clearly remains a tail of resolved objects, so we have flagged
sources for which the FWHM normalised by seeing exceeds a value of
1.25 as probable \hii\ regions in the final catalogue. 

When examining the spatial distribution of the objects flagged by this
process, it was found that they largely lie in a ring around M31's
centre. A further clump of 35 objects was also found in a single
field close to the centre of M31.  This field was the only one close
to the galaxy's centre that was observed under good seeing conditions
($\sim$\,0\farcs8: see Figure~\ref{seeing}). With the seeing this good,
the variation in instrumental focus across the field and differences
between the two PN.S images may become significant, so it is possible
that these apparently extended objects are actually unresolved PNe.
However, for consistency we also flag them as non-PNe; as discussed
previously, we in any case do not expect the survey to be complete in these
crowded high surface-brightness fields.

\begin{figure}
  \includegraphics[width=0.475\textwidth]{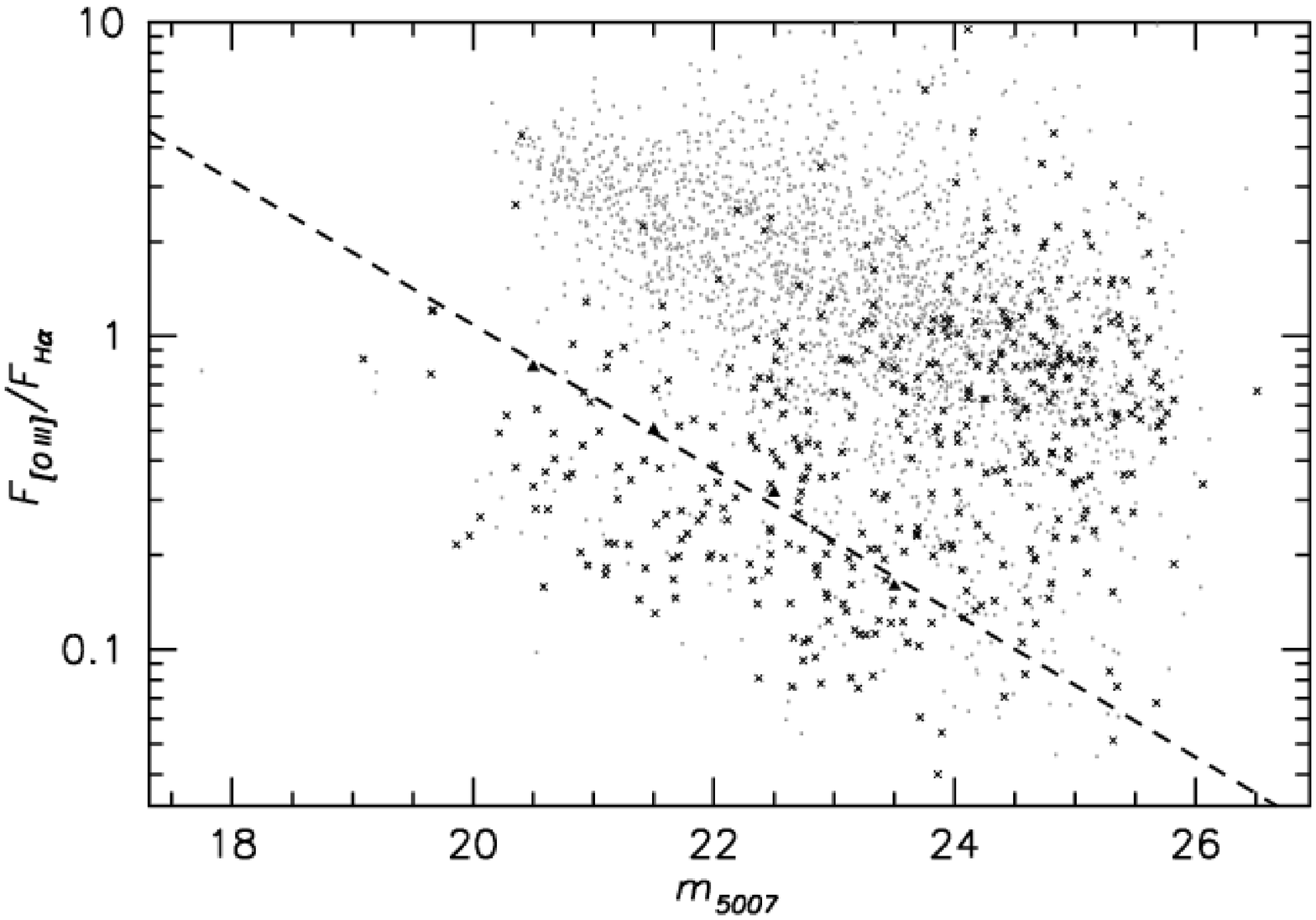}
  \includegraphics[width=0.475\textwidth]{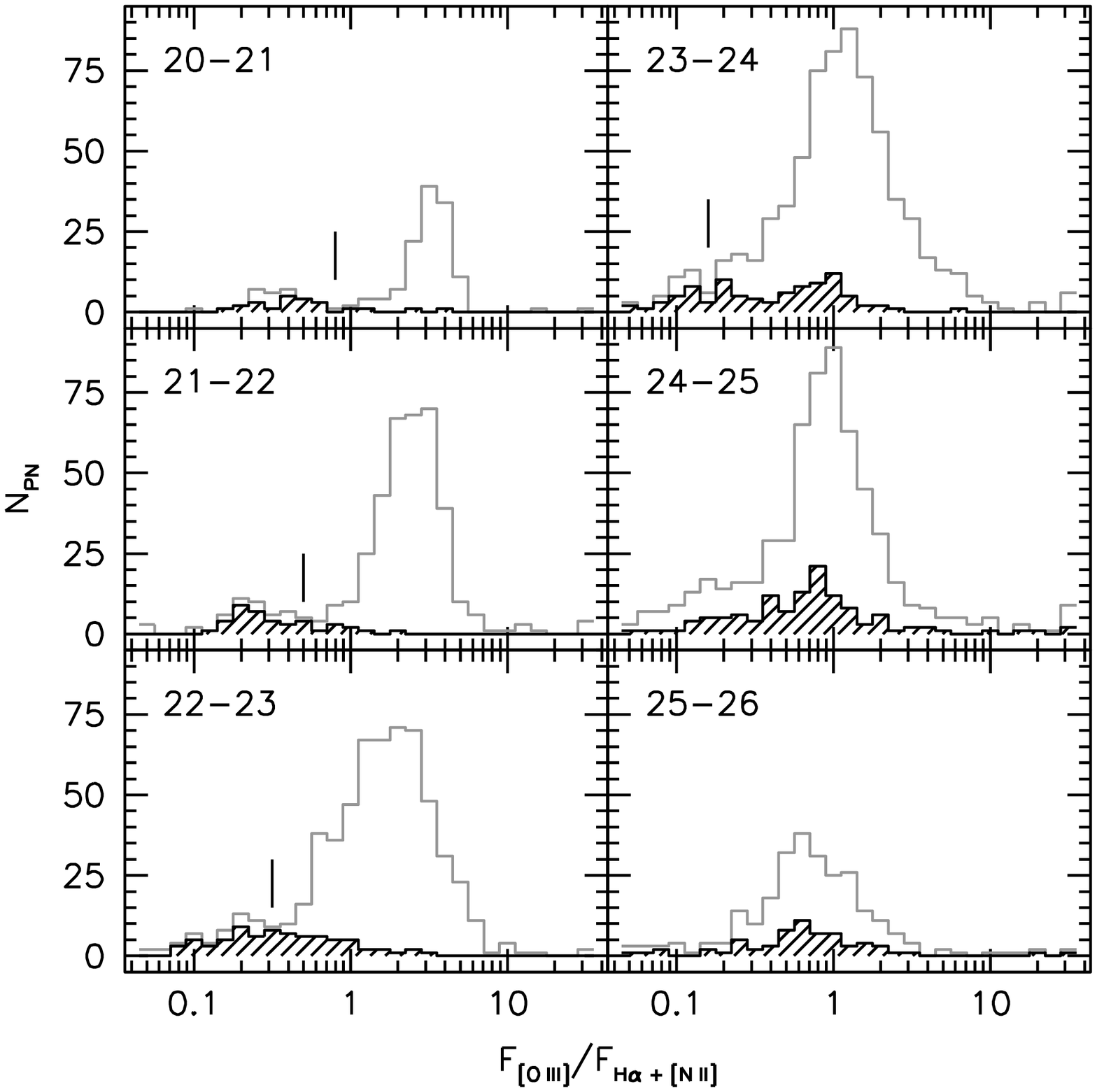}
  \caption{The flux ratio of \oiii\ to (\halpha\,+\,\nii) versus
    magnitude.  The upper panel shows all the data, with extended
    objects shown as black crosses.  The lower panel shows the same
    data as histograms in different magnitude bins, with the extended
    sources shaded.  The vertical line in each histogram shows a
    suggested division between PNe and \hii\ regions based on these
    distributions.  These lines have been plotted as triangles in the
    upper panel, and the dashed line connects them.
  }\label{fluxratio} 
\end{figure}

Presumably, this cut on angular size will still miss the most compact
\hii\ regions and in the central regions where the seeing was very poor
objects with diameters of up to $\sim15$~pc will be classified as compact by
this cut. Fortunately, we do have some further information that we 
can use to try to identify these remaining contaminants.
Specifically, we have measured fluxes from both the \oiii\ and
(\halpha\ + \nii) lines from the M02 data, and the ratio of these
fluxes, $R$, differs between PNe and \hii\ regions \citep{ciardullo04}. In
particular, since PNe give out most of their light in the \oiii\ line,
we would expect $R$ to be large for these objects.  Attempts have been
made to quantify this criterion, leading to the suggestion that for
bright PNe one should place a cut at $R \sim 1$ -- 2
\citep{ciardullo02, ciardullo04, magrini00}.  However, it became clear
when we started investigating this quantity for the data in the PN.S
survey that the appropriate value for this cut varies significantly
with magnitude.  As Figure~\ref{fluxratio} illustrates, the optimum
cut in $R$ to exclude \hii\ regions (as calibrated using the
identified extended sources) decreases at fainter magnitudes.  We have
therefore applied the cut shown as a dashed line in
Figure~\ref{fluxratio}, with all sources below the line flagged as
\hii\ regions, regardless of their sizes.  Although the distribution
of extended sources on this plot would indicate that some fraction of
the sources above this line are \hii\ regions and this cut is
lower than the ones made by \citet{ciardullo02, ciardullo04} and
\citet{magrini00}, we have been chosen to be conservative by only
selecting for exclusion the area of the plot that seems virtually
devoid of PNe.  

\begin{figure}
  \includegraphics[width=0.475\textwidth]{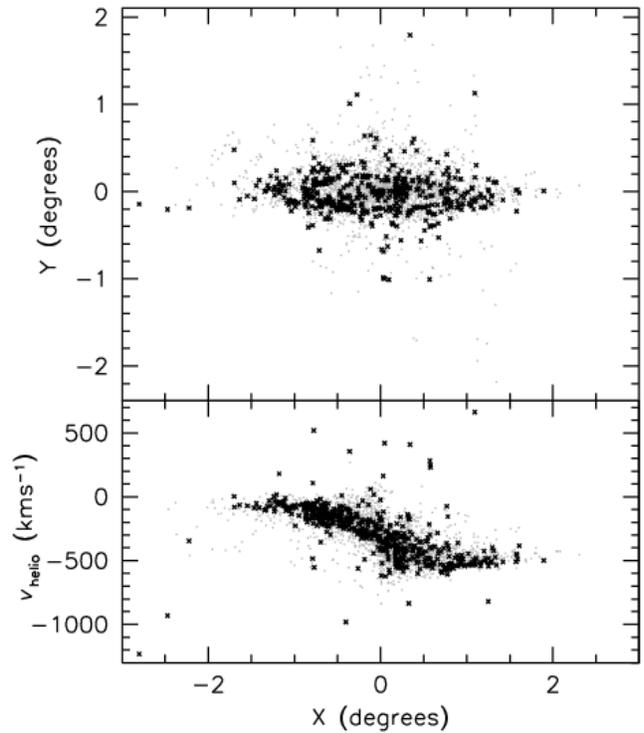}
  \caption{Locations in position and velocity of sources identified as
  probable \hii\ regions.  The whole sample is shown as small dots,
  with the probable \hii\ regions as larger crosses.}
  \label{hii}
\end{figure}

Figure~\ref{hii} picks out the locations in the \{$X$,$Y$\} and
\{$X$,$v$\} planes of the objects that these two criteria have
identified as non-PNe.  Although the criteria were designed to flag up
probable \hii\ regions, it is clear that they also exclude most of the
objects with velocities that are inconsistent with membership of M31.
Closer inspection shows that the majority of these sources have
emission lines extended in the dispersion direction, indicative of 
Doppler broadening of the emission from massive background galaxies.
We could, for example, be observing the redshifted $\lambda 3727$~\AA\
{\sc[O\,ii]} emission from star-forming galaxies at redshifts $\sim
0.34$.  Clearly, more extensive conventional spectroscopy would be
required to confirm any such identifications.  Figure~\ref{hii}
also shows the structure in the distribution of \hii\ regions
associated with spiral arms as well as the ring mentioned above.  At a
radius of 0.8\degr, the ring coincides with the star-forming ring
discussed by \citet{devereux94}, where an excess of \hii\ regions
might be expected.

As a final check on our reliability in identifying contaminating
sources, we have compared our identifications with the catalogue of
1312 emission-line objects that \citet[hereafter MLA93]{meyssonnier93}
detected and classified in M31 using broader-band slitless
spectroscopy.  We find 856 of these objects to be within 4\arcsec\ of
an object in our survey, the great majority of which are classified as
PN by both MLA93 and ourselves.  Only 94 of these objects are listed
by MLA93 as non-PNe, and we have successfully flagged two-thirds of
these as non-PNe as well.  Of the remaining 31, MLA93 identify 25 as
possible Wolf--Rayet stars.  Since Wolf--Rayet stars are at a late stage
in stellar evolution rather similar to PNe, and the two are sometimes
even found together \citep{gorny95}, it is not clear that they should
be considered as contaminants at all. One object classified as a
possible QSO or Wolf-Rayet star is discussed further in
Section~\ref{Ssats}.  The last five objects are of uncertain classification
in the MLA93 catalogue and will be left in this catalogue.  The 456
objects that were not found presumably do not exhibit \oiii\ emission.

\begin{figure}
  \includegraphics[width=0.475\textwidth]{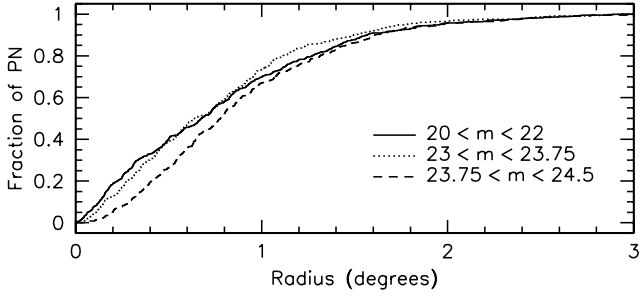}
  \caption{Cumulative distribution of radii in the disk plane for
  bright (solid line), intermediate-luminosity (dotted line) and faint
  (dashed line) PNe. A Kolmogorov-Smirnov test shows the distribution
  of bright and intermediate-luminosity PNe to be statistically
  indistinguishable at the 95\% confidence level. There is a dearth of
  faint PNe near to the centre of M31 resulting from poor seeing
  conditions and a high surface brightness.  }  \label{cum_rad_ks}
\end{figure}

The other side to this coin is to ask how many PNe we might have
missed.  A simple test of completeness can be made by comparing the
distribution of PNe with radius in different magnitude bins.  As
Figure~\ref{cum_rad_ks} shows, bright and intermediate-luminosity PNe
show spatial distributions that are statistically indistinguishable, but
the faint PNe display a dearth of objects at small radii.  As discussed
above, such incompleteness is expected given a combination of the high
surface brightness of the bulge region and the poor seeing conditions
during observations of the central fields.  Overall, we can
tentatively conclude that the sample is complete to $m_{5007}=23$ over
the entire survey, to $m_{5007}=23.75$ at projected disk radii larger
than 0.2\degr, and to $m_{5007}\sim 25$ beyond 1\degr; we will revisit
this issue when we look at the PN luminosity function in
Section~\ref{Spnlf}.  In terms of spatial coverage, as
Figure~\ref{pns_survey} shows, M31 has been completely mapped out to a
deprojected disk radius of 1.5\degr. Beyond that radius, the major and
minor axes are somewhat unevenly sampled from one side to the other,
but still with reasonable spatial coverage.  Thus, although not
spatially complete, useful kinematic information can be gleaned out to
2\degr (27.4~kpc).

\section{The Catalogue}
\label{Scat}

Having calibrated the data spectroscopically, astrometrically and
photometrically, both internally and against external results, we are
now in a position to present the complete database from the survey.
The full final version of the resulting data table of 3300 emission
line objects including 2730 probable PNe is included in the
electronic version of this paper, but Table~\ref{tab.cat} gives a
small sample of the tabulated data listing coordinates, magnitudes,
heliocentric velocities, whether the source is a probable PN, whether
it lies in M31 (see Section~\ref{Ssats}), when the data were obtained,
and cross-identifications with other catalogues.

\begin{table*}
\centering
\begin{minipage}{0.95\textwidth}
\caption{Emission line objects identified in the M31 survey. See
  electronic version for full table.} 
\label{tab.cat} 
\begin{tabular}{c c c c c c c c c c c c c c c}
  \hline
  ID & RA & Dec & $m_{5007}$ & $v_{\rm helio}$ & Note & Host   & Date & Field & H06     &        C89 &       HK04 &       MLA93 \\
  & J2000.0 & J2000.0 &      & \kms            & $^a$ & $^b$   &      & Number& c       &        c   &       c    &       c     \\
  \hline
     1 &  0:41:01.7 & 42:07:58.6 & 25.48 &  -283.3 &     - &        - &   2002.10.12 &    1 &         - &         - &        - &        -   \\ 
     2 &  0:40:35.4 & 42:10:32.0 & 22.16 &  -133.9 &     - &        - &   2002.10.10 &    2 &         - &         - &        - &        -   \\ 
     3 &  0:39:35.9 & 42:09:39.5 & 25.51 &   268.9 &     - &   2MASXi &   2002.10.10 &    3 &         - &         - &        - &        -   \\ 
     4 &  0:39:37.3 & 42:09:59.6 & 23.52 &   354.5 &     E &   2MASXi &   2002.10.10 &    3 &         - &         - &        - &        -   \\ 
     5 &  0:39:31.6 & 42:11:56.6 & 23.28 &  -343.3 &     - &        - &   2002.10.10 &    3 &         - &         - &        - &        -   \\ 
     6 &  0:38:54.0 & 42:09:39.3 & 25.36 &  -266.7 &     E &        - &   2002.10.10 &    4 &         - &         - &        - &        -   \\ 
     7 &  0:37:54.3 & 42:14:49.3 & 23.37 &  -328.1 &     - &        - &   2002.10.10 &    5 &         - &         - &        - &        -   \\ 
     8 &  0:43:11.6 & 42:07:13.0 & 22.31 &  -194.0 &     - &        - &   2002.10.12 &    7 & PN\_10\_3\_1 &         - &        - &        -   \\ 
     9 &  0:43:10.5 & 42:11:33.3 & 23.58 &  -286.1 &     - &        - &   2002.10.12 &    7 & PN\_10\_3\_2 &         - &        - &        -   \\ 
    10 &  0:43:09.5 & 42:14:24.2 & 24.87 &  -314.2 &     - &        - &   2002.10.12 &    7 &         - &         - &        - &        -   \\ 
    11 &  0:42:49.6 & 42:15:03.7 & 24.38 &   107.3 &     E &        - &   2002.10.12 &    7 &         - &         - &        - &        -   \\ 
    12 &  0:44:08.6 & 42:06:00.0 & 25.62 &  -202.5 &     - &        - &   2002.10.13 &    8 &         - &         - &        - &        -   \\ 
    13 &  0:44:09.8 & 42:06:40.2 & 22.99 &  -108.0 &     - &        - &   2002.10.13 &    8 & PN\_10\_3\_7 &         - &        - &        -   \\ 
    14 &  0:43:31.0 & 42:09:25.7 & 24.20 &  -136.6 &     - &        - &   2002.10.13 &    8 &         - &         - &        - &        -   \\ 
    15 &  0:43:39.8 & 42:13:46.3 & 25.07 &  -210.8 &     - &        - &   2002.10.13 &    8 &         - &         - &        - &        -   \\ 
    16 &  0:43:37.6 & 42:06:58.6 & 24.65 &  -554.4 &     E &  Stream? &   2002.10.13 &    8 &         - &         - &        - &        -   \\ 
    17 &  0:44:59.8 & 42:07:44.8 & 21.76 &  -193.6 &     - &        - &   2002.10.12 &    9 &  PN\_9\_3\_4 &         - &        - &     1091   \\ 
    18 &  0:45:00.9 & 42:08:44.8 & 23.53 &   -72.2 &     - &        - &   2002.10.12 &    9 &         - &         - &        - &        -   \\ 
    19 &  0:44:30.5 & 42:08:55.2 & 22.36 &   -81.9 &     - &        - &   2002.10.12 &    9 & PN\_10\_3\_8 &         - &        - &     1005   \\ 
    20 &  0:44:51.4 & 42:09:34.2 & 23.81 &  -148.4 &   E,R &        - &   2002.10.12 &    9 &         - &         - &        - &        -   \\ 
   \vdots & \vdots & \vdots & \vdots & \vdots  & \vdots & \vdots & \vdots & \vdots & \vdots & \vdots & \vdots & \vdots  \\ 
 \hline
\end{tabular}
\flushleft
$^a$ Probable \hii\ regions (or background galaxies): E -- extended in
the PN.S data, R -- Flux ratio $R$ below cutoff value (see
Figure~\ref{fluxratio}), other -- excluded as a PN due to flux . \\   
$^b$ Sources of non-M31 objects:
M32 / M32?  -- Objects associated with and in the vicinity of M32. 
NGC205 / NGC205? -- Objects associated with and in the vicinity of NGC205.
AndIV -- Objects associated with the galaxy Andromeda IV.
2MASXi -- Non-M31 source in the 2MASS catalogue.
MLA93 -- Non-M31 source listed in \citet{meyssonnier93}.
NS -- Objects in the region of the Northern Spur.
Stream? -- Objects referred to in \citet{merrett03} as being a possible
extension of the Southern Stream.\\
$^c$ cross-identifications from other catalogues: H06 -- \citet{halliday06}; C89 --
\citet{ciardullo89}; HK04 -- \citet{hurleyk04} (format: table\_ID); MLA93 -- \citet{meyssonnier93}.
\end{minipage}
\end{table*}


\section{Satellite galaxies and other objects in the PN.S survey area}
\label{Ssats}

Within the area of this survey are a number of galaxies besides M31.
Some of the objects in our catalogue which reside in these systems can
be identified in the database because their velocities are far from
that of M31.  As we have seen above, most such sources have already
been flagged as non-PNe because of their extended nature or low
\oiii/\halpha\ ratio, and are probably in quite distant faint
background systems.  In some cases, such as the sources in the
less-distant background galaxy Andromeda IV, there may be a few
genuine PNe.  Of more immediate interest are the satellite galaxies
around M31.  In these systems, we do detect quite a number of PNe,
which means we can obtain some crude measure of the internal
kinematics of these systems, but we must take some care in
flagging these sources so that they do not compromise the use of this
data set to model the dynamics of M31.

\begin{figure}
  \includegraphics[width=0.475\textwidth]{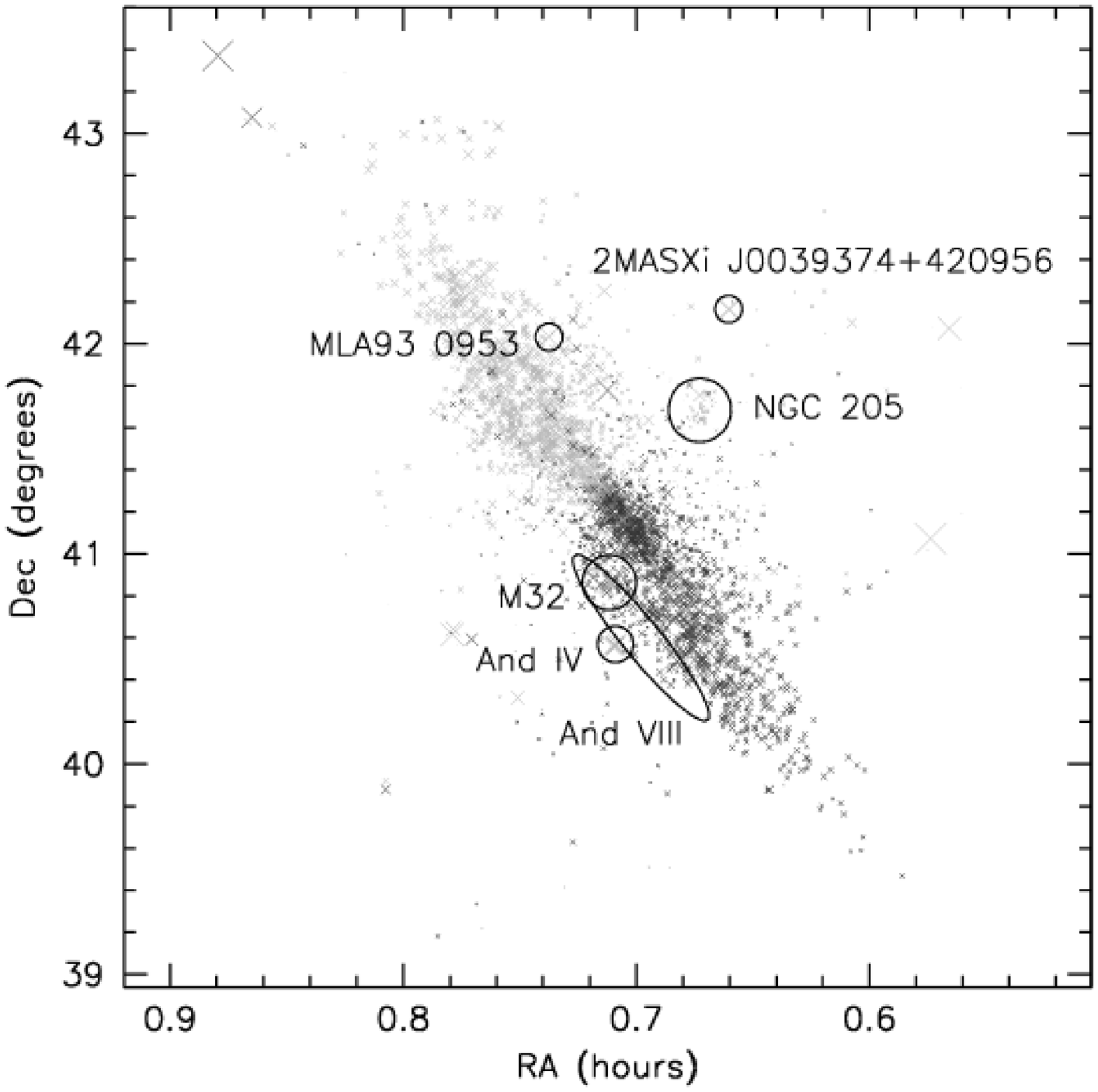}
  \caption{Positions of other galaxies and the the detected emission
  line objects. An object's size
  and colour reflects its velocity with respect to M31's system
  velocity: grey objects are receding and black ones are approaching.
  }
  \label{sats}
\end{figure}

The locations of the known galaxies in the survey field are indicated
in Figure~\ref{sats}, and we now consider them in turn, to try to
identify the sources that lie within them, both on the basis of their
positions and their velocities.  Probable non-members of M31
identified in this way are annotated as such in the full database (see
Section~\ref{Scat}).

\begin{figure*}
  \includegraphics[width=0.49\textwidth]{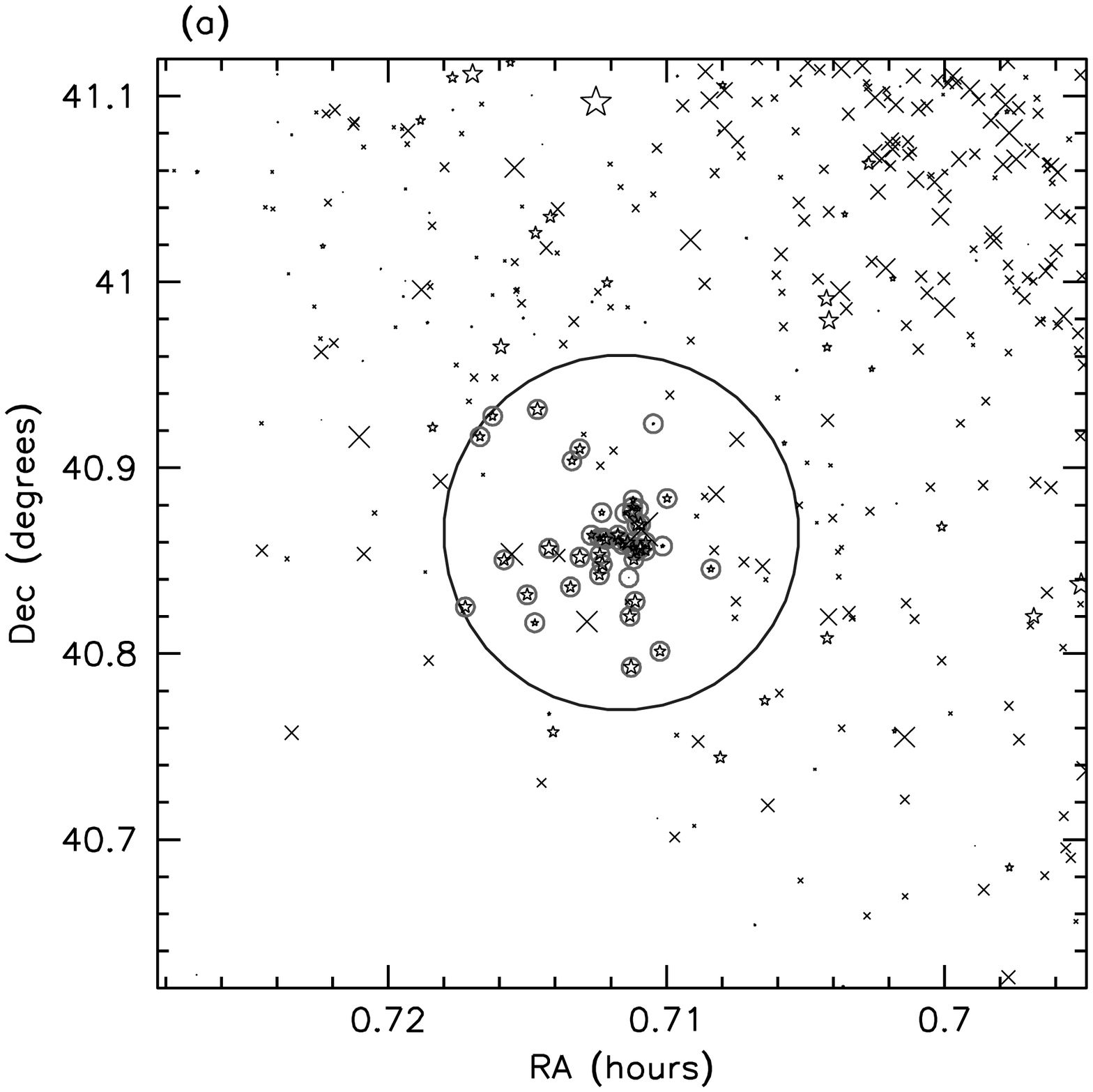}
  \includegraphics[width=0.49\textwidth]{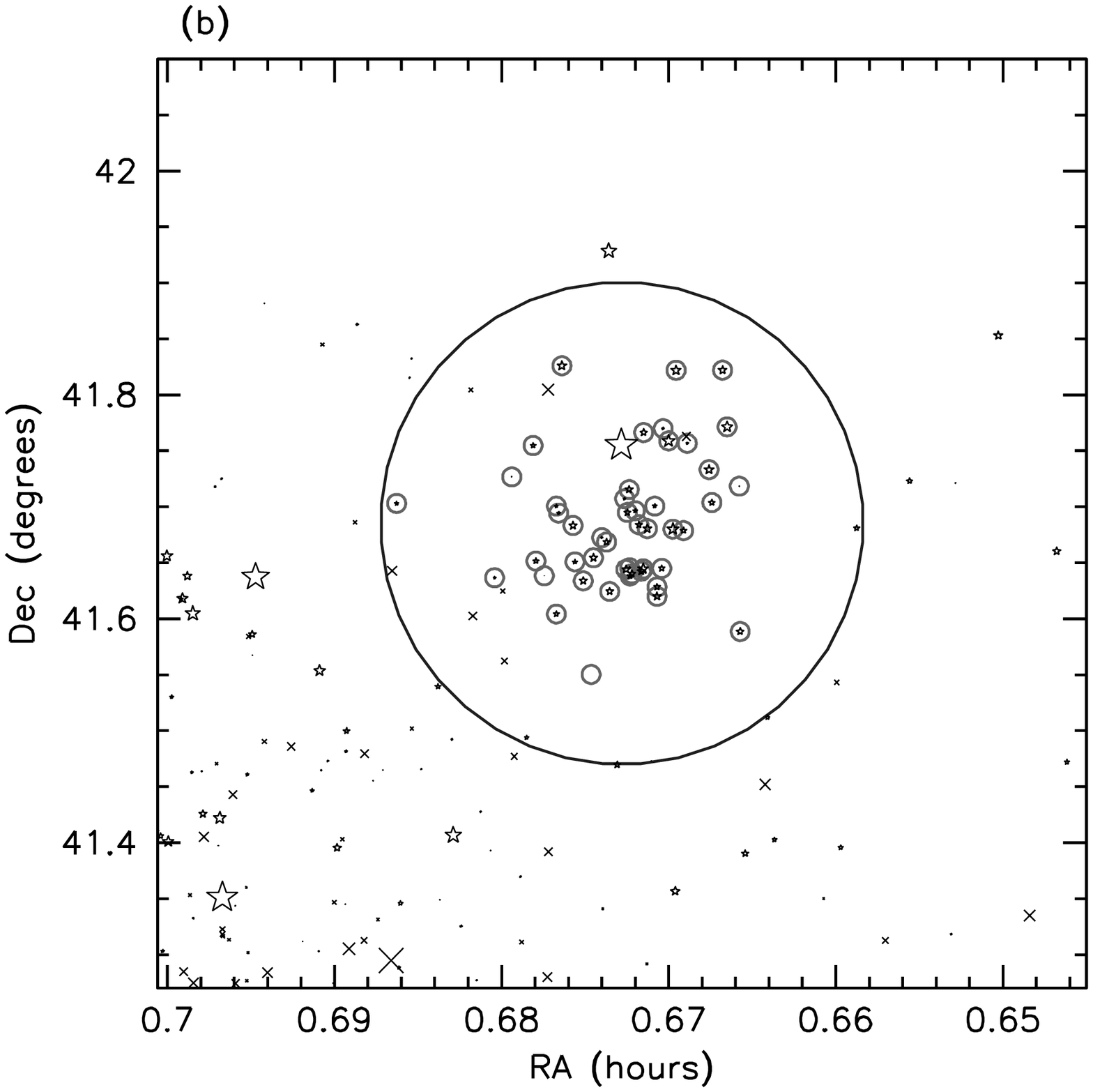}
  \includegraphics[width=0.49\textwidth]{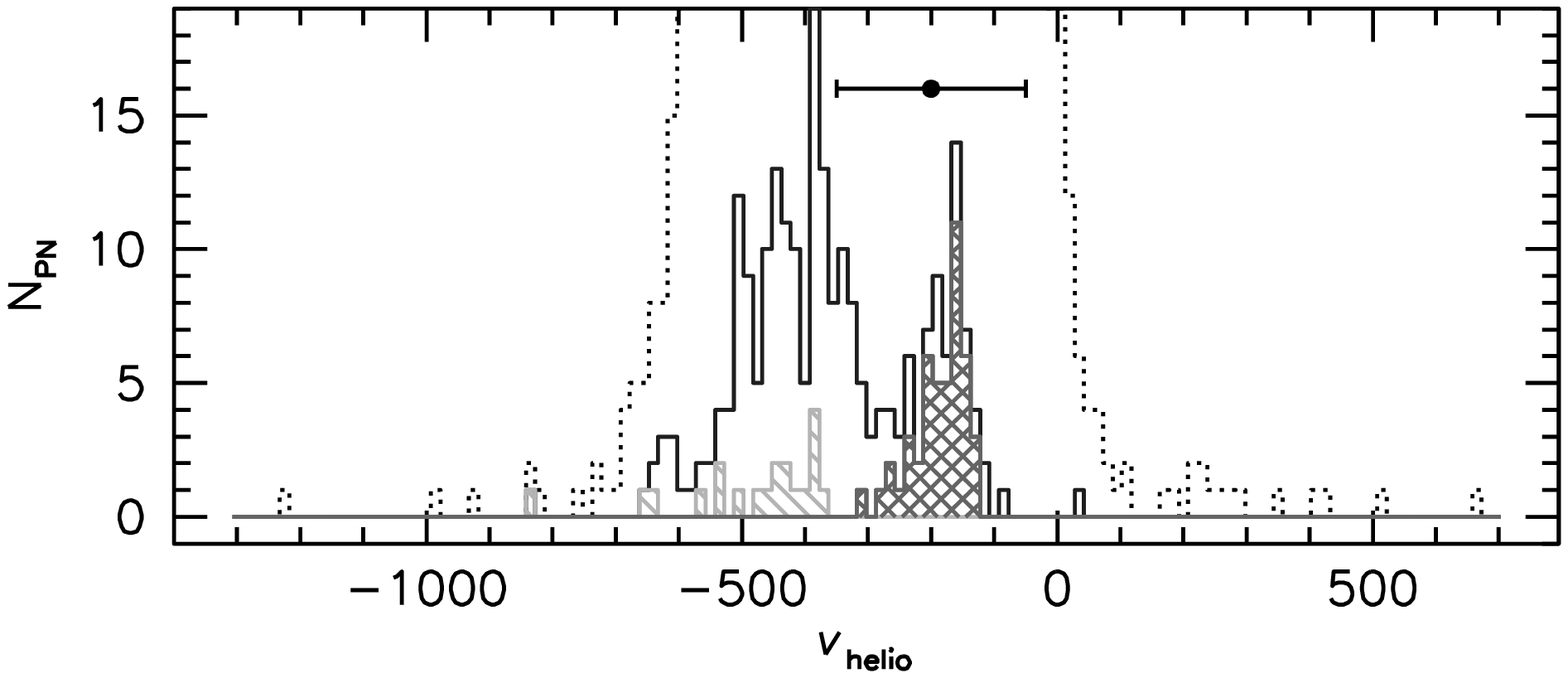}
  \includegraphics[width=0.49\textwidth]{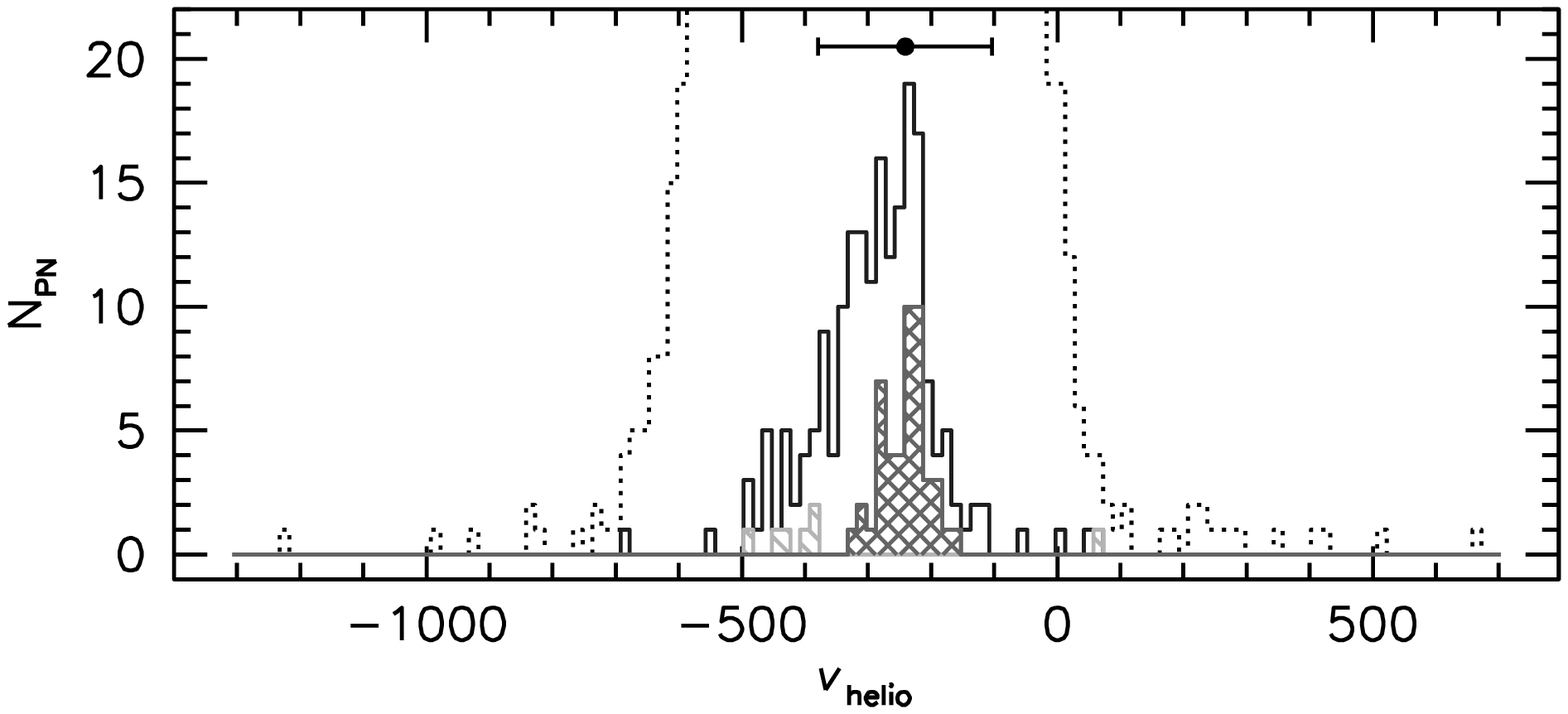}
  \includegraphics[width=0.32\textwidth]{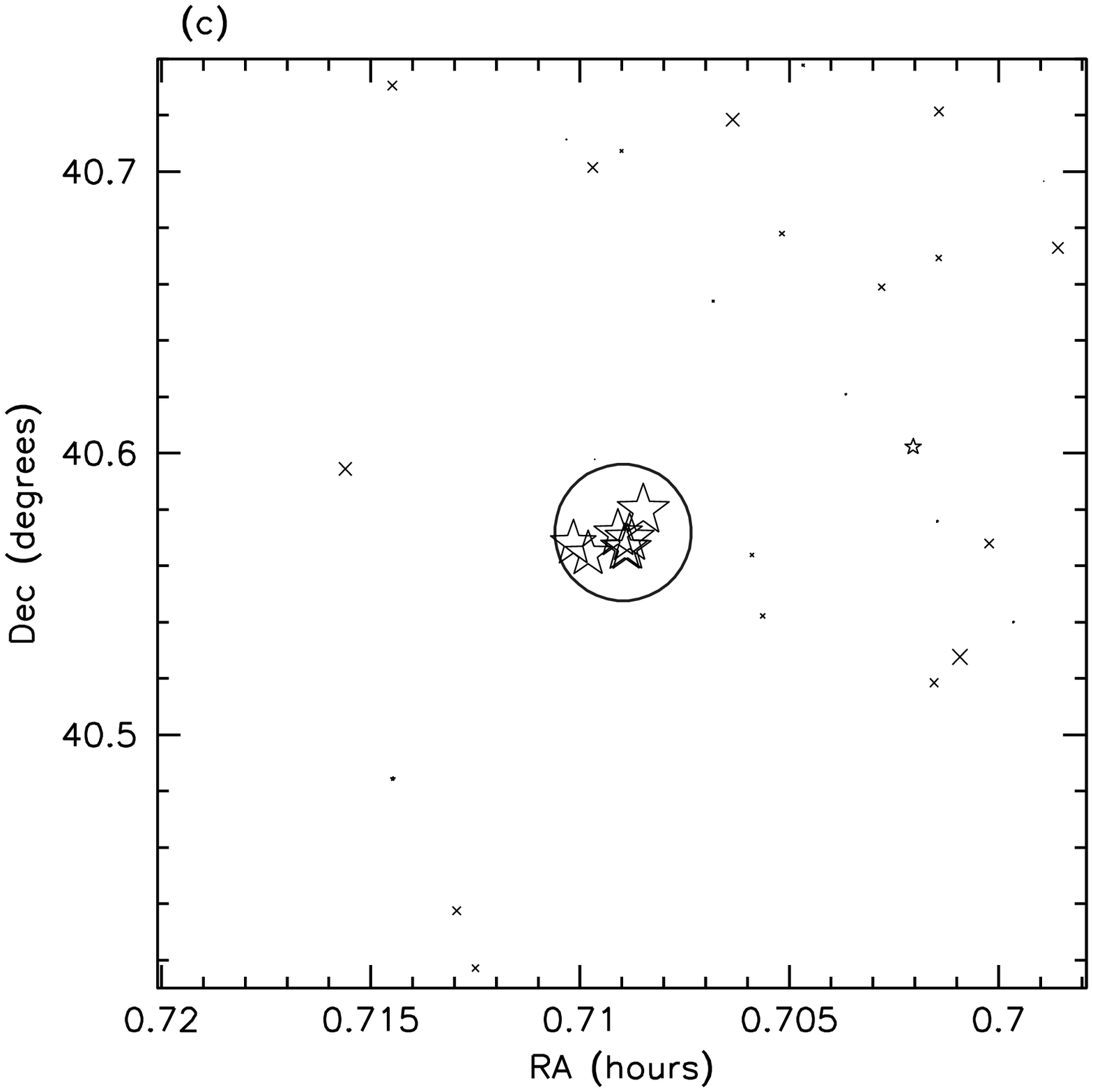}
  \includegraphics[width=0.32\textwidth]{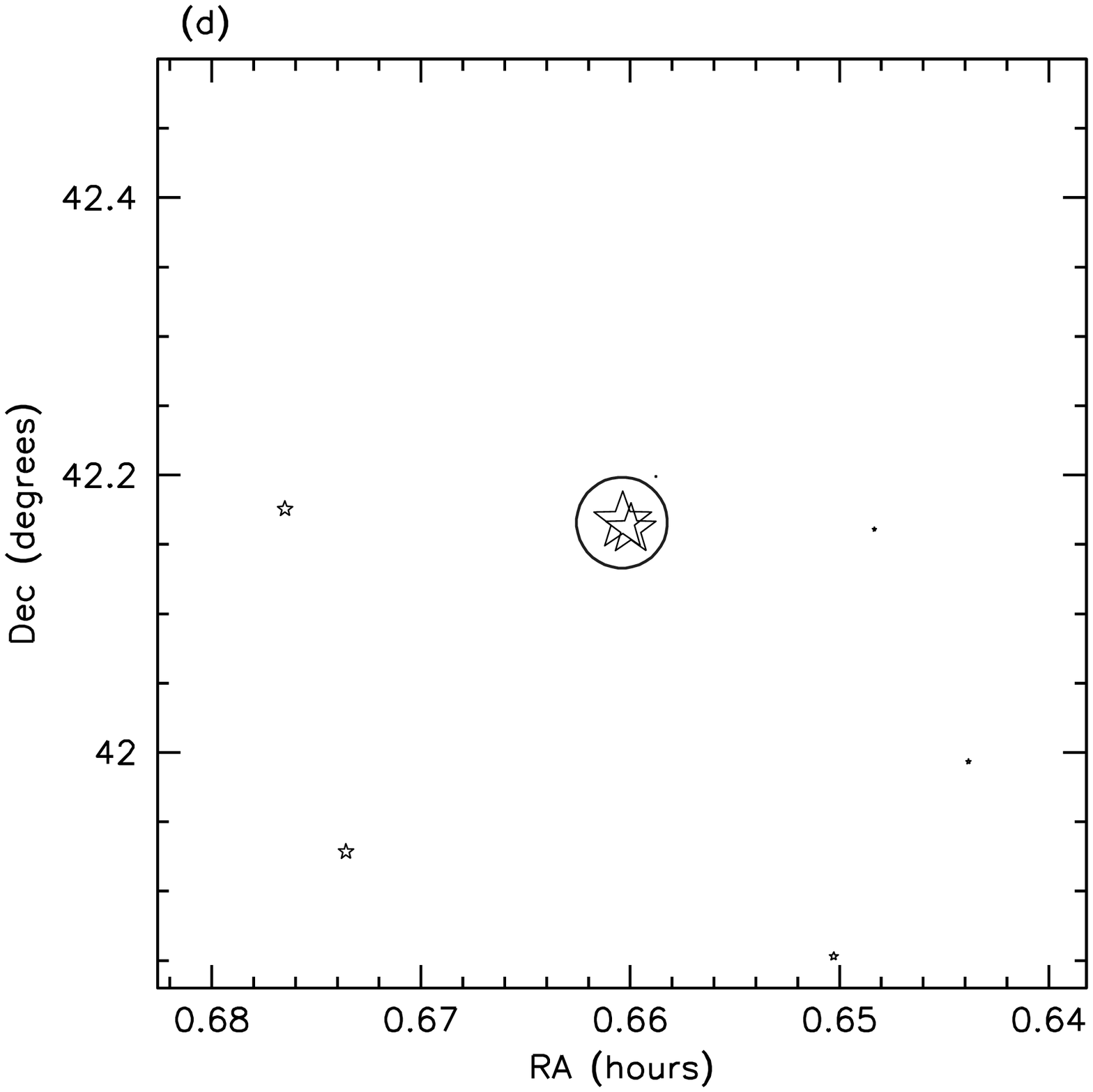}
  \includegraphics[width=0.32\textwidth]{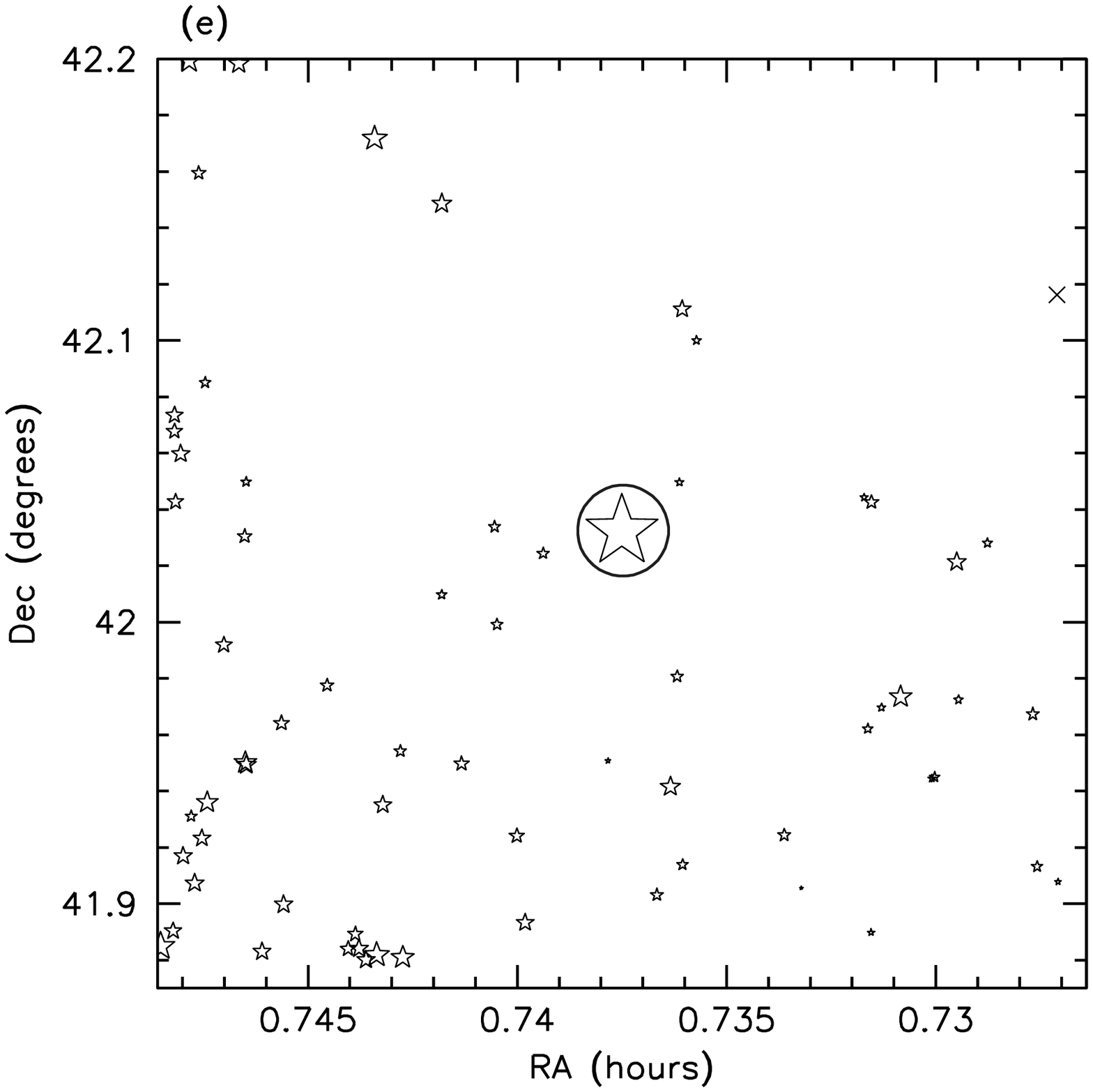}
  \includegraphics[width=0.32\textwidth]{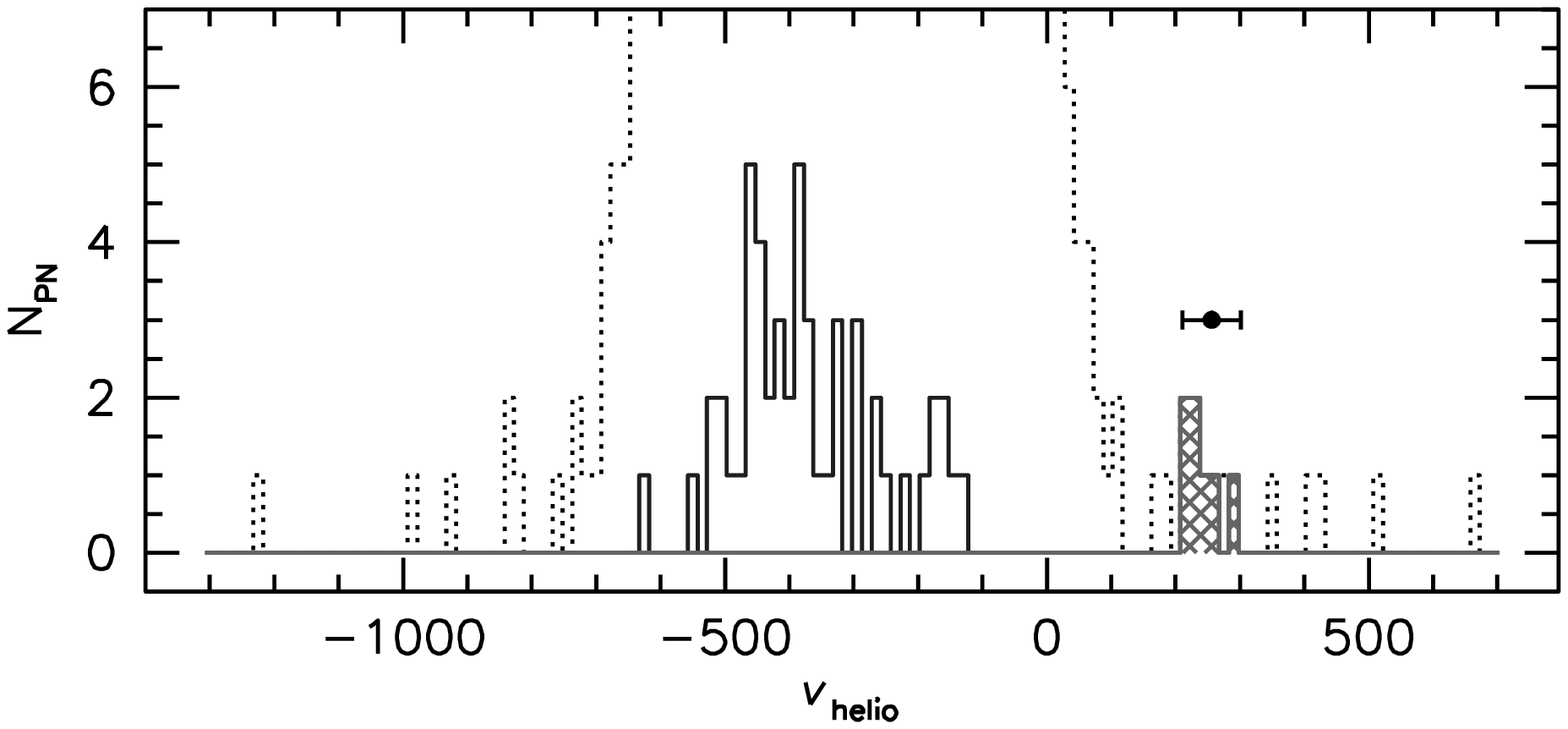}
  \includegraphics[width=0.32\textwidth]{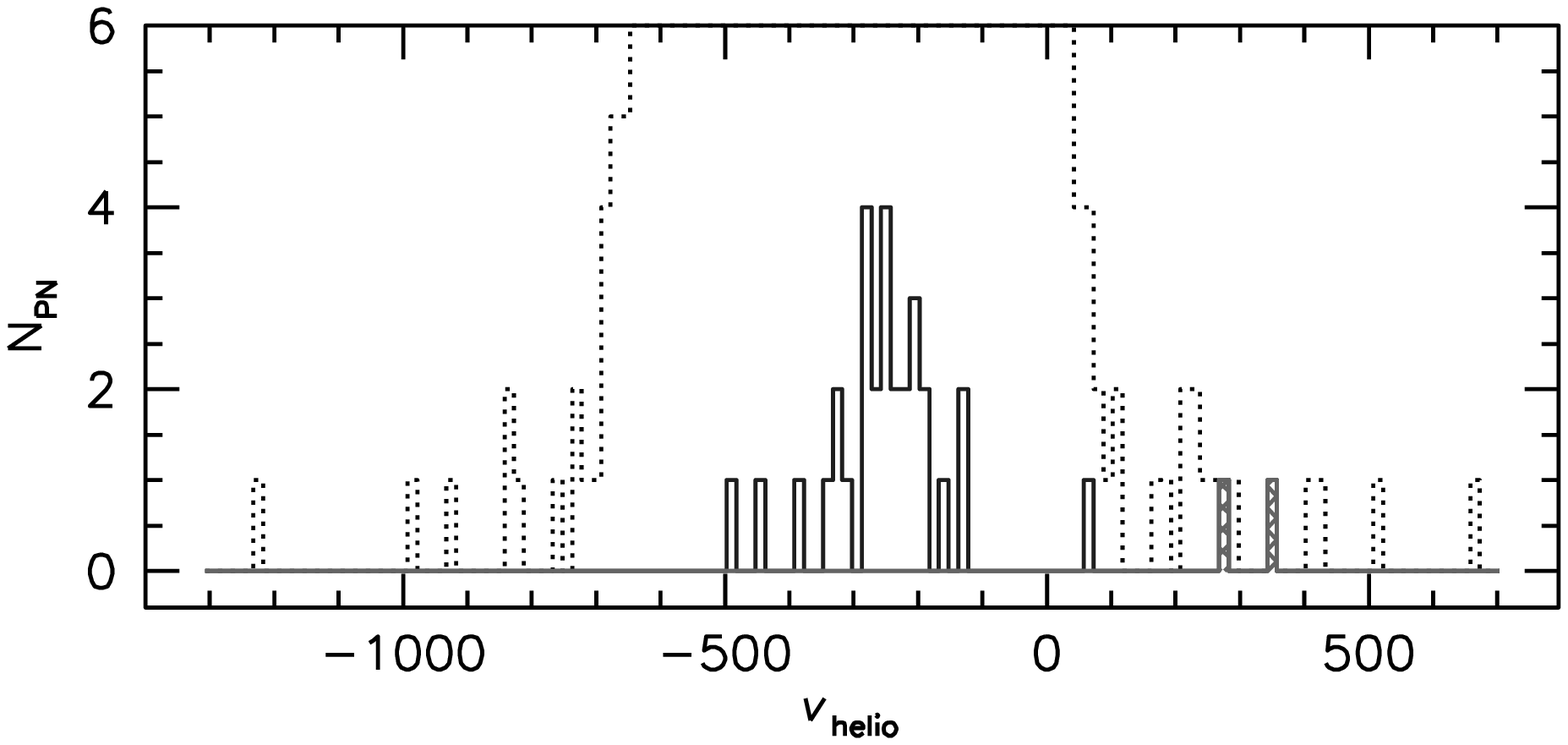}
  \includegraphics[width=0.32\textwidth]{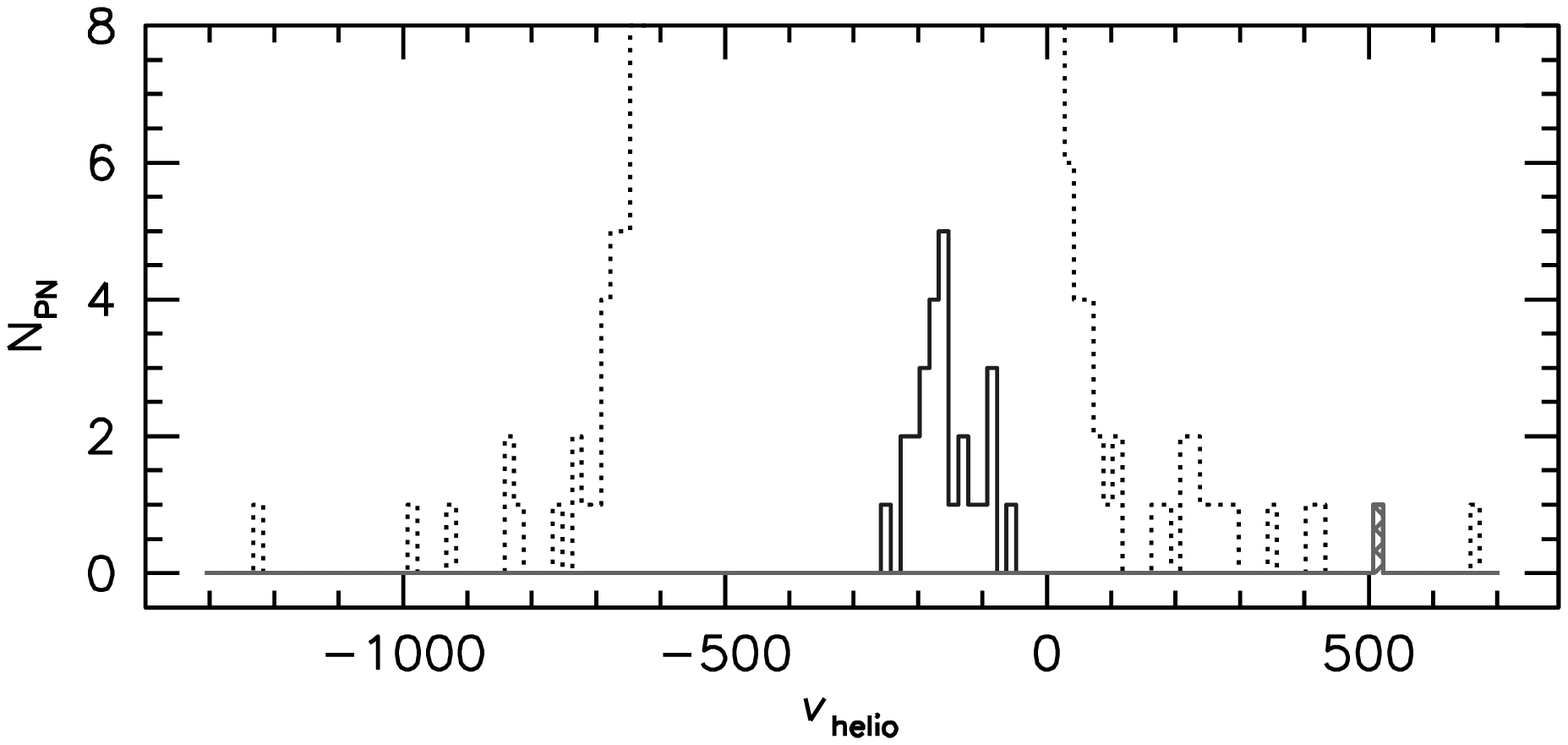}
  \caption{PN.S data in the vicinity of: (a) M32, (b) NGC 205, (c) And
    IV, (d) 2MASXi J0039374+420956, (e) MLA93 0953.  The area around
    each minor galaxy is shown in the upper panels.  The sizes of each
    symbol indicate an emission-line object's velocity with respect to
    M31, with crosses showing negative values and stars showing
    positive values.  The large circles indicate the approximate
    spatial extent of M32 and NGC 205 (1.5 times the mean 25.0 B-mag
    arcsec$^{-2}$ radius) and the location of the smaller galaxies.
    For M32 and NGC 205 PNe with circles around them have both
    position and velocity consistent with membership of that galaxy.
    The lower panels shows velocity histograms, the dotted line being
    the whole PN.S sample, the solid line indicating an area including
    the minor galaxy but extending well beyond its optical extent, the
    hatched histograms showing the PNe within the area of galaxy, and
    the cross-hatching highlighting PNe that agree in velocity as well
    as position; the points indicate the smaller galaxy's systemic
    velocity (from the literature, if known), with $3\sigma_v$
    errorbars.  
  }  \label{sats_ind}
\end{figure*}

\subsection{M32}

M32 (NGC 221), is a compact dwarf elliptical galaxy (cE2) projected in
line with the disk of M31. Its major and minor axes measure $\sim$
8\farcm7 and 6\farcm75 in diameter respectively to a B-band magnitude
level of 25.0 mag arcsec$^{-2}$ \citep{devauc91}.  It has a system
velocity of $-200\pm6$~\kms\ \citep{huchra99}, and a velocity
dispersion of $50\pm10$~\kms, rising to $\sim80$~\kms\ at its centre
\citep{simien02}. 

We have considered all the emission-line objects in the PN.S survey
within a 5\farcm7 radius (1.5 times the mean 25.0 mag arcsec$^{-2}$
B-band radius) of the  
centre of M32 as candidate members of the satellite (see
Figure~\ref{sats_ind}).  Using the PNe at somewhat larger distances from
M32 to characterise the kinematics of M31's disk in this region, we
find that a velocity cut at $-350$~\kms\ (three times the measured
dispersion from M32's systemic velocity) yields a reasonably clean
division between the two systems.  However, it is possible that a few
of the PNe on the M31 side of the cut may be objects with the most
extreme velocities in M32.  To warn of this possibility, objects in
the catalogue are flagged as ``M32'' if they meet the cut in velocity
as well as position, and ``M32?'' if they are simply coincident with
M32's location.

The sample of likely M32 PNe has only 46 members, so there is
not much that we can do in terms of detailed dynamical modelling.  A
simple Gaussian fit to the velocity distribution yields a mean
velocity of $-174\pm6$~\kms\ and a velocity dispersion of
$36\pm5$~\kms. The mean velocity is not in particularly good agreement
with the usually-adopted value; the difference seems to arise from the
statistical coincidence that most of the PNe in M32 reside on one side
of the galaxy's minor axis (see Figure~\ref{sats_ind}), so the measured
mean velocity contains an element of the system's rotation; the
somewhat low value for the velocity dispersion is also consistent with
this interpretation.

\subsection{NGC 205}

NGC 205 (M 110) is a dwarf elliptical galaxy (dE5) projected
$\sim$35\arcmin\ from the centre of M31, close to its minor axis.  Its
major- and minor-axis diameters are $\sim$21\farcm9 and
$\sim$11\farcm0 respectively at a position angle of 170\degr\
\citep{devauc91}. It has a mean velocity of $v_{\rm NGC
205}=-241\pm3$~\kms\ \citep{bender91}, and a mean velocity dispersion
of $\sigma_{\rm NGC 205}=46\pm8$~\kms\ dipping to $\sim 20$~\kms\ at
the very centre of the galaxy \citep{mateo98,carter90,simien02}.

To identify PNe in NGC~205, we have selected the objects within
12\farcm3 of NGC 205's centre (1.5 times the mean 25.0 mag
arcsec$^{-2}$ B-band radius), and applied velocity cuts at
$-375$~\kms\ and $-102$~\kms\ to seek to isolate the galaxy's members.
These velocity selections lie at approximately $v_{\rm NGC 205} \pm
3\sigma_{\rm NGC 205}$, although the lower bound is placed slightly
high to avoid excluding M31 PNe that are seen to have such velocities
at slightly larger distances from NGC 205.  As with M32, the possible
overlap at these velocities means that there is unavoidably some
ambiguity in the identification of satellite members, so
Table~\ref{tab.cat} annotates those objects that meet the velocity 
constraints as ``NGC205,'' while those that are just spatially
coincident are marked as ``NGC205?''  A Gaussian fit to the velocity
distribution of likely members yields a mean velocity of
$-235\pm6$~\kms and a velocity dispersion of $35\pm9$~\kms, in
agreement with the published values.

\subsection{Andromeda IV} 

\begin{table*}
\centering
\begin{minipage}{0.85\textwidth}
\caption{Emission-line objects found in the vicinity of And IV, and
  comparison to the \citet{ferguson00} data set.} \label{tab.and4}  
\begin{tabular}{ccccccccl}
  \hline
  \multicolumn{6}{l}{PN.S survey$^a$} & \multicolumn{2}{l}{\citet{ferguson00}$^b$} & Notes\\
  ID   & RA     & Dec & r & $m_{5007}$ & $v_{\rm helio}$ & ID & $v_{\rm helio}$ & \\
  \hline
  2110 &  0:42:20.3 & 40:32:31.8 & 174 &  25.27 & -386.7 & F7 &            &  Below flux ratio cutoff, Belongs to M31 \\
  2111 &  0:42:35.3 & 40:33:51.0 &  44 &  24.14 &  219.2 &    &            &  Not found by \citet{ferguson00}\\
  2112 &  0:42:21.2 & 40:33:50.0 & 130 &  21.74 & -372.5 & F8 &            &  Belongs to M31   \\
  2113 &  0:42:32.2 & 40:33:58.8 &  20 &  22.53 &  236.2 & F3 & $244\pm15$ &  Extended     \\
  2114 &  0:42:36.6 & 40:34:04.8 &  51 &  23.47 &  213.3 & F2 &            &              \\
  2115 &  0:42:31.7 & 40:34:11.1 &  10 &  23.13 &  253.9 & F4 & $250\pm13$ &  Extended     \\
  2117 &  0:42:32.7 & 40:34:17.0 &   5 &  25.20 &  248.8 & F5 &            &        \\
  2119 &  0:42:30.6 & 40:34:47.3 &  35 &  24.43 &  285.7 & F6 & $273\pm19$ &  Extended     \\
  2128 &  0:42:31.9 & 40:33:58.7 &  20 &  24.79 &  228.5 &    &            &  Not found by \citet{ferguson00}\\
       &  0:42:37.6 & 40:38:12   & 241 &       &         & F1 &            &  Not found in PN.S survey   \\
  \hline
\end{tabular}
\flushleft
$^a$ ID, RA, Dec, $m_{5007}$ and $v_{\rm helio}$ from
Table~\ref{tab.cat} except object F1, which was not detected. $r$ is
the objects radial distance from And IV measured in arcseconds. \\
$^b$ ID and velocity from \citet{ferguson00} Tables 1 and 2.\\
\end{minipage}
\end{table*}

\begin{figure}
  \centering
  \includegraphics[width=0.35\textwidth]{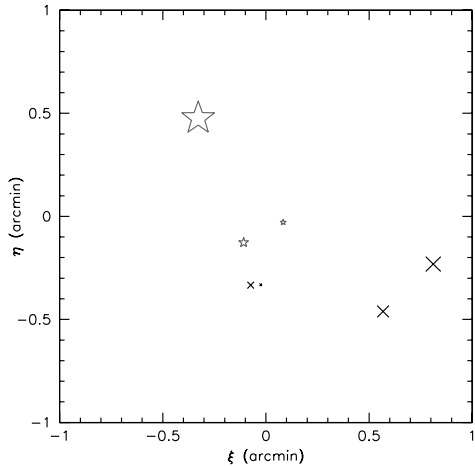}
  \caption{Kinematics of emission-line sources in Andromeda IV.  The
  coordinates are defined relative to the centre of the satellite.
  Point sizes are proportional to velocity with respect to And~IV, 
  with grey stars and black crosses respectively representing positive
  and negative values respectively.  }  \label{andiv_kin}
\end{figure}

Andromeda IV is a background dwarf irregular galaxy lying a projected
distance of 40\arcmin\ from the centre of M31. \citet{ferguson00} have
previously identified a number of emission-line objects associated
with this galaxy, and we list the cross-identifications in the current
survey in Table~\ref{tab.and4}. The only object that we fail to
recover from the \citet{ferguson00} study is their ``F1'' source,
which was originally detected in \halpha, but has no \oiii\
counterpart in the magnitude range of our survey.  The
\citet{ferguson00} objects F7 and F8 are located a little away from
the galaxy centre, and our measurements show that their velocities are
consistent with the disk of M31, so they do not appear to be members of 
And IV, and they are excluded from the remaining
analysis. Table~\ref{tab.and4} also adds two new emission-line sources
to the list of objects with positions and velocities consistent with 
membership of And IV.  

The resulting list of seven likely And~IV members has a mean velocity
of $241\pm10$~\kms, in agreement with the published value of
$256\pm9$~\kms, as measured from spectra of only three sources
\citep{ferguson00}.  As Figure~\ref{andiv_kin} illustrates, the
velocities do not appear to be randomly distributed through the
system, but indicate a line-of-sight component of rotation of $\sim
35$~\kms\ about a centre of $\rm RA_0 \approx 0^h 42^m 32\fs3$, $\rm
Dec_0 \approx 40\degr 34\arcmin18'$.

\subsection{2MASXi J0039374+420956}

Two PN.S emission-line objects that exhibit velocities inconsistent
with M31, one of which is extended, were found in the vicinity of this
extended infrared 2MASS source (marked as 2MASXi in
Table~\ref{tab.cat}). The two objects lie at projected radii of
2\farcs8 and 24\farcs2 from the source centre. With only one emission
line detected, we cannot unequivocally confirm its nature, but if we
assume it originates from \oiii, then this system would be placed at a
heliocentric velocity of $\sim$300~\kms. As this velocity is similar
to that derived for And IV, it is possible that both these galaxies
belong to the same nearby background group.  However, further
observations would be required to confirm that this is the case.

\subsection{MLA93 0953}

One object with a velocity inconsistent with M31 was also listed in
the \citet{meyssonnier93} catalogue of M31 emission-line objects.
They identified it as either a Wolf--Rayet star or a background QSO.
Given our velocity measurement, it seems most likely that the object
is a background QSO with an emission line that happens to lie within
the \oiii\ bandpass.

\subsection{Andromeda VIII}

\begin{figure}
  \includegraphics[width=0.475\textwidth]{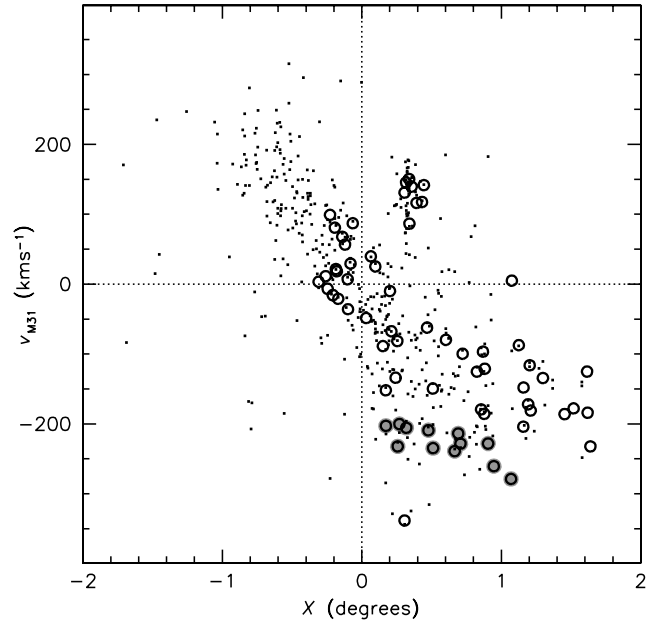}
  \caption{Plot of major-axis coordinate versus line-of-sight
  velocity, highlighting the postulated location of the And~VIII
  satellite.  The small dots show the PN.S survey data with $Y$ coordinates
  between $-0.175\degr$ and $-0.450\degr$, the region in which the 
  known And VIII candidate PNe lie. The open circles are the
  \citet{hurleyk04} data set within the same slice in $Y$, with the
  highlighted points indicating the suggested members of And~VIII.
  }\label{satand8}
\end{figure}

\begin{figure}
  \includegraphics[width=0.475\textwidth]{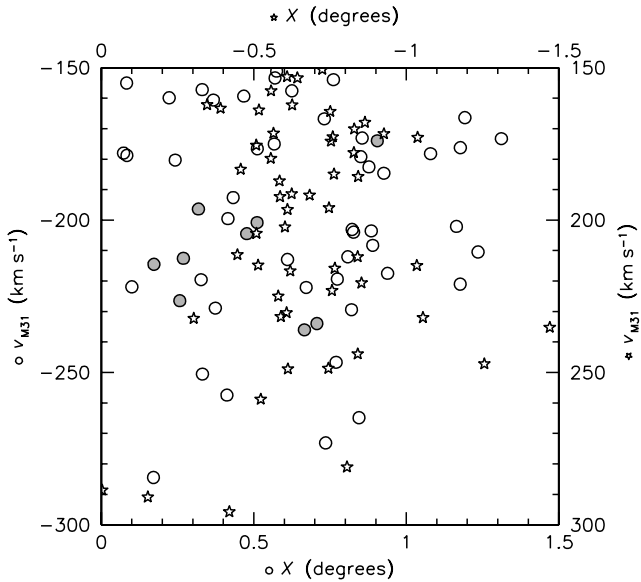}
  \caption{Overlay of the major-axis coordinate versus line-of-sight
  velocity for the PN.S data in the location of And~VIII (circles) and
  the equivalent phase-space location on the opposite side of M31
  (stars). The PNe previously identified as possible And~VIII members are
  highlighted in grey.
  }\label{satand8_b}
\end{figure}

\citet{morrison03} pointed out a coincidence in position and velocity
of a number of globular clusters and PNe from the \citet{hurleyk04}
survey.  These objects appeared to be quite well separated in
velocity from both the modelled kinematics of M31's thin disk in this
region and the nearest other PNe, leading \citet{morrison03} to
suggest that these sources might be providing the signature of a new
tidally-distorted satellite galaxy in M31, Andromeda~VIII.

However, as Figure~\ref{satand8} illustrates, with our larger sample
of PNe this gap in velocities between the disk of M31 and And~VIII is
no longer devoid of PNe. The errors on the new velocities are somewhat
larger than those previously published, but not sufficiently so to
fill in the apparent $\sim50$~\kms\ gap in the original smaller data
set.  Perhaps more tellingly, Figure~\ref{satand8_b} shows that the
equivalent region on the opposite side of M31 contains a very similar
distribution of PNe. In fact the equivalent regions in all four
quadrants of M31 are similarly populated by PNe. We therefore conclude
that while these PNe may not be members of M31's thin disk, they do
probably constitute a normal component of the system's PN
distribution, and it is not necessary to invoke the external
explanation of a tidally-distorted dwarf companion.

\subsection{Other objects flagged in the PN.S catalogue}

On the basis of the analysis in this section, all objects whose
kinematics are inconsistent with M31 have either been identified with
known satellites, or they have been previously flagged as non-PNe on
the basis of their extended nature or line ratios (see
Section~\ref{Sconcom}).  The only other source that we flag as likely
not an 
M31 PN is object 2654.  Although this source's kinematics and position
are consistent with membership of M31, it is nearly 0.4 magnitudes
brighter than any other PN in M31.  Unfortunately, it lies outside the
region mapped by \citet{massey02}'s imaging, so we do not have the
line ratio diagnostic as a further discriminant.  However, given the
very sharp cut-off in the PN luminosity function (see Section~\ref{Spnlf}),
it is most unlikely that this object is a PN, so we flag it in the
catalogue and exclude it from the following analysis.

 
\section{The planetary nebula luminosity function}\label{Spnlf}

\begin{figure}
  \includegraphics[width=0.475\textwidth]{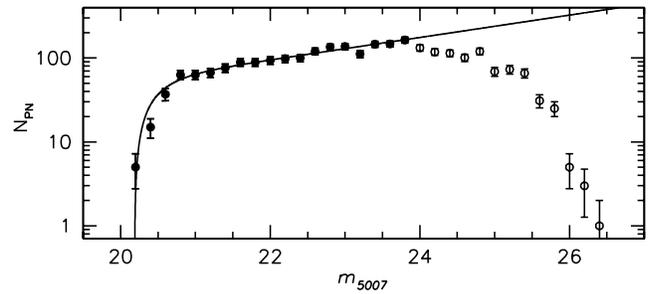}
  \caption{The planetary nebula luminosity function of M31 as derived
  from the current survey. Filled circles are plotted up to the
  magnitude completeness limit; open circles are shown beyond this
  point.  }  \label{pnlf_m31}
\end{figure}

Planetary nebulae are observed to follow a well-defined luminosity
function of the form
\begin{equation}
  N(M_{5007}) \propto e^{0.307 M_{5007}}\left(1-e^{3(M_{5007}^*-M_{5007})}\right)
\label{eqPNLF}
\end{equation}
\citep{ciardullo89}, with an equivalent formula for apparent
magnitudes for any individual system.  The absolute magnitude of the
bright-end cutoff, $M^*_{5007}$, has a surprisingly universal value of
$-4.5$, varying only weakly with the host galaxy metallicity
\citep{ciardullo02}.  Figure~\ref{pnlf_m31} shows the PNe from this
survey, with non-members of M31 removed, fitted by this function.  As
the figure shows, this functional form fits very well all the way to
our completeness limit of $m_{5007}=23.75$.  The apparent magnitude of
the bright end cutoff, for the sample is $m^*_{5007} = 20.2 \pm 0.1$.
This result is in excellent agreement with the value published by
\citet{ciardullo89} of 20.17.

\begin{figure}
  \includegraphics[width=0.475\textwidth]{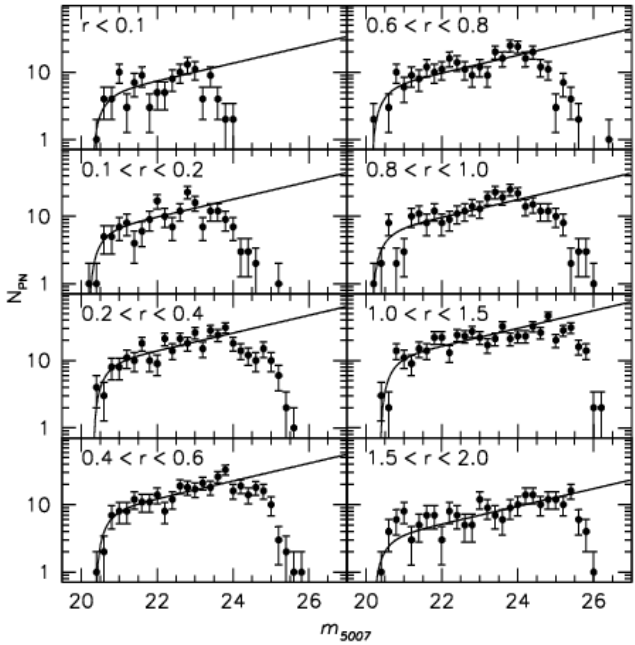}
  \caption{The planetary nebula luminosity function at different
  deprojected disk radii in M31.} \label{pnlf_m31_rad}
\end{figure}

The large number of PNe in our survey means we can divide up the
planetary nebula luminosity function (PNLF) and examine it at
different points within the galaxy to test the universality of its
form.  PNLFs for sections at different deprojected disk radii are shown in
Figure~\ref{pnlf_m31_rad}.  The positional variation in the completeness
of the survey is clearly shown in this figure: as discussed above,
the high surface brightness and poor seeing of the central fields mean
that these data are only complete to $m_{5007}\sim23$.  For radii
between 0.2\degr\ and 1.0\degr, incompleteness becomes apparent at
$m_{5007}\sim24$, while at the largest radii, for which we used the
best seeing conditions (see Figure~\ref{seeing}), we are close to
complete at $m_{5007}\sim25$.

Recent observations of PNe in the SMC \citep{jacoby02} and M33
\citep{ciardullo04} have detected deviations from the simple monotonic
function of equation~(\ref{eqPNLF}). In these objects, dips have been
observed in the PNLF, whose origins have been proposed as resulting
from a recent star formation episode.  
The only indication of a dip in the PN.S data set is the
point at 23.2~mag in Figure~\ref{pnlf_m31} which lies $\sim2.5\sigma$
from the model value.  It is interesting to note from
Figure~\ref{pnlf_m31_rad} that most of the signal responsible for this
dip comes from the radial range $0\fdg6<r<0\fdg8$ where the data
dips $\sim2\sigma$ below the model value at 23.2~mag. This is
coincident with the star formation ring where the majority of \hii\
regions are located discussed in Section~\ref{Sconcom}, suggesting
that star formation may be responsible for this marginally-detected
feature as well.

\begin{figure}
  \includegraphics[width=0.475\textwidth]{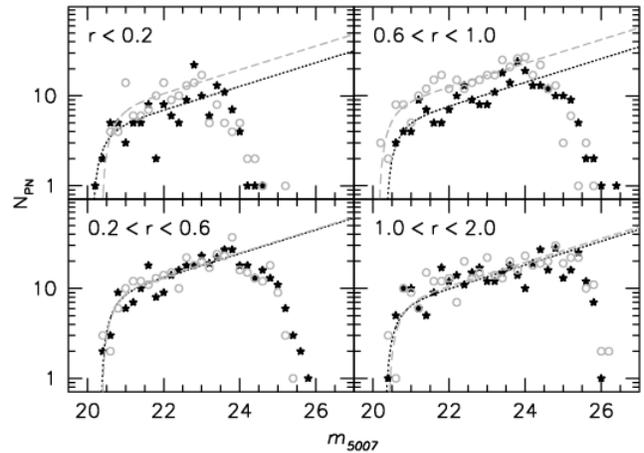}
  \caption{The planetary nebula luminosity function on the near and
  far side of M31 over different radial ranges.  Black stars and the
  dotted line represent the data and fit to the near side of M31 (to
  the north-west), while grey circles and the dashed line show the
  corresponding information for the far side (to the south-east).  }
  \label{pnlf_m31_rad_Y}
\end{figure}

\begin{figure}
  \includegraphics[width=0.475\textwidth]{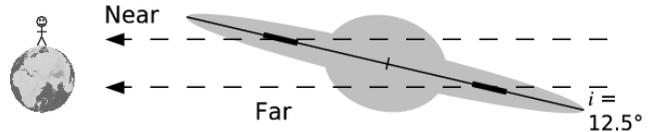}
  \caption{An illustration of the location of an Earth based observer
  with respect to the disk of M31. The dust lane is shown as the thick
  lines in the disk plane.  On the near side of the disk the majority
  of light comes from behind the dust lane while opposite is true for
  the far side.  Not to scale.
  }   \label{disk_ext_mod}  
\end{figure}

More radically, \citet{marigo04} have suggested that the brightest
PNe all originate from intermediate-mass stars, so that the bright end
of the PNLF should vary dramatically between spheroidal and disk
populations.  The absence of any significant variation in M31's PNLF
with radius, as one goes from the bulge-dominated to the
disk-dominated parts of the galaxy, implies that no such effect is
seen in this system. 

As a further test of the universality of the PNLF, and to search for
the effects of obscuration, we have also divided the data according to
whether it comes from the near side (positive $Y$) or far side
(negative $Y$) of the disk.  As Figure~\ref{pnlf_m31_rad_Y} illustrates,
at most radii the PNLFs are statistically indistinguishable.  However,
once again in the region of the ring at $0\fdg6<r<1\degr$, we do see a
small but significant difference, with the PNLF from the near side
systematically fainter than that from the far side: values of the best
fit for $m^*_{5007}$ on the near and far side are 20.34 and 20.15
respectively.  This rather counter-intuitive difference can be
qualitatively explained if the ring of star formation also contains a
modest amount of dust obscuration \citep{walterbos88, dejong05}. As
Figure~\ref{disk_ext_mod} shows, on the far side of the disk, the line
of sight of an observer through a highly inclined system like M31
passes through more of the galaxy in front of the dust layer than
behind, while the opposite is true on the near side.  Hence on the
near side of the disk most of the PNe observed are located behind the
dust layer, dimming the PNLF relative to that seen on the far side.
The lack of similar differences between other parts of M31 suggests
that obscuration is not a major issue across most of the galaxy.


\section{Spatial distribution of planetary nebulae}\label{Sscale}

In order to interpret the kinematics of the PNe in this survey, we
need to know what stellar population or populations these objects are
tracing.  It is generally assumed that PNe all originate from
low-to-intermediate mass stars, so they should mainly trace the old
stellar population, but, as we have discussed in Section~\ref{Spnlf}, it
has been suggested that some fraction of PNe originate from more
massive stars.  As a further test of this possibility, we can compare
the distribution of PNe with that of the red light from the galaxy,
which we would expect to be dominated by relatively low-mass red
giants, and hence would only be expected to follow the distribution of
PNe if they originate from similar low-mass systems. To this end, we
have obtained the radial distribution of the PNe with $m_{5007}<24$
(the completeness limit of most of the survey) along the major and
minor axes of M31. 

\begin{figure}
  \includegraphics[width=0.475\textwidth]{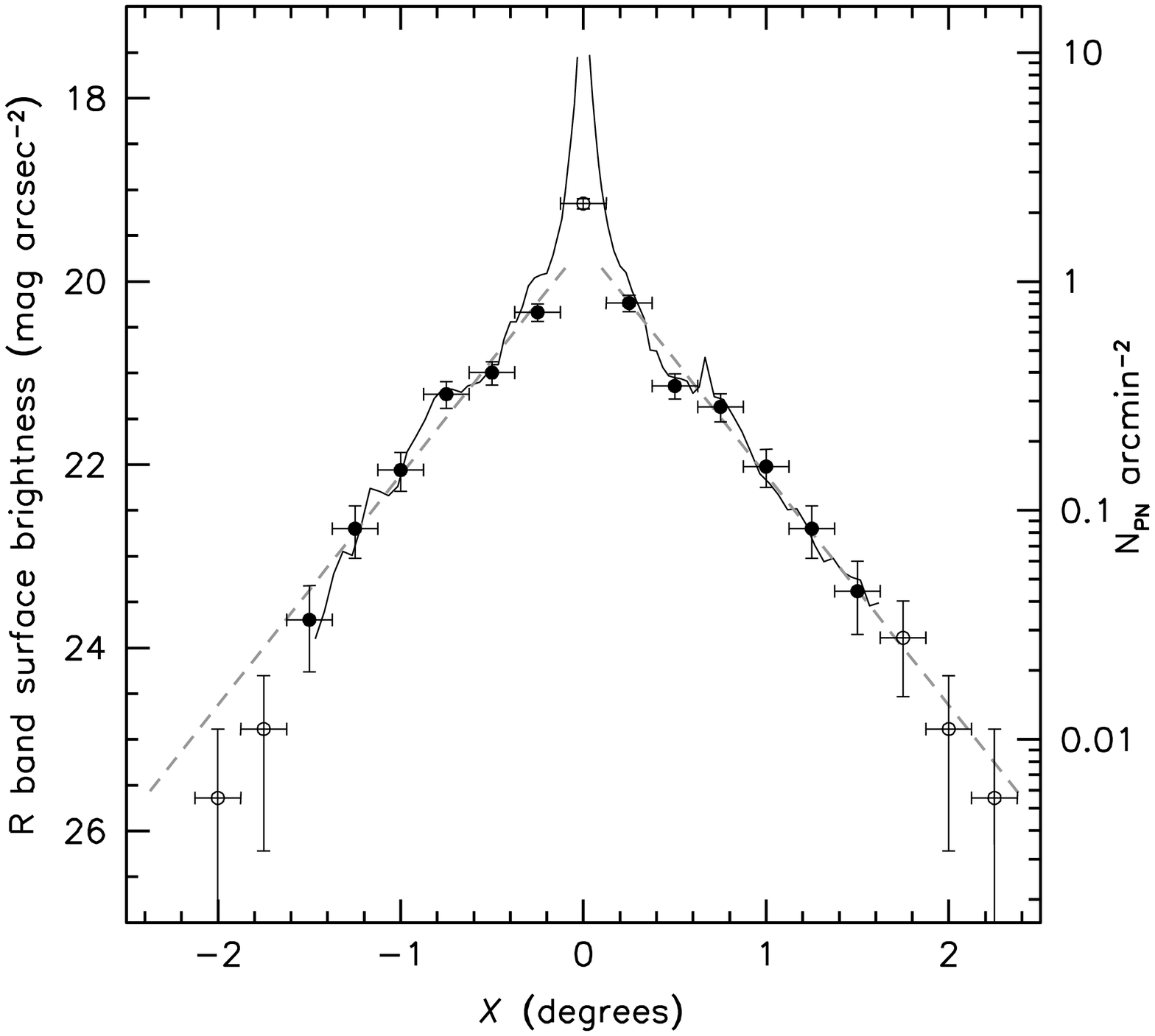}
  \caption{Surface brightness profile and PN number counts along the major
    axis of M31.  The solid line shows the R-band surface brightness
    profile of \citet{walterbos87a}, and the dashed line is an
    exponential profile with a scale length of 0\fdg43
    \citep{walterbos88}.  The number density of PNe with
    $m_{5007}<24$ is indicated by closed circles where the survey is
    complete both spatially and to $m_{5007}=24$; open circles are
    used elsewhere.} \label{scale_length_maj}
\end{figure}
Along the major axis, Figure~\ref{scale_length_maj} shows excellent
agreement between the number counts of PNe and the R-band photometry of
\citet{walterbos87a}.  Fitting an exponential function to the
distribution of PNe beyond the bulge and out to 1.6\degr\ yields a
scale-length of $R_d = 0.43\pm0.02$\degr\ ($5.9\pm0.4$~kpc), also in
excellent agreement with the R-band value of $0.43\pm0.02$\degr
\citep{walterbos88}.  There are no indications of any cut-off in the
exponential disk out to the limit of the survey at $\sim 5 R_d$.

\begin{figure}
  \includegraphics[width=0.475\textwidth]{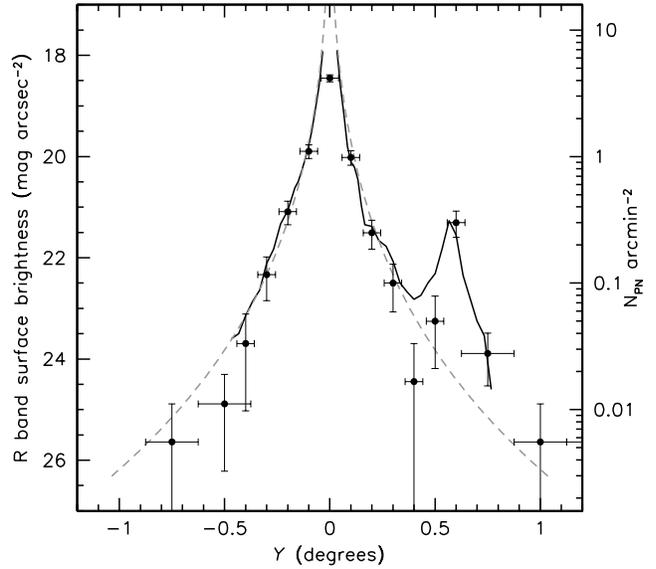}
  \caption{Surface brightness profile and PN number counts along the
    minor axis of M31.  The solid line is the R-band surface
    brightness profiles of \citet{walterbos87a}, and the dashed lines
    is the $R^{1/4}$ profile fit to the photometry by \citet{irwin05}.
    The number density of PNe with $m_{5007}<24$ is indicated by the
    filled circles.  } \label{scale_length_min}
\end{figure}
Figure~\ref{scale_length_min} shows a similar plot for the minor axis.
Once again, there is a good match to the published R-band photometry
outside the central region, including reproducing the bump at $Y \sim
0.6\degr$ due to NGC~205.  The PNe also follow the published $R^{1/4}$
profile fit to the photometry with a bulge effective radius of $R_{\rm
  eff}=0\fdg10$ (1.4~kpc) \citep{irwin05}.  Even at $\sim 10 R_{\rm
  eff}$, there are no signs of an excess over this fit, and hence no
evidence for a separate halo population even at these very faint
surface brightness levels (equivalent to $\sim 26$ mag in the R band).

It thus appears that the PNe trace the same stellar population as the
old stars seen in the galaxy's red light.  They also seem to broadly
fit a simple two-component model comprising an $R^{1/4}$ central bulge
and an exponential disk out to the very faint surface brightness
levels accessible using such discrete dynamically-selected tracers.


\section{The luminosity-specific PN density}\label{Salpha}

\begin{figure}
  \includegraphics[width=0.475\textwidth]{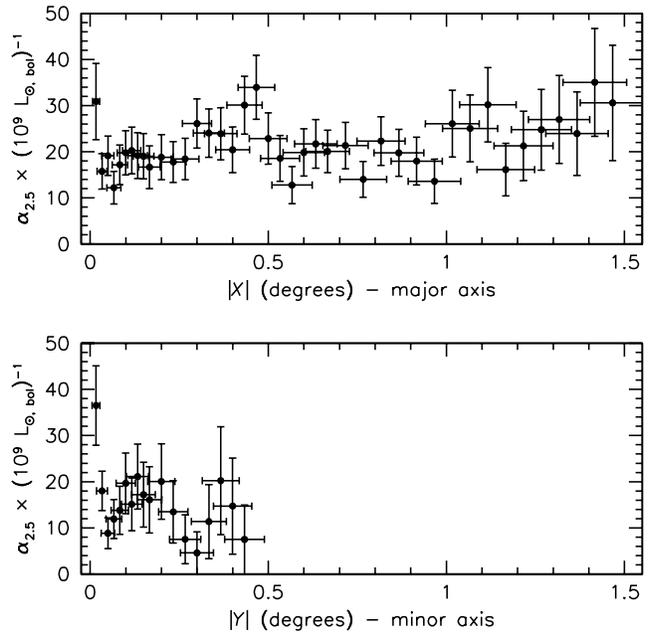}
  \caption{Variation of $\alpha_{2.5}$ with distance along the major
  and minor axes. $\alpha_{2.5}$ is estimated at positions along the two axes
  where \citet{walterbos87a} list a value for the surface brightness,
  the number density of PNe at each location is estimated from the
  number of PNe in a box whose dimensions are equal to the horizontal
  error bar. 
  }\label{alpha_2.5}
\end{figure}

Having established the proportionality between the stellar continuum
emission and the distribution of PNe, we now turn to the constant of
proportionality.  This quantity, the luminosity-specific PN number
density, is usually usually parametrised by the number of PNe in the
top 2.5~magnitudes of the PN luminosity function per unit bolometric
solar luminosity, and is given the symbol $\alpha_{2.5}$

The bolometric luminosity of any part of M31 can be calculated via the
standard relations
\begin{equation}
M_{bol}=M_V+BC_V
\end{equation}
and
\begin{equation}
\frac{L_{M31,bol}}{L_{\odot,bol}} = 10^{-0.4 \left( M_{M31,bol} -
  M_{\odot,bol} \right)},
\end{equation}
where $BC_V$ is the V-band bolometric correction which has a value of
$-0.80$~mag for M31 \citep{ciardullo89}. The solar absolute bolometric
magnitude, $M_{\odot,bol}$, is taken to be $+4.74$~mag \citep{cox00}.

The number density of PNe along the major axis and the V-band
photometry of \citet{walterbos87a} [correcting to absolute magnitudes
using $A_V=0.206$ \citep{schlegel98} and a distance modulus of 24.47]
give an average value of $\alpha_{2.5} = (21 \pm 1)\times
(10^9~L_{\odot,bol}) ^{-1}$.  As Figure~\ref{alpha_2.5} shows, outside
the very uncertain central bin there is no strong evidence for any
gradient in the value of $\alpha_{2.5}$ with position, as we would
expect on the basis of the close proportionality determined in the
previous section.  The same calculation on the minor axis, where the
light profile is dominated by the bulge, gives $\alpha_{2.5} = (15 \pm
2) \times (10^9~L_{\odot,bol}) ^{-1}$, once again with no evidence for a
gradient (see Figure~\ref{alpha_2.5}).  This value agrees very well
with the calculation by \citet{ciardullo89} for the PNe in M31's
bulge, from which they derived a value of $\alpha_{2.5} = 16.3 \times
(10^9~L_{\odot,bol})^{-1}$.  

The modest difference between the mean values for the minor and major
axes is consistent with the existing evidence that $\alpha_{2.5}$ is
correlated with galaxy colour, with red ellipticals being poorer in
PNe per unit luminosity than blue spiral galaxies \citep{peimbert90,
hui93}; here for the first time we are seeing evidence for such a
difference between spheroidal and disk components within a single
galaxy.  Fortunately, such a mild deficit in the number of PNe from
the bulge component relative to those from the disk will have little
impact on the dynamical modelling of these data, since it merely
changes the effective normalisation of the number density of tracers
in these two discrete components.  We can therefore proceed to make a
preliminary study of the PN kinematics in M31.  


\section{Planetary nebula kinematics}\label{Skin}

\begin{figure}
  \includegraphics[width=0.475\textwidth]{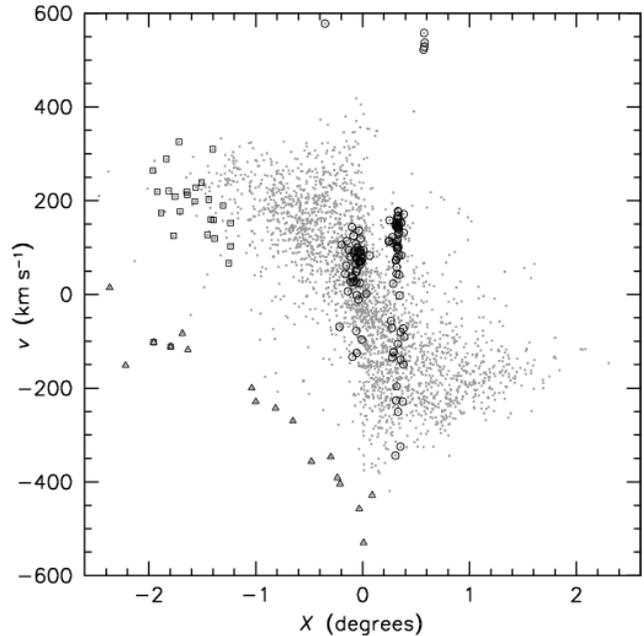}
  \caption{Scatter diagram of position along the major axis of M31
    versus velocity for all the survey PNe. The highlighted objects
    are those in the region of satellite galaxies (circles; see
    Section~\ref{Ssats}); those in the region of the Northern Spur
    (squares); and those identified as forming a continuation of the
    Southern Stream (triangles).  }
  \label{vels}
\end{figure}

In Figure~\ref{vels}, we show a plot of PN velocity against major axis
position.  A number of fairly subtle dynamical features are apparent
in this plot.  For example, many of the satellite galaxies identified
in Section~\ref{Ssats} (shown here as circles) stand out quite
clearly.  Similarly, the PNe identified by \citet{merrett03} as a
possible continuation of the merging Southern Stream of stars
\citep{ibata01, ibata04, mcconnachie03}, shown as triangles, are
apparent.  Interestingly, the other strong photometric structure, the
Northern Spur \citep{ferguson02}, does not stand out in the plot.  The
kinematics of PNe in this region (shown as squares in
Figure~\ref{vels}) are indistinguishable from those of the disk,
despite the Northern Spur's location some distance from the plane.  It
would thus appear that this feature is related to the disk, perhaps
indicative of a warp.  After these minor features have been excluded,
Figure~\ref{vels} shows that the kinematics are dominated by rotation
at large radii, but that at small radii there is a larger random
component, attributable to a bulge population.  This result indicates
that the majority of the PNe form a relatively simple system of
rotationally-supported disk and randomly-supported bulge, in agreement
with the conclusion based on photometry in Section~\ref{Sscale}.

\begin{figure}
  \includegraphics[width=0.475\textwidth]{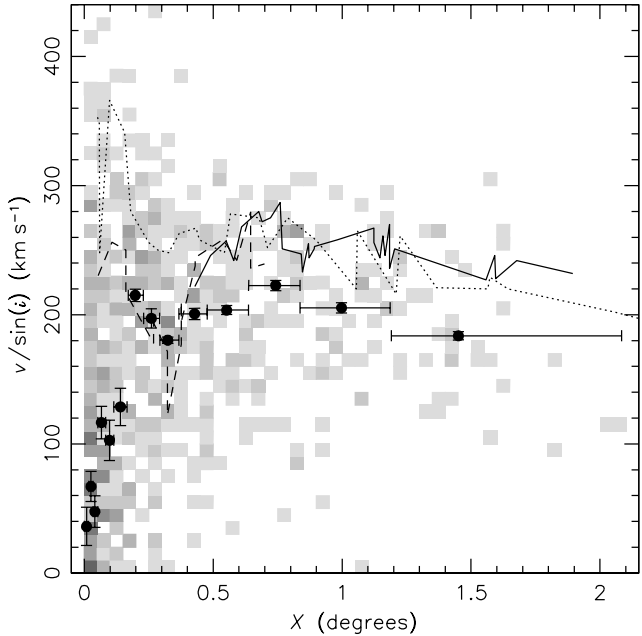}
  \caption{Plot of position versus velocity for PNe close to the major
  axis of M31 ($|Y|<0\fdg04$). The data from both sides of the galaxy
  have been combined, and the velocities have been corrected for the
  galaxy's inclination.  The grey-scale shows the density of PNe in
  this phase-space projection.  The circles show the mean velocities of
  these data in a number of bins -- the horizontal error bars show the
  extent of each bin, and the vertical bars the uncertainty in mean
  velocity.  The dotted line shows M31's \hi\ rotation curve
  \citep{braun91}; the dashed line shows the rotation curve derived
  from CO measurements \citep{loinard95}; and the solid line shows a
  rotation curve derived from \hii\ region velocities
  \citep{kent89a}.} 
  \label{rot}
\end{figure}
A full dynamical model fit to the entire dataset is beyond the scope
of this paper, but we can obtain some interesting insights by
considering the PNe that lie close to M31's major axis.  For a
highly-inclined galaxy like M31, the observable kinematics in this
region are dominated by the rotational motion, so we can essentially
isolate this component of velocity.  Accordingly, we have
selected a sample of 690 PNe with $|Y|<0\fdg04$, close to the major axis.
Figure~\ref{rot} shows a plot of this subsample's line-of-sight
velocities versus major axis position, together with the mean
velocities derived by radially binning these PNe into groups of 50
objects. A system velocity of $-309$~\kms\ has been adopted as
discussed previously.

The distribution of PN velocities in Figure~\ref{rot} has an upper
envelope that matches the rotation curves derived from various tracers
(also shown in this figure) reasonably closely at large radii.  At
small radii, the PNe show a wide spread in velocities, indicative of a
contribution from the bulge.  It is also interesting to note the dip
in the mean rotation speed of PNe at $\sim 0\fdg3$.  This feature,
which shows up even more clearly in the CO rotation curve, has been
attributed by \citet{loinard95} to non-circular motions associated
with a barred bulge.  Its appearance in the PNe kinematics confirms
that they are dominated by disk objects even at this fairly small
radius, but also serves as a warning that large non-circular motions
may complicate the observed dynamics at these radii.  At larger radii,
the mean velocities of the PNe trace the rotation curve relatively
well, except that the mean velocity lies systematically below the
circular speed.  We will return to the origins of this asymmetric
drift in the mean velocities of PNe in Section~\ref{Sasymd}, but first
we look in more detail at their random motions.

\begin{figure}
  \includegraphics[width=0.475\textwidth]{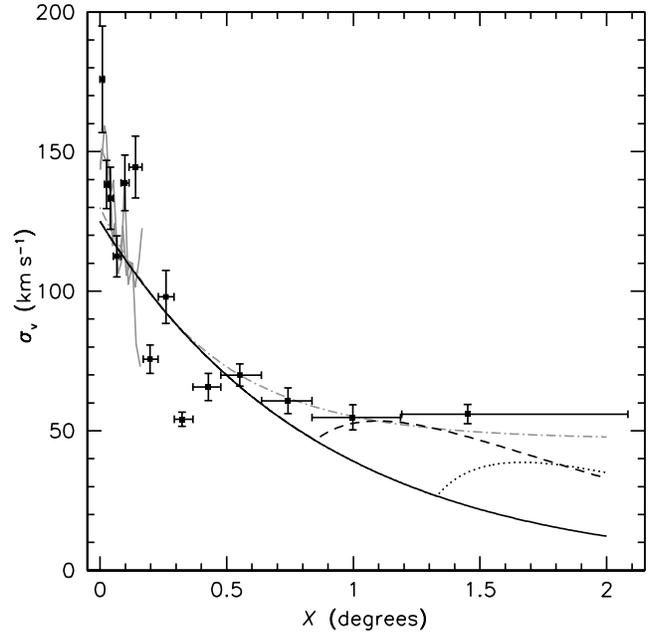}
  \caption{Velocity dispersion profile for the major-axis PN sample
    shown in Figure~\ref{rot}.  Points with error bars show the binned
    PN dispersion profile, corrected in quadrature for the uncertainty
    in the PNe velocities.  The solid grey line is the inner stellar
    velocity dispersion profile of \citet{mcelroy83}.  The curves are
    for a number of models fit to the PN data with a $\chi^2$
    minimisation: the grey dot--dash line shows the empirical fit of 
    equation~(\ref{eqdispfit}); the solid black line shows a simple
    exponential decrease; the dotted black line is for a flaring disk
    model with parameters from \citet{brinks84b}; and the dashed black
    line shows a flaring disk model with parameters matched to the PN
    velocity dispersion profile.  }
    \label{vdisp}
\end{figure}

\subsection{Velocity dispersion}\label{Svdisp}
Figure~\ref{vdisp} shows the observed velocity dispersion as a function of
position along the major axis obtained from the PNe in Figure~\ref{rot},
binned radially into groups of 50 objects.  Near the centre, the
calculated PN velocity dispersions agree well with the declining
profile derived by \citet{mcelroy83} from stellar absorption lines
(also shown in the figure), providing further confidence in the PN
results.  However, the stellar absorption-line data only reach to
$\sim0\fdg2$, while the PN profile goes out to $\sim 2\degr$ (27.4~kpc).  It is
in these previously-unexplored outer regions that the velocity
dispersion starts to behave a little strangely.  First, there is a dip
in the observed dispersions at $\sim 0.3\degr$, suggesting that the
bar's influence shows up in the dispersion profile as well as the mean
rotational velocity.  Second, the dispersion profile seems to flatten
out at larger radii to a constant value of $\sim 55$~\kms.  Ignoring
the bar-related dip, a simple empirical fit to the large-scale
profile, plotted as a grey dot--dash line in Figure~\ref{vdisp}, is
given by
\begin{equation}\label{eqdispfit}
\sigma_\phi = \sigma_A + \sigma_B \exp\left(-\frac{R}{R_d}\right).
\end{equation}
with $\sigma_A=47$~\kms, $\sigma_B=84$~\kms\ and $R_d=0\fdg43$.
Other recent stellar kinematic data from large radii also reveal
significant dispersions: \citet{ibata05} find velocity dispersions of
RGB stars of $\sim 30$~\kms\ from small fields at even larger
distances along the major axis.

These results are not consistent with the expectations of the simplest
disk models.  By applying the tensor virial theorem to a
one-dimensional sheet approximation to the vertical distribution of
stars in a disk, one can show that the vertical velocity dispersion,
$\sigma_z$, should be given by an equation of the form
\begin{equation}
\sigma_z^2=2\pi G\Sigma z_0 \label{eqs1} 
\end{equation}
where $z_0$ is the scale height and $\Sigma$ the surface density of
the disk \citep[p.727]{binney98}.  If the surface density of stars
drops off exponentially with radius with a scale-length $R_d$, and the
scale-height stays constant with radius, then $\sigma_z$ should drop
exponentially to zero with a scale-length of $2R_d$.  If the velocity
ellipsoid remains roughly constant in shape, then the other components,
including the observed $\phi$ component, should decrease in the same
way.  As the solid line in Figure~\ref{vdisp} shows, this is not what we
find in M31.

\begin{figure}
  \includegraphics[width=0.475\textwidth]{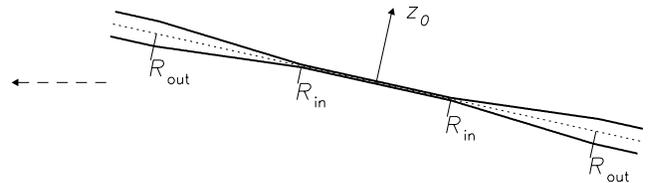}
  \caption{Schematic diagram of the flaring disk of M31. The disk
  scale height is constant out to a radius $R_{\rm in}$ it then
  increases linearly with radius to $R_{\rm out}$, at which point it
  becomes constant again.  } \label{flaremod}
\end{figure}
So, where have we gone wrong?  Well, it is known that the outer parts
of M31's disk are not a simple constant-thickness flat sheet.  The
outer \hi\ disk is significantly warped \citep{brinks84a}, and
\citet{brinks84b} have suggested that the thickness of the \hi\ layer
also flares to greater scale-heights at large radii.  Their model for
the flaring, shown schematically in Figure~\ref{flaremod}, assumes
that the \hi\ scale-height is constant at $z_{\rm in}=0\fdg01$
(136~pc) out to $1\fdg33$ (18~kpc), then increases linearly out to
$R_{\rm out}= 2\fdg54$ (34.5~kpc) where $z_{\rm out}= 0\fdg14$
(1.9~kpc).  Beyond this radius the \hi\ becomes undetectable, but
\citet{ibata05} have assumed the disk scale height becomes constant
from this point onwards to account for their stellar observations.

If the scale-height increases in this way, then equation~(\ref{eqs1})
implies that, for a given mass density, the vertical velocity
dispersion also increases.  If we insert the above prescription for
the scale-height into this equation, but continue to assume that the
disk density drops exponentially with radius, then we find the
velocity dispersion profile shown as a dotted line in
Figure~\ref{vdisp}.  Although a step in the right direction, the fit to
the observed profile is still not very good.  Of course, there is no
reason why the stellar disk, having been built up over an extended
period and undergone secular heating since it formed, should flare in
the same way as the \hi\ distribution.  Indeed, it is often assumed
that the stellar scale-heights of disk galaxies remain constant with
radius.  However, recent analysis has shown that existing photometric
data do not rule out significant variation in scale-height with radius
\citep{narayan02}.  Further, if a disk comprises two components of
different constant scale-heights, with the ``thick disk'' having a
longer scale-length than the ``thin disk,'' then one would expect a
variation with radius in their combined effective scale-height similar
to this simple flaring model.  Whether such components should be
combined or should be considered as discrete entities is largely a
matter of semantics unless one can establish some other feature that
distinguishes their constituent stars.  If we use this toy
varying-scale-height model, but move the point at which the flaring
starts to $R_{\rm in}= 0\fdg85$ (11.6~kpc) while leaving all the other
parameters unaltered, we obtain the dashed line in Figure~\ref{vdisp},
in respectable agreement with the data.  Clearly, by altering the way
in which the layer flares, or by varying the mass-to-light ratio of
the disk with radius, we could improve the fit still further, but such
detailed modelling goes beyond what is justified by the simplifying
approximations of a one-dimensional sheet of stars in hydrostatic
equilibrium.

\subsubsection{Variation with magnitude}\label{Svdispmag}
This analysis of the velocity dispersion allows us to make a further
search for any systematic variation in the properties of PNe with
luminosity.  Such a test is motivated by knowledge of the Solar
neighbourhood of the Milky Way, where we know that the tangential
components of the velocity dispersions of different types of star vary
by more than a factor of two, with the youngest, most massive stars
having the lowest dispersions \citep{binney98}.  If the PNe in M31
that populate different parts of the PNLF had progenitor stars with
very different masses, then we might expect similar variations in
their velocity dispersions.  Not only would such a variation fit with
the suggestion of \citet{marigo04} that the bright end of the PNLF is
populated by more massive stars, but it would also resonate with the
recent claim by \citet{sambhus06} that the PNe in NGC~4697 can be
divided on the basis of their luminosities into two discrete
populations with distinct kinematics.

\begin{figure}
  \includegraphics[width=0.475\textwidth]{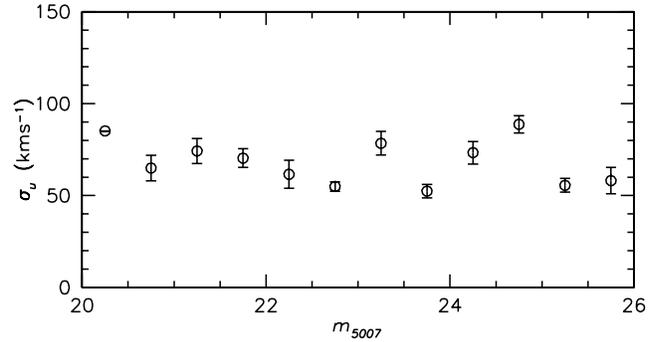}
  \caption{Variation in velocity dispersion with apparent magnitude
  for the major-axis data from the disk region at $X > 0.3\degr$.}
  \label{dispmag}
\end{figure}
To test for such an effect, we have extracted from the major-axis
subsample those objects for which $X > 0.3\degr$: since the global
velocity dispersion is approximately constant at these radii (see
Figure~\ref{vdisp}), we can combine data across this range to form a
meaningful average.  Figure~\ref{dispmag} shows the velocity
dispersions determined from these data combined into half-magnitude
bins.  Clearly, there is no evidence for an decrease in dispersion for
brightest objects, adding further weight to the conclusion that PNe of
all magnitudes form a relatively homogeneous population of old stars.

\subsection{Asymmetric drift}\label{Sasymd}

The presence of significant random velocities at all radii means that
the support against gravitational collapse does not come entirely from
circular motions even at large distances from the centre of M31.  We
would therefore expect the observed mean rotational streaming of the
stars, $\overline{v}_\phi$, to lag behind the circular speed, $v_c$.
As noted above, such an asymmetric drift is seen, but having
measured the velocity dispersion profile we can now quantify whether
the predicted velocity difference is dynamically consistent with the
observations. 

The relationship between $\overline{v}_\phi$ and $v_c$ at any radius
$R$ is given by the appropriate axisymmetric Jeans equation
\citep[equation 4-33]{binney87}, which we can rewrite as
\begin{equation}\label{eqad1}
\overline{v}_\phi = \sqrt{ v_c^2 -\sigma_\phi^2 + \overline{v_R^2} +
  \frac{R}{\nu} \frac{\partial ( \nu \overline{v_R^2} )}
  {\partial R} + R \frac{\partial ( \overline{v_R v_z})}
  {\partial z} }, 
\end{equation}
where we have adopted a cylindrical coordinate system, $\{R,\phi,z\}$,
and measure stellar velocities $\{v_R,v_\phi,v_z\}$ and number density
$\nu(R, z)$ in this reference frame.  To solve this equation, we have
to make some further assumptions.  First, the final term, which arises
from any tilt in the velocity ellipsoid, is unobservable. However, for
all plausible dynamical models it remains small and has little impact
on an analysis of this kind \citep{gerssen97}, so we can safely
neglect it. Second, we must assume something about the shape of the
velocity ellipsoid of random motions.  The simplest assumption is that
it remains constant with position, so that we can write
\begin{equation}\label{eqalpha}
\overline{v_R^2} = \alpha\sigma_\phi^2,
\end{equation}
where $\alpha$ is a constant. In the solar neighbourhood of the Milky
Way, $\alpha$ is found to be $\sim 2$ \citep{binney98}, and in the
absence of any other constraint we set it to this value at all radii
in M31 (varying $\alpha$ by $\pm0.5$ makes little difference to
  the final result). If we make this approximation, use the \hi\ data
to obtain a 
simple linear approximation to the rotation curve $v_c(R)$, take the
number counts of PNe in Section~\ref{Sscale} to give the exponential
disk distribution for $\nu(R)$, and use the smooth approximation to
$\sigma_\phi(R)$ of equation~(\ref{eqdispfit}), then we can substitute
these quantities into equation~(\ref{eqad1}) to predict the value of
$\overline{v}_\phi$.

To compare this value to the observed mean streaming of stars, we must
make a couple of corrections to the model.  The first is due to the
inclination of the galaxy, but M31 is sufficiently close to edge-on
that this effect is relatively minor and has already been accounted
for in the data of Figure~\ref{rot}.  The second effect is more
significant: although we have selected a sample of PNe whose positions
lie close to the major axis of M31, the cut at $|Y|<0\fdg04$ does
sample a range of azimuths, particularly at small radii.  We have
allowed for this effect by averaging the line-of-sight component of
$\overline{v}_\phi$ in the model over the range of azimuths admitted
at each position $X$ along the major axis.

\begin{figure}
  \includegraphics[width=0.475\textwidth]{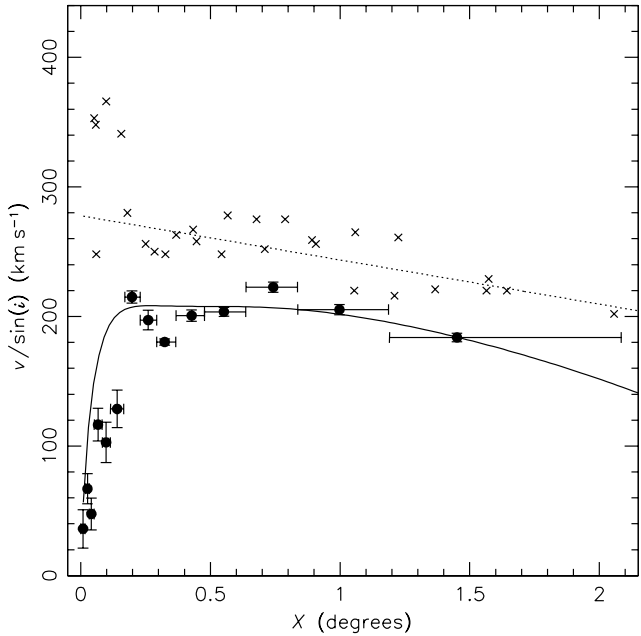}
  \caption{Predicted and observed rotational motions of PNe in M31.
  The filled circles show the observed streaming of stars, as in
  Figure~\ref{rot}.  The crosses give M31's \hi\ rotation curve as
  derived by \citet{braun91}, with the straight line fit to the data
  used in this analysis shown by the dotted line. The solid line shows
  the predicted rotational velocity obtained by solving
  equation~(\ref{eqad1}).} \label{asymdrift}
\end{figure}
The net result of this calculation is shown in Figure~\ref{asymdrift}.
At small radii, the detailed match between model and observation is
not perfect, but discrepancies here are not surprising since many of
the PNe in this region will belong to the bulge rather than the disk
as modelled here; even PNe from the disk will have their kinematics
affected by the bar, which also has not been accounted for in this
simple axisymmetric picture.  At larger radii, however, the agreement
is remarkably good.  It should be stressed that this model has not
been fitted to the data: the amplitude of the asymmetric drift between
$\overline{v}_\phi$ and $v_c$ is entirely fixed by the Jeans equation
once we have determined the other quantities in this equation as
described.  Note also that although we have constructed this dynamical
model by assuming the scale-height of M31's disk population increases
with radius, the absence of $\sigma_z$ in equation~(\ref{eqad1}) means
that the agreement is not dependent on this specific model.  All that
is required is that the radial and tangential kinematics are coupled
via equation~(\ref{eqalpha}); if $\sigma_z$ is not coupled to these
motions, then any scale-height can be accommodated.  The generic
agreement between model and data therefore provides a healthy
consistency check on this analysis, and leaves us with a relatively
simple self-consistent picture of the dynamics of M31's surprisingly
warm disk.

\section{Conclusions}\label{Sconc}

Using the Planetary Nebula Spectrograph, we have surveyed a wide area
over the Andromeda Galaxy. Some 3300 emission-line objects have
been detected and their velocities measured.  Cross-checks against
other smaller data sets have confirmed the accuracy of these
measurements, and we have also analysed the detected objects in order
to separate the PNe from other emission-line objects.  Although this
paper is primarily intended just to present the data, we can draw a
number of intriguing conclusions from even the initial analysis of these
observations.  Specifically:
\begin{itemize}
\item A number of M31's satellites are detected in the survey, but the
  existence of Andromeda~VIII is brought into question by this larger
  data set.  
\item The luminosity function is well reproduced by the usual simple
  functional form, with no strong evidence for spatial variations that
  cannot be attributed to modest amounts of obscuration.  It has been
  suggested that the bright end of the luminosity function is
  dominated by relatively massive stars, so one might expect the PNLF
  to vary significantly between the spheroid and disk components.  The
  absence of such effects suggests that all PNe in M31 arise from old
  low-mass stars. 
\item The number counts of PNe matches that of the R-band photometry,
  further indicating that both arise from the same old population.
  With these dynamically-selected number counts of discrete objects,
  we can also go much deeper than the conventional photometry of M31.
  There is no evidence for a cut-off in its exponential disk
  population to beyond four scale-lengths, and there are no signs of a
  halo population that is distinct from the bulge out to ten effective
  bulge radii.
\item Somewhat unexpectedly, the velocity dispersion of M31 does not
  drop to zero at large radii, and there is a significant asymmetric
  drift between the rotational motion and the local circular speed at
  all radii.  These data can be self-consistently reproduced by a
  simple disk model which is partially supported by random motions
  even at large radii.  If the shape of the velocity ellipsoid remains
  approximately constant, then M31's disk must increase in scale-height
  at large radii or its mass-to-light ratio must increase significantly.  
\item There is no evidence for any systematic variation in disk PN
  velocity dispersion with magnitude, again suggesting that PNe of all
  luminosities arise from a relatively homogeneous old stellar
  population.  
\end{itemize}

Clearly, this quick pass through the database has only scratched the
surface of what can be done with a kinematic survey of this size.  We
look forward to more detailed dynamical modelling of the full
three-dimensional $\{X,Y,v\}$ phase-space data set, as well as tagging
the PNe with extra dimensions such as their magnitudes and line
ratios.  Such analysis will allow us to dig deeper into this rich
parameter space in order to further understand the dynamical
properties of our nearest large neighbour and its planetary nebula
population.


\section*{Acknowledgements}
This research is based on data obtained using the William Herschel
Telescope operated by the Isaac Newton Group in La Palma; the support
and advice of the ING staff are gratefully acknowledged. We would like
to thank Phil Massey and the Local Group Survey team for their
assistance.  MRM is currently supported by a PPARC Senior fellowship,
for which he is most grateful. We would also like to thank the referee
for a timely reading of the paper and a number of helpful
suggestions.  

\bibliographystyle{mn2e}

\begin{thebibliography}{}

\bibitem[\protect\citeauthoryear{{Bender}, {Paquet} \& {Nieto}}{{Bender}
  et~al.}{1991}]{bender91}
{Bender} R.,  {Paquet} A.,    {Nieto} J.-L.,  1991, \aap, 246, 349

\bibitem[\protect\citeauthoryear{{Bertin} \& {Arnouts}}{{Bertin} \&
  {Arnouts}}{1996}]{bertin96}
{Bertin} E.,  {Arnouts} S.,  1996, \aaps, 117, 393

\bibitem[\protect\citeauthoryear{{Binney} \& {Merrifield}}{{Binney} \&
  {Merrifield}}{1998}]{binney98}
{Binney} J.,  {Merrifield} M.,  1998, {Galactic astronomy}.
Princeton, NJ : Princeton University Press, 1998.

\bibitem[\protect\citeauthoryear{{Binney} \& {Tremaine}}{{Binney} \&
  {Tremaine}}{1987}]{binney87}
{Binney} J.,  {Tremaine} S.,  1987, {Galactic dynamics}.
Princeton, NJ, Princeton University Press, 1987.

\bibitem[\protect\citeauthoryear{{Braun}}{{Braun}}{1991}]{braun91}
{Braun} R.,  1991, \apj, 372, 54

\bibitem[\protect\citeauthoryear{{Brinks} \& {Burton}}{{Brinks} \&
  {Burton}}{1984}]{brinks84b}
{Brinks} E.,  {Burton} W.~B.,  1984, \aap, 141, 195

\bibitem[\protect\citeauthoryear{{Brinks} \& {Shane}}{{Brinks} \&
  {Shane}}{1984}]{brinks84a}
{Brinks} E.,  {Shane} W.~W.,  1984, \aaps, 55, 179

\bibitem[\protect\citeauthoryear{{Cardelli}, {Clayton} \& {Mathis}}{{Cardelli}
  et~al.}{1989}]{cardelli89}
{Cardelli} J.~A.,  {Clayton} G.~C.,    {Mathis} J.~S.,  1989, \apj, 345, 245

\bibitem[\protect\citeauthoryear{{Carter} \& {Sadler}}{{Carter} \&
  {Sadler}}{1990}]{carter90}
{Carter} D.,  {Sadler} E.~M.,  1990, \mnras, 245, 12P

\bibitem[\protect\citeauthoryear{{Ciardullo}, {Durrell}, {Laychak}, {Herrmann},
  {Moody}, {Jacoby} \& {Feldmeier}}{{Ciardullo} et~al.}{2004}]{ciardullo04}
{Ciardullo} R.,  {Durrell} P.~R.,  {Laychak} M.~B.,  {Herrmann} K.~A.,  {Moody}
  K.,  {Jacoby} G.~H.,    {Feldmeier} J.~J.,  2004, \apj, 614, 167

\bibitem[\protect\citeauthoryear{{Ciardullo}, {Feldmeier}, {Jacoby}, {Kuzio de
  Naray}, {Laychak} \& {Durrell}}{{Ciardullo} et~al.}{2002}]{ciardullo02}
{Ciardullo} R.,  {Feldmeier} J.~J.,  {Jacoby} G.~H.,  {Kuzio de Naray} R.,
  {Laychak} M.~B.,    {Durrell} P.~R.,  2002, \apj, 577, 31

\bibitem[\protect\citeauthoryear{{Ciardullo}, {Jacoby}, {Ford} \&
  {Neill}}{{Ciardullo} et~al.}{1989}]{ciardullo89}
{Ciardullo} R.,  {Jacoby} G.~H.,  {Ford} H.~C.,    {Neill} J.~D.,  1989, \apj,
  339, 53

\bibitem[\protect\citeauthoryear{{Cox}}{{Cox}}{2000}]{cox00}
{Cox} A.~N.,  2000, {Allen's astrophysical quantities}.
Allen's astrophysical quantities, 4th ed.~Publisher: New York: AIP Press;
  Springer, 2000.~Edited by Arthur N.~Cox.~ ISBN: 0387987460

\bibitem[\protect\citeauthoryear{{de Jong} et~al.,}{{de Jong}
  et~al.}{2005}]{dejong05}
{de Jong} J.~T.~A.,  et~al., 2005, \aap, 446, 855

\bibitem[\protect\citeauthoryear{{de Vaucouleurs}}{{de
  Vaucouleurs}}{1958}]{devauc58}
{de Vaucouleurs} G.,  1958, \apj, 128, 465

\bibitem[\protect\citeauthoryear{{de Vaucouleurs}, {de Vaucouleurs}, {Corwin},
  {Buta}, {Paturel} \& {Fouque}}{{de Vaucouleurs} et~al.}{1991}]{devauc91}
{de Vaucouleurs} G.,  {de Vaucouleurs} A.,  {Corwin} H.~G.,  {Buta} R.~J.,
  {Paturel} G.,    {Fouque} P.,  1991, {Third Reference Catalogue of Bright
  Galaxies}.
Springer-Verlag Berlin Heidelberg New York

\bibitem[\protect\citeauthoryear{{Dehnen} \& {Binney}}{{Dehnen} \&
  {Binney}}{1998}]{dehnen98}
{Dehnen} W.,  {Binney} J.~J.,  1998, \mnras, 298, 387

\bibitem[\protect\citeauthoryear{{Devereux}, {Price}, {Wells} \&
  {Duric}}{{Devereux} et~al.}{1994}]{devereux94}
{Devereux} N.~A.,  {Price} R.,  {Wells} L.~A.,    {Duric} N.,  1994, \aj, 108,
  1667

\bibitem[\protect\citeauthoryear{{Douglas} et~al.,}{{Douglas}
  et~al.}{2002}]{douglas02}
{Douglas} N.~G.,  et~al., 2002, \pasp, 114, 1234

\bibitem[\protect\citeauthoryear{{Ferguson}, {Gallagher} \& {Wyse}}{{Ferguson}
  et~al.}{2000}]{ferguson00}
{Ferguson} A.~M.~N.,  {Gallagher} J.~S.,    {Wyse} R.~F.~G.,  2000, \aj, 120,
  821

\bibitem[\protect\citeauthoryear{{Ferguson}, {Irwin}, {Ibata}, {Lewis} \&
  {Tanvir}}{{Ferguson} et~al.}{2002}]{ferguson02}
{Ferguson} A.~M.~N.,  {Irwin} M.~J.,  {Ibata} R.~A.,  {Lewis} G.~F.,
  {Tanvir} N.~R.,  2002, \aj, 124, 1452

\bibitem[\protect\citeauthoryear{{Gerssen}, {Kuijken} \&
  {Merrifield}}{{Gerssen} et~al.}{1997}]{gerssen97}
{Gerssen} J.,  {Kuijken} K.,    {Merrifield} M.~R.,  1997, \mnras, 288, 618

\bibitem[\protect\citeauthoryear{{G{\'o}rny} \& {Stasi{\'n}ska}}{{G{\'o}rny} \&
  {Stasi{\'n}ska}}{1995}]{gorny95}
{G{\'o}rny} S.~K.,  {Stasi{\'n}ska} G.,  1995, \aap, 303, 893

\bibitem[\protect\citeauthoryear{{Halliday} et~al.,}{{Halliday}
  et~al.}{2006}]{halliday06}
{Halliday} C.,  et~al., 2006, Submitted to MNRAS

\bibitem[\protect\citeauthoryear{{Huchra}, {Brodie} \& {Kent}}{{Huchra}
  et~al.}{1991}]{huchra91}
{Huchra} J.~P.,  {Brodie} J.~P.,    {Kent} S.~M.,  1991, \apj, 370, 495

\bibitem[\protect\citeauthoryear{{Huchra}, {Vogeley} \& {Geller}}{{Huchra}
  et~al.}{1999}]{huchra99}
{Huchra} J.~P.,  {Vogeley} M.~S.,    {Geller} M.~J.,  1999, \apjs, 121, 287

\bibitem[\protect\citeauthoryear{{Hui}, {Ford}, {Ciardullo} \& {Jacoby}}{{Hui}
  et~al.}{1993}]{hui93}
{Hui} X.,  {Ford} H.~C.,  {Ciardullo} R.,    {Jacoby} G.~H.,  1993, \apj, 414,
  463

\bibitem[\protect\citeauthoryear{{Hurley-Keller}, {Morrison}, {Harding} \&
  {Jacoby}}{{Hurley-Keller} et~al.}{2004}]{hurleyk04}
{Hurley-Keller} D.,  {Morrison} H.~L.,  {Harding} P.,    {Jacoby} G.~H.,  2004,
  \apj, 616, 804

\bibitem[\protect\citeauthoryear{{Ibata}, {Chapman}, {Ferguson}, {Irwin},
  {Lewis} \& {McConnachie}}{{Ibata} et~al.}{2004}]{ibata04}
{Ibata} R.,  {Chapman} S.,  {Ferguson} A.~M.~N.,  {Irwin} M.,  {Lewis} G.,
  {McConnachie} A.,  2004, \mnras, 351, 117

\bibitem[\protect\citeauthoryear{{Ibata}, {Chapman}, {Ferguson}, {Lewis},
  {Irwin} \& {Tanvir}}{{Ibata} et~al.}{2005}]{ibata05}
{Ibata} R.,  {Chapman} S.,  {Ferguson} A.~M.~N.,  {Lewis} G.,  {Irwin} M.,
  {Tanvir} N.,  2005, \apj, 634, 287

\bibitem[\protect\citeauthoryear{{Ibata}, {Irwin}, {Lewis}, {Ferguson} \&
  {Tanvir}}{{Ibata} et~al.}{2001}]{ibata01}
{Ibata} R.,  {Irwin} M.,  {Lewis} G.,  {Ferguson} A.~M.~N.,    {Tanvir} N.,
  2001, \nat, 412, 49

\bibitem[\protect\citeauthoryear{{Irwin}, {Ferguson}, {Ibata}, {Lewis} \&
  {Tanvir}}{{Irwin} et~al.}{2005}]{irwin05}
{Irwin} M.~J.,  {Ferguson} A.~M.~N.,  {Ibata} R.~A.,  {Lewis} G.~F.,
  {Tanvir} N.~R.,  2005, \apjl, 628, L105

\bibitem[\protect\citeauthoryear{{Jacoby}}{{Jacoby}}{1989}]{jacoby89}
{Jacoby} G.~H.,  1989, \apj, 339, 39

\bibitem[\protect\citeauthoryear{{Jacoby} \& {De Marco}}{{Jacoby} \& {De
  Marco}}{2002}]{jacoby02}
{Jacoby} G.~H.,  {De Marco} O.,  2002, \aj, 123, 269

\bibitem[\protect\citeauthoryear{{Kalirai} et~al.,}{{Kalirai}
    et~al.}{2005}]{kalirai06}
  {Kalirai} J.~S., {Guhathakurta} P., {Gilbert} K.~M., {Reitzel} D.~B.,
  {Majewski} S.~R., {Rich} R.~M., {Cooper} M.~C., 2005.
  arXiv:astro-ph/0512161

\bibitem[\protect\citeauthoryear{{Kent}}{{Kent}}{1989}]{kent89a}
{Kent} S.~M.,  1989, \pasp, 101, 489

\bibitem[\protect\citeauthoryear{{Loinard}, {Allen} \& {Lequeux}}{{Loinard}
  et~al.}{1995}]{loinard95}
{Loinard} L.,  {Allen} R.~J.,    {Lequeux} J.,  1995, \aap, 301, 68

\bibitem[\protect\citeauthoryear{{Magrini}, {Corradi}, {Mampaso} \&
  {Perinotto}}{{Magrini} et~al.}{2000}]{magrini00}
{Magrini} L.,  {Corradi} R.~L.~M.,  {Mampaso} A.,    {Perinotto} M.,  2000,
  \aap, 355, 713

\bibitem[\protect\citeauthoryear{{Marigo}, {Girardi}, {Weiss}, {Groenewegen} \&
  {Chiosi}}{{Marigo} et~al.}{2004}]{marigo04}
{Marigo} P.,  {Girardi} L.,  {Weiss} A.,  {Groenewegen} M.~A.~T.,    {Chiosi}
  C.,  2004, \aap, 423, 995

\bibitem[\protect\citeauthoryear{{Massey}, {Hodge}, {Holmes}, {Jacoby}, {King},
  {Olsen}, {Smith} \& {Saha}}{{Massey} et~al.}{2002}]{massey02}
{Massey} P.,  {Hodge} P.~W.,  {Holmes} S.,  {Jacoby} J.,  {King} N.~L.,
  {Olsen} K.,  {Smith} C.,    {Saha} A.,  2002, Bulletin of the American
  Astronomical Society, 34, 1272

\bibitem[\protect\citeauthoryear{{Mateo}}{{Mateo}}{1998}]{mateo98}
{Mateo} M.~L.,  1998, \araa, 36, 435

\bibitem[\protect\citeauthoryear{{McConnachie}, {Irwin}, {Ferguson}, {Ibata},
  {Lewis} \& {Tanvir}}{{McConnachie} et~al.}{2005}]{mcconnachie05}
{McConnachie} A.~W.,  {Irwin} M.~J.,  {Ferguson} A.~M.~N.,  {Ibata} R.~A.,
  {Lewis} G.~F.,    {Tanvir} N.,  2005, \mnras, 356, 979

\bibitem[\protect\citeauthoryear{{McConnachie}, {Irwin}, {Ibata}, {Ferguson},
  {Lewis} \& {Tanvir}}{{McConnachie} et~al.}{2003}]{mcconnachie03}
{McConnachie} A.~W.,  {Irwin} M.~J.,  {Ibata} R.~A.,  {Ferguson} A.~M.~N.,
  {Lewis} G.~F.,    {Tanvir} N.,  2003, \mnras, 343, 1335

\bibitem[\protect\citeauthoryear{{McElroy}}{{McElroy}}{1983}]{mcelroy83}
{McElroy} D.~B.,  1983, \apj, 270, 485

\bibitem[\protect\citeauthoryear{{Merrett} et~al.,}{{Merrett}
  et~al.}{2003}]{merrett03}
{Merrett} H.~R.,  et~al., 2003, \mnras, 346, L62

\bibitem[\protect\citeauthoryear{{Meyssonnier}, {Lequeux} \&
  {Azzopardi}}{{Meyssonnier} et~al.}{1993}]{meyssonnier93}
{Meyssonnier} N.,  {Lequeux} J.,    {Azzopardi} M.,  1993, \aaps, 102, 251

\bibitem[\protect\citeauthoryear{{Monet} et~al.,}{{Monet}
  et~al.}{2003}]{monet03}
{Monet} D.~G.,  et~al., 2003, \aj, 125, 984

\bibitem[\protect\citeauthoryear{{Morrison}, {Harding}, {Hurley-Keller} \&
  {Jacoby}}{{Morrison} et~al.}{2003}]{morrison03}
{Morrison} H.~L.,  {Harding} P.,  {Hurley-Keller} D.,    {Jacoby} G.,  2003,
  \apjl, 596, L183

\bibitem[\protect\citeauthoryear{{Narayan} \& {Jog}}{{Narayan} \&
  {Jog}}{2002}]{narayan02}
{Narayan} C.~A.,  {Jog} C.~J.,  2002, \aap, 390, L35

\bibitem[\protect\citeauthoryear{{Nolthenius} \& {Ford}}{{Nolthenius} \&
  {Ford}}{1984}]{nolthenius84}
{Nolthenius} R.~A.,  {Ford} H.~C.,  1984, \baas, 16, 456

\bibitem[\protect\citeauthoryear{{Oke}}{{Oke}}{1974}]{oke74}
{Oke} J.~B.,  1974, \apjs, 27, 21

\bibitem[\protect\citeauthoryear{{Oke}}{{Oke}}{1990}]{oke90}
{Oke} J.~B.,  1990, \aj, 99, 1621

\bibitem[\protect\citeauthoryear{{Oke} \& {Gunn}}{{Oke} \&
  {Gunn}}{1983}]{oke83}
{Oke} J.~B.,  {Gunn} J.~E.,  1983, \apj, 266, 713

\bibitem[\protect\citeauthoryear{{Peimbert}}{{Peimbert}}{1990}]{peimbert90}
{Peimbert} M.,  1990, Revista Mexicana de Astronomia y Astrofisica, 20, 119

\bibitem[\protect\citeauthoryear{{Reitzel} \& {Guhathakurta}}{{Reitzel} \&
  {Guhathakurta}}{2002}]{reitzel02}
{Reitzel} D.~B.,  {Guhathakurta} P.,  2002, \aj, 124, 234

\bibitem[\protect\citeauthoryear{{Sambhus}, {Gerhard} \&
  {M{\'e}ndez}}{{Sambhus} et~al.}{2006}]{sambhus06}
{Sambhus} N.,  {Gerhard} O.,    {M{\'e}ndez} R.~H.,  2006, \aj, 131, 837

\bibitem[\protect\citeauthoryear{{Schlegel}, {Finkbeiner} \&
  {Davis}}{{Schlegel} et~al.}{1998}]{schlegel98}
{Schlegel} D.~J.,  {Finkbeiner} D.~P.,    {Davis} M.,  1998, \apj, 500, 525

\bibitem[\protect\citeauthoryear{{Sevenster}, {Chapman}, {Habing}, {Killeen} \&
  {Lindqvist}}{{Sevenster} et~al.}{1997}]{sevenster97}
{Sevenster} M.~N.,  {Chapman} J.~M.,  {Habing} H.~J.,  {Killeen} N.~E.~B.,
  {Lindqvist} M.,  1997, \aaps, 122, 79

\bibitem[\protect\citeauthoryear{{Simien} \& {Prugniel}}{{Simien} \&
  {Prugniel}}{2002}]{simien02}
{Simien} F.,  {Prugniel} P.,  2002, \aap, 384, 371

\bibitem[\protect\citeauthoryear{{van Dokkum}}{{van
  Dokkum}}{2001}]{vandokkum01}
{van Dokkum} P.~G.,  2001, \pasp, 113, 1420

\bibitem[\protect\citeauthoryear{{Walterbos} \& {Kennicutt}}{{Walterbos} \&
  {Kennicutt}}{1987}]{walterbos87a}
{Walterbos} R.~A.~M.,  {Kennicutt} R.~C.,  1987, \aaps, 69, 311

\bibitem[\protect\citeauthoryear{{Walterbos} \& {Kennicutt}}{{Walterbos} \&
  {Kennicutt}}{1988}]{walterbos88}
{Walterbos} R.~A.~M.,  {Kennicutt} R.~C.,  1988, \aap, 198, 61

\end{thebibliography}

\label{lastpage}

\end{document}